\documentclass[english,english,aps,prd,amsmath,amssymb,superscriptaddress]{revtex4}
\usepackage[T1]{fontenc}
\usepackage[latin9]{inputenc}
\usepackage{amsmath}
\usepackage{amssymb}
\usepackage{graphicx}
\usepackage{esint}

\makeatletter

\providecommand{\tabularnewline}{\\}

\@ifundefined{textcolor}{}
{%
 \definecolor{BLACK}{gray}{0}
 \definecolor{WHITE}{gray}{1}
 \definecolor{RED}{rgb}{1,0,0}
 \definecolor{GREEN}{rgb}{0,1,0}
 \definecolor{BLUE}{rgb}{0,0,1}
 \definecolor{CYAN}{cmyk}{1,0,0,0}
 \definecolor{MAGENTA}{cmyk}{0,1,0,0}
 \definecolor{YELLOW}{cmyk}{0,0,1,0}
 }

\usepackage{slashed}
\usepackage{babel}
\usepackage{bbm}

\makeatother

\usepackage{babel}
\begin{document}

\title{Some recent progress on quark pairings in dense quark and nuclear
matter}

\author{Jin-cheng Wang}

\affiliation{Department of Modern Physics, University of Science and Technology
of China, Anhui 230026, People's Republic of China}

\author{Jin-yi Pang}

\affiliation{Department of Modern Physics, University of Science and Technology
of China, Anhui 230026, People's Republic of China}

\author{Qun Wang}

\affiliation{Department of Modern Physics, University of Science and Technology
of China, Anhui 230026, People's Republic of China}
\begin{abstract}
In this review article we give a brief overview on some recent progress
in quark pairings in dense quark/nuclear matter mostly developed in
the past five years. We focus on following aspects in particular:
the BCS-BEC crossover in the CSC phase, the baryon formation and dissociation
in dense quark/nuclear matter, the Ginzburg-Landau theory for three-flavor
dense matter with $U_{A}$(1) anomaly, and the collective and Nambu-Goldstone
modes for the spin-one CSC. 
\end{abstract}
\maketitle

\section{\textup{\normalsize Introduction}}

It is well known that fermion pairing is the underlying mechanism
for superconductivity and superfluidity. In a fermionic system, the
weak and attractive interaction between two fermions leads to the
formation of Cooper pairs at low temperatures, which is well described
by the Bardeen-Cooper-Schrieffer (BCS) theory. In strong interaction
quarks and gluons are elementary particles which are described by
quantum chromodynamics (QCD). There are also quark pairings called
\emph{color superconductivity} (CSC) first proposed in cold dense
quark matter by Barrois \cite{Barrois:1977xd}, Bailin and Love \cite{Bailin:1983bm},
and further developed by many others (for reviews, see \cite{Rajagopal:2000wf,Alford:2001dt,Rischke:2003mt,Casalbuoni:2003wh,Schafer:2003vz,Buballa:2003qv,Ren:2004nn,Shovkovy:2004me,Huang:2004ik,Alford:2007xm,Wang:2009xf}).
However color superconducting quark matter may not be relevant in
collisional experiments in the foreseeable future. The reason is that
this exotic phase of matter requires extremely high baryonic densities
and relatively low temperatures. In nature, such conditions may be
realized in the cores of cold compact stars, i.e. in relatively old
remnants of supernova explosions (for an introduction of compact star,
see e.g. \cite{Schmitt:2010pn}). Among some major developments on
quark pairings in the past decade are comprehensive studies of important
CSC phases on the phase diagram \cite{Fukushima:2004zq,Ruester:2005jc,Ruester:2005ib,Blaschke:2005uj,Sedrakian:2009kb,Huang:2010bf}:
the color-flavor-locking (CFL) state \cite{Alford:1998mk,Rajagopal:2000ff},
the two-flavor (2SC) state \cite{Son:1998uk,Pisarski:1999tv,Hong:1999fh,Wang:2001aq,Schmitt:2002sc,Nickel:2006vf},
the single flavor pairing state \cite{Schmitt:2002sc,Schafer:2000tw,Alford:2002rz,Buballa:2002wy,Schmitt:2003xq,Schmitt:2004et,Aguilera:2005tg,Marhauser:2006hy,Brauner:2008ma,Pang:2010wk,Schmitt:2005ee,Schmitt:2005wg,Feng:2008dh,Blaschke:2008br,Feng:2009vt,Brauner:2009df},
mis-matched pairings \cite{Rajagopal:2000ff,Alford:2000ze,Alford:2003fq,Shovkovy:2003uu,Huang:2004bg}
and many other phases \cite{Schafer:2000ew,Miransky:2001tw}. The
construction of a theory for the CSC at weak couplings is another
important progress \cite{Son:1998uk,Pisarski:1999tv,Hong:1999fh,Wang:2001aq,Schmitt:2002sc,Brown:1999aq,Brown:1999yd,Brown:2000eh,Schmitt:2003xq,Hou:2004bn,Giannakis:2004xt}. 

In this review article we will give a brief overview on some recent
progress in quark pairings in dense quark/nuclear matter mostly developed
in the past five years. We will focus on following aspects in particular:
the BCS-BEC crossover in the CSC phase, the baryon formation and dissociation
in dense quark/nuclear matter, the Ginzburg-Landau theory for three-flavor
dense matter with $U_{A}$(1) anomaly, and the collective and Nambu-Goldstone
modes for the spin-one CSC. Random phase approximation and Dyson-Schwinger
equation are used to obtain the propagating modes of diquark pairings.
The BCS-BEC crossover are investigated within the boson-fermion model
and the NJL-type model. The baryon formation and dissociation in different
phases are studied by regarding a baryon as composed of a quark and
a diquark in the NJL-type model. With a nonlocal extended NJL model
one can obtain the constituent quark mass which is momentum dependent.
The bubble diagrams are automatically convergent, providing an effective
confinement mechanism. In a three-flavor NJL model including the axial
anomaly, a low temperature critical point is found due to the coupling
between chiral and diquark condensates. The collective modes in the
spin-1 CSC are analyzed in the Ginzburg-Landau approach.

\section{BEC-BCS crossover in a relativistic boson-fermion model}

\label{sec:bose-ferm}The Cooper pairs are formed by two fermions
at the Fermi surface in weakly attractive channel, therefore they
have spatial extension called the coherence length which is much larger
than the mean inter-particle distance. In a strongly coupling regime,
the Cooper pairs are bound to bosonic molecules and condense in the
ground state to form the so-called Bose-Einstein condensation (BEC).
The Fermi surface disappears and no fermion degree of freedom remains
in this situation. Although the features of BEC and BCS are very different,
there is no phase transition but just a crossover between them. Recently
many experimental advances have been made in cold atom system. With
the help of an applied magnetic field, the effective attractive interaction
between the atoms can be tuned via Feshbach resonances and the BCS-BEC
crossover can be observed. 

The BEC-BCS crossover in the CSC was first studied by Nishida and
Abuki \cite{Nishida:2005ds,Abuki:2006dv} with the NJL model and followed
by others \cite{Deng:2006ed,Sun:2007fc,He:2007yj,Kitazawa:2007im,Kitazawa:2007zs,Brauner:2008td,Chatterjee:2008dr,Deng:2008ah,Sedrakian:2011mq}.
Inspired by those previously used in the context of cold fermionic
atoms, one can directly extend the boson-fermion model to a relativistic
version. In this case both the fermion and difermion channels are
included as fundamental degrees of freedom with total fermion number
density fixed. Tuning the effective interaction strength the crossover
between the BEC and BCS can be studied \cite{Deng:2006ed,Deng:2008ah}.
In a non-relativistic system, people normally use the scattering length
to characterize the strength of the attractive interaction between
fermions. In the relativistic boson-fermion model, a crossover parameter
$x$ is defined by the difference between the square of effective
boson mass and boson chemical potential. The inverse of $x$ play
the similar role of scattering length. A varying magnetic field can
also lead to the relativistic BEC-BCS crossover \cite{Wang:2010uj}.
Starting from any initial state at zero field, with a ultrahigh magnetic
field the system always settles into a pure BCS regime. 

For the NJL type model for the BCS-BEC crossover in the CSC phase
\cite{Nishida:2005ds,Abuki:2006dv,Sun:2007fc,He:2007yj}, since the
fundamental degrees of freedom are fermions, the Cooper pairs are
introduced with the help of bosonization procedures like Hubbard-Stratonovich
transformation. Some nonperturbative tools such as random phase approximation
and Dyson-Schwinger equation are needed to determine the propagating
modes of the boson field. By taking the phase shift to the non-relativistic
limit ($p\rightarrow0$), an effective scattering length between fermions
can be derived in relativistic case.

We review the boson-fermion model for the BCS-BEC crossover with the
total fermion number density fixed \cite{Deng:2006ed,Deng:2008ah}.
The Lagrangian respects the global $U(1)$ symmetry, which consists
of free fermion and boson parts, $\mathcal{L}_{f}$ and $\mathcal{L}_{b}$,
and a Yukawa interaction part $\mathcal{L}_{I}$, 
\begin{equation}
\mathcal{L}=\mathcal{L}_{f}+\mathcal{L}_{b}+\mathcal{L}_{I},\label{lagrangian1}
\end{equation}
with 
\begin{eqnarray}
\mathcal{L}_{f} & = & \bar{\psi}(i\gamma^{\mu}\partial_{\mu}+\gamma^{0}\mu-m)\psi,\nonumber \\
\mathcal{L}_{b} & = & |(\partial_{t}-i\mu_{b})\varphi|^{2}-|\nabla\varphi|^{2}-m_{b}^{2}|\varphi|^{2},\nonumber \\
\mathcal{L}_{I} & = & g(\varphi\bar{\psi}_{C}i\gamma^{5}\psi+\varphi^{*}\bar{\psi}i\gamma^{5}\psi_{C}).\label{lagrangian1-1}
\end{eqnarray}
The fermions and bosons degrees of freedom are described by the spinor
$\psi$ and the complex scalar boson field $\varphi$. The charge
conjugate spinors are defined by $\psi_{C}=C\bar{\psi}^{T}$ and $\bar{\psi}_{C}=\psi^{T}C$
with $C=i\gamma^{2}\gamma^{0}$. The fermion/boson mass is denoted
by $m/m_{b}$. The Lagrangian is invariant under the $U(1)_{B}$ transformation
$\psi\rightarrow e^{-i\alpha}\psi$, $\varphi\rightarrow e^{2i\alpha}\varphi$.
Considering a system with the total $U(1)_{B}$ charge conservation,
the boson chemical potential is chosen to be twice the fermion one,
$\mu_{b}=2\mu.$ Therefore, the system is in chemical equilibrium
with respect to the conversion of two fermions into one boson and
vice verse. This allows one to model the transition from weakly-coupled
Cooper pairs made of two fermions into a molecular di-fermionic bound
state, described as a boson. The bosonic field are separated into
two parts as $\varphi=\varphi_{0}+\phi$, where $\varphi_{0}$ is
the expectation value of the bosonic field in vacuum or the zero mode
and $\phi$ is the nonzero mode. The condensation is denoted as $\Delta=2g\varphi_{0}$.
The fermionic field can be rewritten in the Nambu-Gorkov (NG) basis,
$\Psi=\left(\psi,\psi_{C}\right)^{T}$, $\bar{\Psi}=(\bar{\psi},\bar{\psi}_{C})$.
The bosonic field can also be rewritten as $\left(\phi_{R},\phi_{I}\right)^{T}$,
where $\phi_{R}$ and $\phi_{I}$ are the real and imaginary part
of the complex bosonic field respectively. Then the Lagrangian can
be cast into the following form 
\begin{eqnarray}
\mathcal{L} & = & \frac{1}{2}\bar{\Psi}S^{-1}\Psi+\frac{1}{2}(\phi_{R},\phi_{I})D^{-1}\left(\begin{array}{c}
\phi_{R}\\
\phi_{I}
\end{array}\right)+(\phi_{R},\phi_{I})\bar{\Psi}\left(\begin{array}{c}
\hat{\Gamma}_{R}\\
\hat{\Gamma}_{I}
\end{array}\right)\Psi+\frac{(\mu_{b}^{2}-m_{b}^{2})\Delta^{2}}{4g^{2}}.\label{lagrangian2}
\end{eqnarray}
Here we have re-arranged the boson-fermion interaction in the third
term. The inverse tree level fermionic and bosonic propagators are
denoted as $S^{-1}$ and $D^{-1}$, $\left(\hat{\Gamma}_{R},\hat{\Gamma}_{I}\right)^{T}$
is the boson-fermion vertex with $\hat{\Gamma}_{R}=i\sqrt{2}g\gamma^{5}\sigma_{1}^{NG}$
and $\hat{\Gamma}_{I}=-i\sqrt{2}g\gamma^{5}\sigma_{2}^{NG}$, where
the $\sigma_{1,2}^{NG}$ are the pauli matrices in the NG basis. In
the following the gap $\Delta$ is treated as real for simplicity. 

With the Lagrangian density one can work out the effective potential
in two levels: mean field approximation \cite{Deng:2006ed} and the
two particle irreducible (2PI) approach \cite{Deng:2008ah}. In the
mean field approximation, only the condensate contribution of the
bosonic field is included while the fluctuation of the field is neglected.
In the 2PI approach the calculation is done in the Cornwall-Jackiw-Tomboulis
(CJT) formalism \cite{Cornwall:1974vz}, in which all 2PI diagrams
are counted so that the fluctuation of the bosonic field can be calculated
self-consistently. The effective potential in the mean field approximation
and in the 2PI approach read 
\begin{equation}
\bar{\Gamma}_{MF}=-I(\Delta)+\frac{1}{2}\mathrm{Tr}\ln D^{-1}-\frac{1}{2}\mathrm{Tr}\ln S^{-1},\label{potential-mf}
\end{equation}
\begin{eqnarray}
\bar{\Gamma}_{2PI} & = & -I(\Delta)+\frac{1}{2}\{\mathrm{Tr}\ln\mathcal{D}^{-1}+\mathrm{Tr}(D^{-1}\mathcal{D}-1)\nonumber \\
 &  & -\mathrm{Tr}\ln\mathcal{S}^{-1}-\mathrm{Tr}(S^{-1}\mathcal{S}-1)-2\Gamma_{2PI}(\mathcal{D},\mathcal{S})\},\label{potential-2PI}
\end{eqnarray}
where $I(\Delta)=\frac{(\mu_{b}^{2}-m_{b}^{2})\Delta^{2}}{4g^{2}}$
is the condensate contribution. $\Gamma_{2PI}$ includes all 2PI contributions
to the effective potential, which in the present case is 
\begin{equation}
\Gamma_{2PI}\thickapprox-\frac{1}{4}\mathrm{Tr}\{\mathcal{D}_{ij}^{-1}\mathrm{Tr}[\hat{\Gamma}_{i}\mathcal{S}\hat{\Gamma}_{j}\mathcal{S}]\},\label{diagram-2PI}
\end{equation}
where $\hat{\Gamma}_{i,j}$ are the boson-fermion vertex, $i,j=R,I$
correspond to the real and imaginary components of the bosonic field,
$\mathcal{D}$ and $\mathcal{S}$ are the dressed bosonic and fermionic
propagators derived by the Dyson-Schwinger equations (DSE) for bosons
and fermions. In the CJT formalism the DSE are given by taking derivatives
of the effective potential with respect to propagators,
\begin{eqnarray}
\frac{\delta\bar{\Gamma}_{2PI}}{\delta\mathcal{D}}=0, & \  & \frac{\delta\bar{\Gamma}_{2PI}}{\delta\mathcal{S}}=0,
\end{eqnarray}
which lead to the DSE, 
\begin{eqnarray}
\mathcal{D}^{-1} & = & D^{-1}-2\frac{\delta\Gamma_{2PI}}{\delta\mathcal{D}}=D^{-1}+\Pi(\mathcal{D},\mathcal{S}),\label{dyson-boson}\\
\mathcal{S}^{-1} & = & S^{-1}+2\frac{\delta\Gamma_{2PI}}{\delta\mathcal{S}}=S^{-1}-\Sigma(\mathcal{D},\mathcal{S}),\label{dyson-fermion}
\end{eqnarray}
where the self-energies for bosons and fermions in the CJT formalism
are given by
\begin{eqnarray}
\Pi=\frac{1}{2}\mathrm{Tr}[\hat{\Gamma}_{i}\mathcal{S}\hat{\Gamma}_{j}\mathcal{S}], & \  & \Sigma(\mathcal{D},\mathcal{S})=\mathrm{Tr}[\mathcal{D}_{ij}\hat{\Gamma}_{i}\mathcal{S}\hat{\Gamma}_{j}].\label{self-energy}
\end{eqnarray}
The Rainbow-Ladder truncation is applied to the calculation for self-energies,
i.e. replacing the dressed boson-fermion vertices and propagators
by the bare ones. Substituting the DSE into the 2PI effective potential,
\begin{eqnarray}
\bar{\Gamma}_{2PI} & = & -I(\Delta)+\frac{1}{2}\{\mathrm{Tr}\ln[D^{-1}(1+D\Pi)]-\mathrm{Tr}\ln[S^{-1}(1-S\Sigma)]-\mathrm{Tr}(\Sigma\mathcal{S})\}\nonumber \\
 & \thickapprox & -I(\Delta)+\frac{1}{2}\{\mathrm{Tr}\ln D^{-1}-\mathrm{Tr}\ln S^{-1}+\mathrm{Tr}[D\Pi(\mathcal{D},\mathcal{S})]\}.\label{potential-exp}
\end{eqnarray}
In the last step an expansion in terms of self-energy has been made
leading to a partial loss of self-consistence but reducing the numerical
complexity. From the last line of Eq. (\ref{potential-exp}), by comparing
the mean-field with 2PI effective potential, one can find the beyond-mean-field
contribution to the effective potential is included in $-\Gamma_{2PI}$. 

\begin{figure}
\includegraphics[scale=0.6]{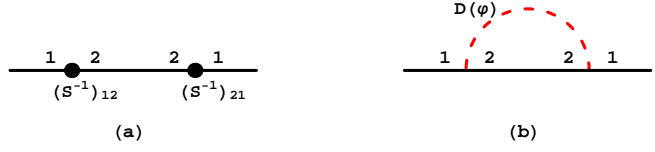}

\caption{\label{fig:Fermionself-energy}Diagrams for the gap and pseudo-gap.
The indices 1 and 2 are for NG space.}
\end{figure}

From the DSE for the fermion, Eq. (\ref{dyson-fermion}), the 11-component
in the NG basis reads
\begin{eqnarray}
\mathcal{S}_{11} & = & [(\mathcal{S}^{-1})_{11}-(\mathcal{S}^{-1})_{12}((\mathcal{S}^{-1})_{22})^{-1}(\mathcal{S}^{-1})_{21}]\nonumber \\
 & = & [(S^{-1})_{11}-\Sigma_{11}-(S^{-1})_{12}((S^{-1})_{22}-\Sigma_{22})^{-1}(S^{-1})_{21}]\nonumber \\
 & \thickapprox & [(S^{-1})_{11}-\Sigma_{11}-(S^{-1})_{12}((S^{-1})_{22})^{-1}(S^{-1})_{21}].\label{fermion-prop-11}
\end{eqnarray}
In the last line $\Sigma_{22}$ is neglected since it is sandwiched
by $(S^{-1})_{12}$ and $(S^{-1})_{21}$ which are proportional to
$\Delta$. $\Sigma_{11}$ is the 11-component of quark self-energy
$\Sigma_{11}=\{\mathrm{Tr}[D_{ij}\hat{\Gamma}_{i}S\hat{\Gamma}_{j}]\}_{11}$.
Then we have 
\begin{eqnarray}
 & -\Sigma_{11} & -(S^{-1})_{12}((S^{-1})_{22})^{-1}(S^{-1})_{21}\nonumber \\
 & \thickapprox & -16g^{2}T\sum_{n}\sum_{e=\pm}\int\frac{d^{3}q}{(2\pi)^{3}}\frac{1}{(p_{0}+\mu_{b})^{2}-(E_{q}^{b})^{2}}\nonumber \\
 &  & \times\frac{1}{p_{0}+q_{0}-\mu-eE_{p+q}}\gamma^{0}\Lambda_{p+q}^{-e}+\sum_{e}\frac{\Delta^{2}}{p_{0}-\mu-eE_{p}}\gamma^{0}\Lambda_{p}^{-e}\nonumber \\
 & = & \sum_{e}\frac{\Delta^{2}+\Delta_{pg}^{2}}{p_{0}-\mu-eE_{p}}\gamma^{0}\Lambda_{p}^{-e}.\label{psudo-gap-1}
\end{eqnarray}
The assumption $q\ll p$ is used in the second step. $E_{q}^{b}=\sqrt{q^{2}+m_{b}^{2}}$
is the boson energy in vacuum. $\Delta_{pg}$ is the pseudo-gap defined
as 
\begin{eqnarray}
\Delta_{pg}^{2} & = & -16g^{2}T\sum_{n}\int\frac{d^{3}q}{(2\pi)^{3}}\frac{1}{(p_{0}+\mu_{b})^{2}-(E_{q}^{b})^{2}}\nonumber \\
 & = & 16g^{2}\int\frac{d^{3}q}{(2\pi)^{3}}\frac{1+f_{B}(E_{q}^{b}-\mu_{b})+f_{B}(E_{q}^{b}+\mu_{b})}{2E_{q}^{b}}\nonumber \\
 & \thickapprox & 16g^{2}\int\frac{d^{3}q}{(2\pi)^{3}}\frac{f_{B}(E_{q}^{b}-\mu_{b})+f_{B}(E_{q}^{b}+\mu_{b})}{2E_{q}^{b}}.\label{psudo-gap-2}
\end{eqnarray}
In the last step, the pseudo-gap are renormalized by directly removing
the divergence part $1/(2E_{q}^{b})$ which comes from vacuum. From
the fermionic self-energy Eq. (\ref{psudo-gap-1}), the lowest order
contribution of the fluctuation is to add a pseudo-gap term $\Delta_{pg}^{2}$
to $\Delta^{2}$. Here we define the pseudo-gap as the correction
to the fermion self-energy at the static limit, see Fig. \ref{fig:Fermionself-energy}.
In the region near and above the superconductivity critical temperature
$T_{c}$, the di-fermion fluctuation will contribute to the fermionic
spectral density near Fermi surface by bringing in two bumps structure,
but it is not a real gap since the fermionic excitation is not forbidden
between the bumps. It has analytical structure from which one can
compute the density of states from the di-fermion fluctuation. In
the region below $T_{c}$ both the gap and pseudo-gap contribute.
Here the pseudo-gap effects are approximated by $\Delta_{pg}^{2}$
for numerical simplicity. The dressed fermion propagator is also approximated
by replacing $\Delta^{2}$ with $\Delta^{2}+\Delta_{pg}^{2}$. 

From Eq. (\ref{self-energy}), the bosonic self-energy have four components
corresponding to $i,j=I,R$. There are simple relations among them:
$\Pi_{RR/II}=\Pi_{0}\pm\Pi_{1}$, $\Pi_{IR}=\Pi_{RI}=0$, where $\Pi_{0}=-8g^{2}\mathrm{Tr}[\gamma^{5}\mathcal{S}_{11}\gamma^{5}\mathcal{S}_{22}]$
is the self-energy in the normal phase ($\Delta=0$), and $\Pi_{1}=-8g^{2}\mathrm{Tr}[\gamma^{5}\mathcal{S}_{21}\gamma^{5}\mathcal{S}_{21}]$
is proportional to $\Delta^{2}$. The momentum integrals are implied
in $\mathrm{Tr}$. Also we have $D_{RR}=D_{II}$ for the bare bosonic
propagator. Inserting the self-energy to Eq. (\ref{diagram-2PI})
we obtain $\Gamma_{2PI}$ as 
\begin{eqnarray}
\Gamma_{2PI} & = & -\frac{1}{2}\mathrm{Tr}[D_{RR}\Pi_{RR}]-\frac{1}{2}\mathrm{Tr}[D_{II}\Pi_{II}]=-\mathrm{Tr}(D_{RR}\Pi_{0}).\label{gamma-2pi}
\end{eqnarray}
Then the effective potential in the mean field approximation and the
2PI approach are given by 
\begin{eqnarray}
\bar{\Gamma}_{MF} & = & -\sum_{e=\pm}\int\frac{d^{3}k}{(2\pi)^{3}}[\epsilon_{k}^{e}+2T\ln(1+e^{-\epsilon_{k}^{e}/T})]+\frac{(m_{b}^{2}-\mu_{b}^{2})\Delta^{2}}{4g^{2}}\nonumber \\
 &  & +\frac{1}{2}\sum_{e=\pm}\int\frac{d^{3}k}{(2\pi)^{3}}[\omega_{k}^{e}+2T\ln[1-e^{-\omega_{k}^{e}/T}],\label{effective-potential-mf}\\
\bar{\Gamma}_{2PI} & = & \bar{\Gamma}_{MF}-\Gamma_{2PI},\label{effectiv-potential-2pi}
\end{eqnarray}
where $\xi_{k}^{e}=\sqrt{k^{2}+m^{2}}-e\mu$ and $\epsilon_{k}^{e}=\sqrt{(\xi_{k}^{e})^{2}+\Delta^{2}}$
are fermionic excitation energies in normal and condensed phases respectively.
In the following we use $\Omega$ to denote the thermal dynamic potential
to replace the effective potential $\bar{\Gamma}$. The total charge
density $n=-\frac{\partial\Omega}{\partial\mu}$ is fixed to a constant.
The fermion number density for fermions, condensed/excited bosons,
and the 2PI component are 
\begin{eqnarray}
\rho_{F} & = & \frac{n_{f}}{n}=\frac{2}{n}\sum_{e=\pm}\int\frac{d^{3}k}{(2\pi)^{3}}\frac{e\xi_{k}^{e}}{2\epsilon_{k}^{e}}[f_{F}(\epsilon_{k}^{e})-f_{F}(-\epsilon_{k}^{e})],\nonumber \\
\rho_{b0} & = & \frac{2\mu\Delta^{2}}{ng^{2}},\nonumber \\
\rho_{b} & = & \frac{2}{n}\sum_{e=\pm}\int\frac{d^{3}k}{(2\pi)^{3}}ef_{B}(E_{k}^{b}-e\mu_{b}),\nonumber \\
\rho_{\Gamma_{2}} & = & \frac{1}{n}\frac{\partial\Gamma_{2PI}}{\partial\mu},\label{density-frac}
\end{eqnarray}
which satisfies 
\begin{equation}
\rho_{F}+\rho_{b0}+\rho_{b}+\rho_{\Gamma_{2}}=1.\label{density-equation}
\end{equation}

The gap equation is settled with the saddle point condition of the
free energy density $F=\Omega+\mu n$, that is
\begin{equation}
\frac{\partial F}{\partial\Delta}=\frac{\partial\Omega}{\partial\Delta}+\frac{\partial\Omega}{\partial\mu}\frac{\partial\mu}{\partial\Delta}+n\frac{\partial\mu}{\partial\Delta}=\frac{\partial\Omega}{\partial\Delta}=0.\label{free-energy-saddle}
\end{equation}
The gap $\Delta$, the chemical potential $\mu$ and the pseudo-gap
$\Delta_{pg}$ can be solved simultaneously from the density and the
gap equations (\ref{density-equation},\ref{free-energy-saddle})
and the pseudo-gap equation (\ref{psudo-gap-2}). In the mean field
approximation $\Gamma_{2PI}$ and $\Delta_{pg}$ are set to zero and
the number of the equations is reduced to two. 

\begin{figure}
\includegraphics[scale=0.45]{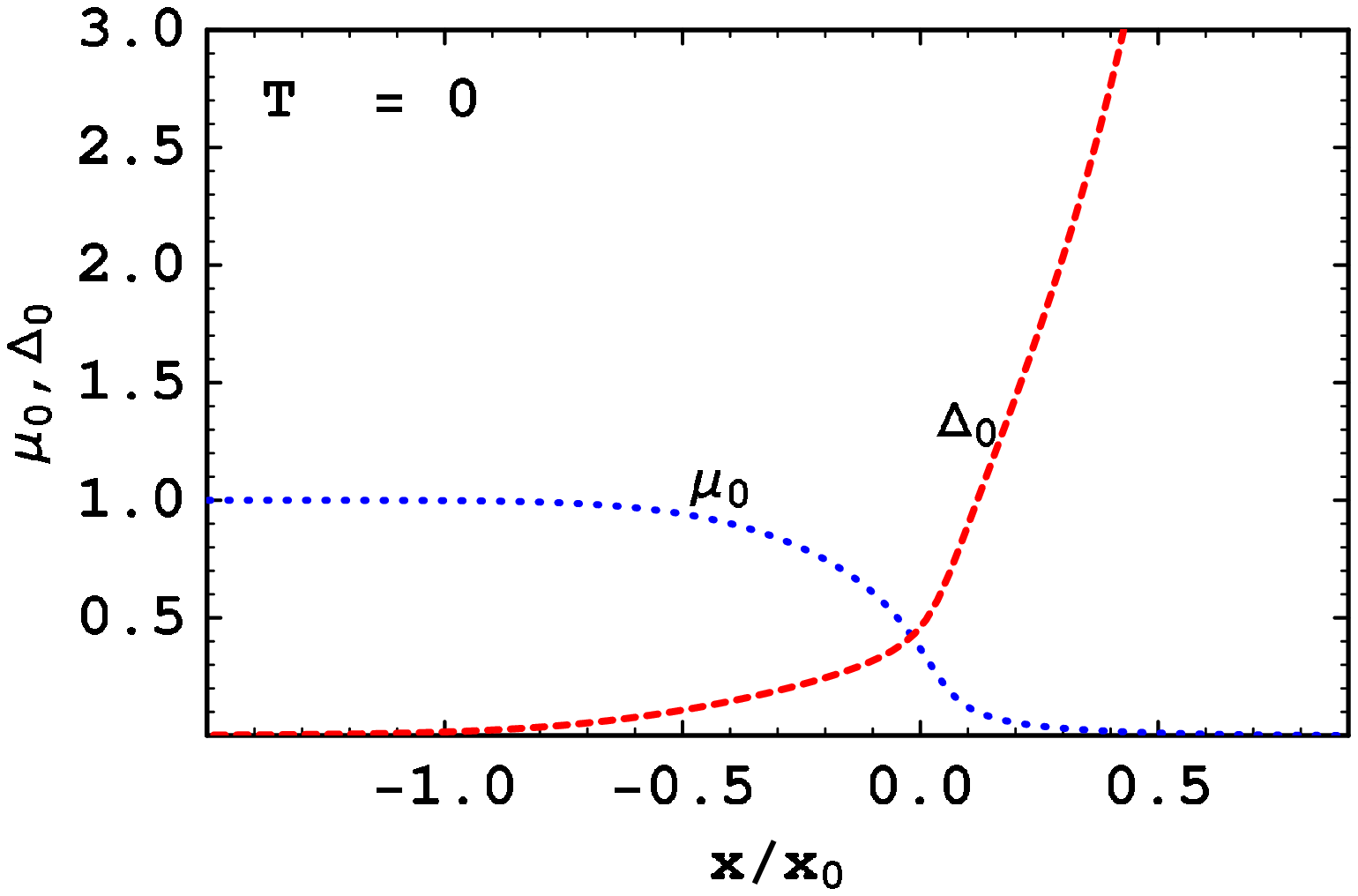}\includegraphics[scale=0.43]{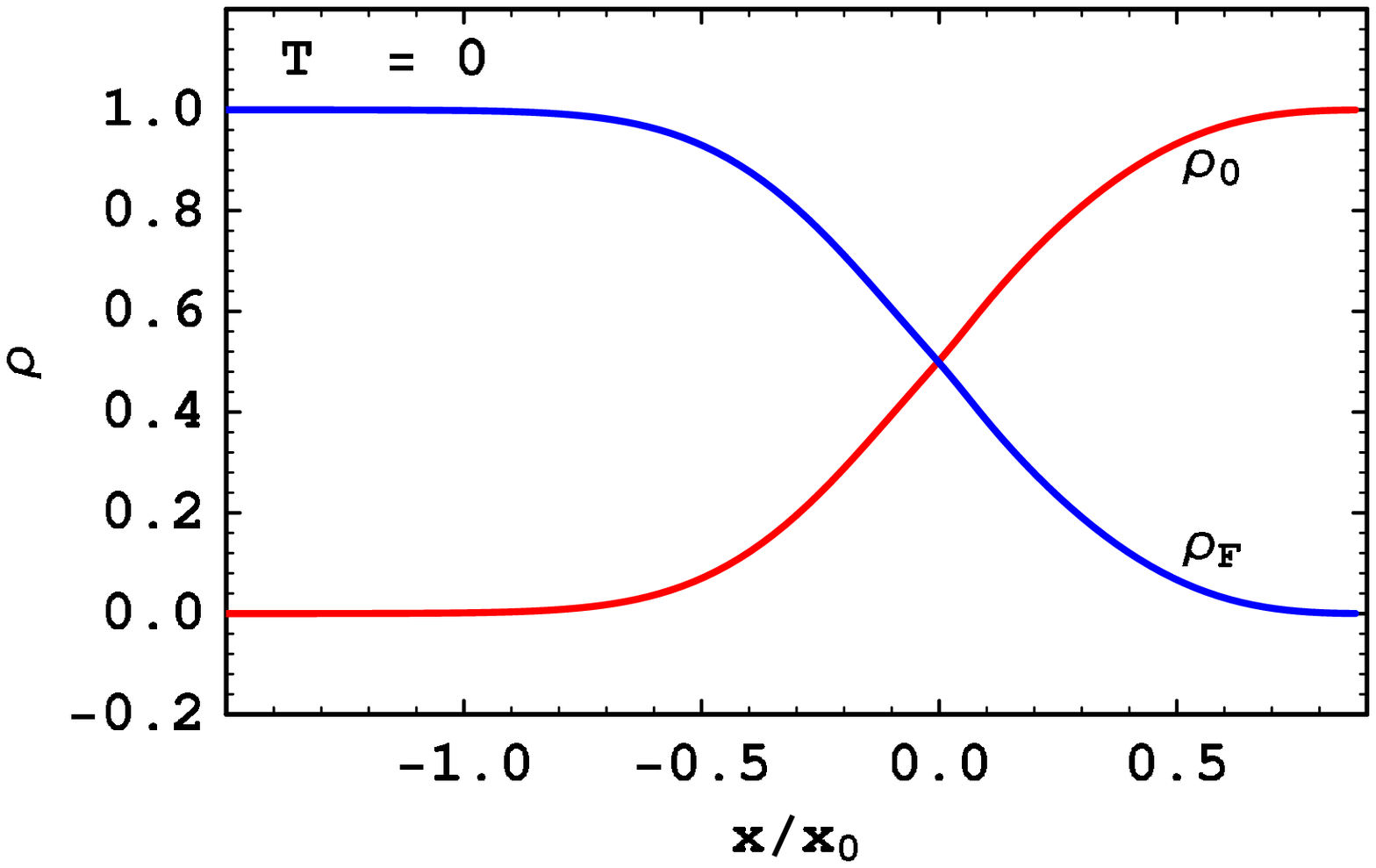}

\caption{Left panel: the chemical potential $\mu$ (blue dotted line) and the
gap $\Delta$ (red dashed line) at zero temperature as functions of
$x$. Right panel: the condensed boson and fermion fractions with
$x$. The units for $\mu_{0}$, $\Delta_{0}$ and $T$ are GeV. \label{fig:crossover-mb-T0}}
\end{figure}

In the mean field approximation, the renormalized boson mass $m_{br}$
can be obtained as 
\begin{equation}
m_{b,r}^{2}=4g^{2}\frac{\partial\Omega}{\partial\Delta^{2}}|_{\Delta=T=\mu=0}=m_{b}^{2}-4g^{2}\int\frac{d^{3}k}{(2\pi)^{3}}\frac{1}{\sqrt{k^{2}+m^{2}}}.\label{mbr-mean-field}
\end{equation}
In the CJT formalism to calculate the renormalized bosonic mass $m_{br}$
one should work with the bosonic DSE (\ref{dyson-boson}) for self-consistency.
Then the pole equation in the homogeneous limit with gap set to zero
is given by 
\begin{equation}
\mathrm{det}\mathcal{D}^{-1}|_{p=\Delta=\Delta_{pg}=0}=\mathrm{det}[D^{-1}+\Pi(p_{0},\mathbf{p})]|_{\mathbf{p}=\Delta=\Delta_{pg}=0}=0,\label{pole-eq}
\end{equation}
with which the dressed boson mass $m_{br}$ is determined as 
\begin{eqnarray}
m_{br} & = & (p_{0}+\mu_{b})^{2}=m_{b}^{2}+\mathrm{Re}\Pi_{0}(p_{0}-i\frac{\eta}{2}),\label{mbr}
\end{eqnarray}
\begin{equation}
-(p_{0}+\mu_{b})\eta-\mathrm{Im}\Pi_{0}(p_{0}-i\frac{\eta}{2})=0,\label{im-pole-eq}
\end{equation}
where $p_{0}$ is the pole position and $\eta$ is the width of the
dressed boson propagator determined by the imaginary part of the pole
equation (\ref{pole-eq}) . The condition $\eta=0$ defines the bosonic
dissociation boundary ($T^{*},\mu^{*}$). By solving the real and
imaginary parts of the pole equation simultaneously, the dissociation
boundary turns out to be very simple $m_{br}=2m$. In the present
model the ($T^{*},\mu^{*}$) are functions of the bare boson mass
$m_{b}$. From Eq. (\ref{lagrangian2}), with a fixed boson-fermion
coupling constant $g$, $m_{b}$ is equivalent to fix the coupling
constant in a pure fermionic model. Assuming the boson is stable,
then we get $\eta=0$ and Eq. (\ref{mbr}) becomes 
\begin{equation}
m_{br}^{2}=m_{b}^{2}+\Pi_{0}(p_{0}),\label{mbr-eq1}
\end{equation}
$m_{br}$ together with the bosonic chemical potential serve as the
crossover parameter $x=-\frac{m_{br}-\mu_{b}^{2}}{2g^{2}}$. The parameter
$x$ can be varied from negative values with large modulus (BCS) to
large positive values (BEC). In between, $x_{0}$ is the unitary limit.
Therefore, $x$ behaves as the scattering length. 

In Fig. \ref{fig:crossover-mb-T0}, the calculation is done in mean
field approximation for $\Delta$, $\mu$ and bosonic/fermionic charge
fractions as functions of $x$. The parameters are set to $T=0$ GeV,
$m=0.2$ GeV, $g=4$ and $\Lambda=1$ GeV. $x_{0}=\int\frac{d^{3}k}{(2\pi)^{3}}\frac{1}{\sqrt{k^{2}+m^{2}}}$
is an upper limit of $x$ which ensures non-negative bosonic occupation
number. In the right panel of Fig. \ref{fig:crossover-mb-T0}, the
charge fraction of the thermal boson is always equal to zero since
the temperature is zero. From the negative value to positive value
of $x$, the system goes through a crossover from the BCS side to
the BEC side.

\begin{figure}
\includegraphics[scale=0.44]{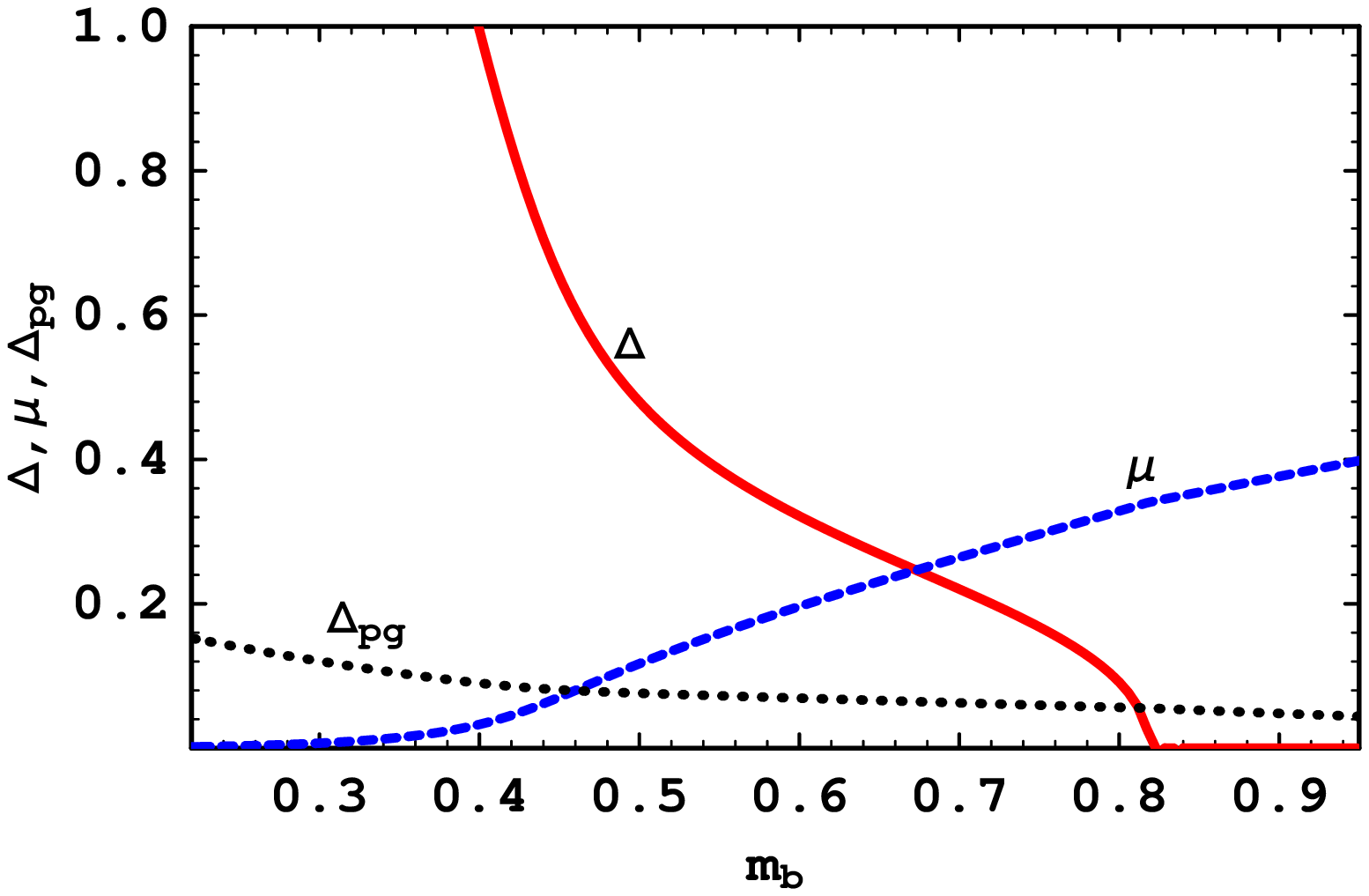}\includegraphics[scale=0.43]{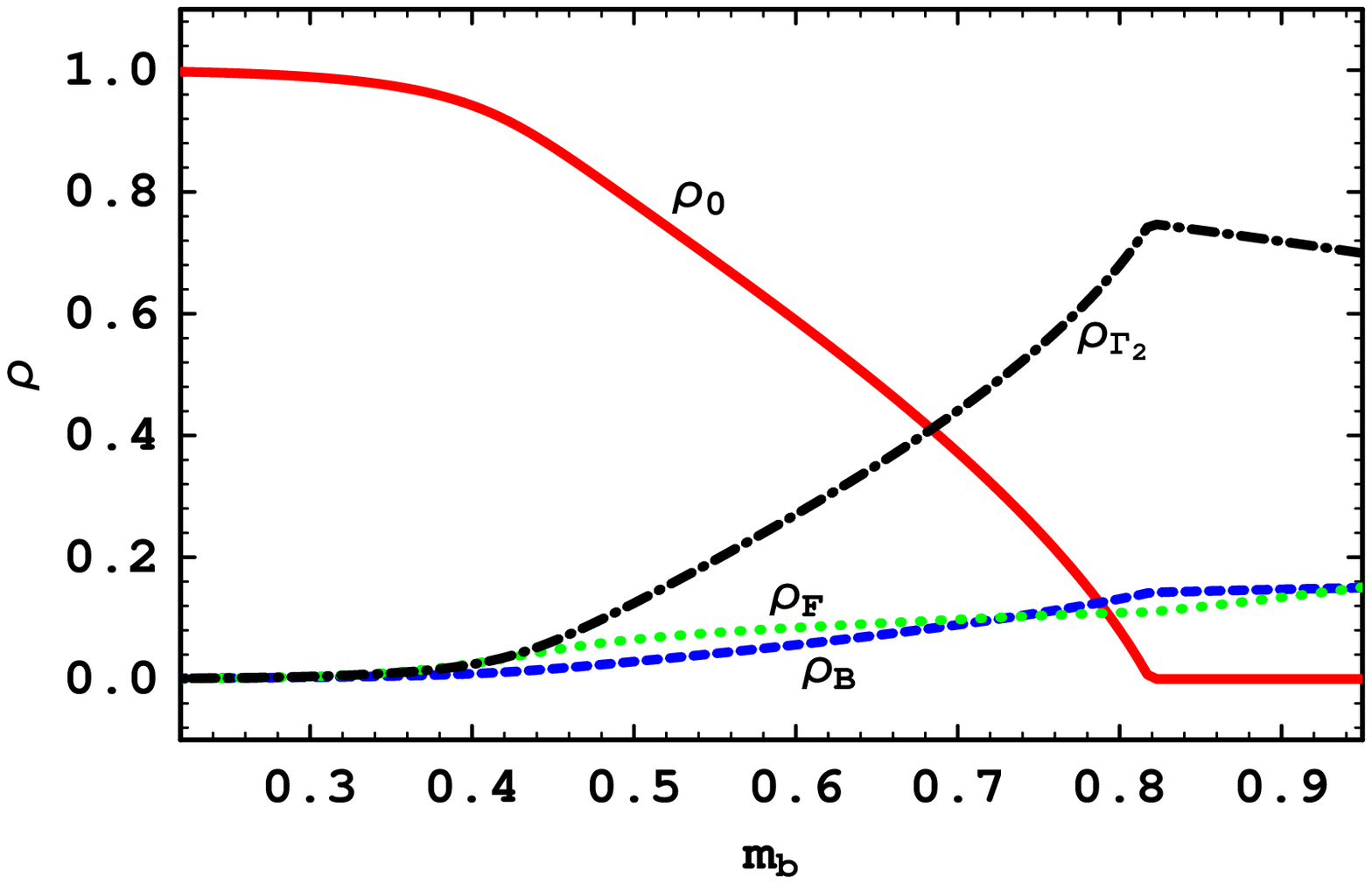}

\caption{The fermionic chemical potential $\mu$ (blue dashed line), the gap
$\Delta$ (red solid line) and pseudo-gap (black dotted line) as functions
of $m_{b}$. Right panel: fractions of fermions, condensed/thermal
bosons and 2PI component. The units for $m_{b}$ $\mu$, $\Delta$
and $\Delta_{pg}$ are GeV. \label{fig:crossover-mb-2pi}}
\end{figure}

At high temperatures, both the condensed and thermal bosons should
be considered. The results for the calculation up to 2PI are shown
in Fig. \ref{fig:crossover-mb-2pi}. The parameters are chosen to
be $T=0.14$ GeV, $m=0.28$ GeV, $g=1.8$ and $\Lambda=1$ GeV. With
a small boson bare mass $m_{b}$ the system is in a strong coupling
regime where the fermion chemical potential $\mu$ is low and the
condensed bosons are dominant. When $m_{b}$ goes larger, both the
gap $\Delta$ and pseudo-gap $\Delta_{pg}$ decrease while the chemical
potential $\mu$ increases. The pseudo-gap $\Delta_{pg}$ is small
due to a low temperature chosen here. The fraction of condensed bosons
becomes smaller and the fermionic degree of freedom becomes more important.
From the left panel of Fig. \ref{fig:crossover-mb-2pi}, the 2PI contribution
is dominant in the large $m_{b}$ region indicating the strong interaction
between fermions and bosons. 

The fluctuations may change the superconducting phase transition to
be of first order, such as the intrinsic fluctuating magnetic field
in normal superconductors \cite{Halperin:1973jh}, or the gauge field
fluctuations in color superconductors \cite{Giannakis:2004xt}, due
to the fact that fluctuations bring a cubic term of the condensate
to the effective potential making the Landau theory of continuous
phase transition invalid. In the present case, a nonzero cubic term
of $\Delta$ is generated in the bosonic fluctuation $\Gamma_{2PI}$,
leading to a first order phase transition.

\section{Relativistic BCS-BEC crossover in Magnetic field}

It is well known that an applied magnetic field can tune the BEC-BCS
crossover in cold atom system, because the magnetic field can adjust
the effective interaction between fermions via Feshbach resonance.
In a relativistic fermion system, by applying a magnetic field with
a magnitude near the energy scale of the interaction, the properties
of the system will also be affected. The magnetic fields on the surface
of pulsars are about $10^{12}\sim10^{13}$ G , and for magnetars they
are about $10^{14}\sim10^{15}$ G \cite{Paczynski:1992zz,Duncan:1992hi,Thompson:1996pe}.
In the core the magnetic field can be even stronger. In heavy ion
collision experiments at RHIC and LHC, the background magnetic field
generated in non-central collisions can be about $10^{18}-10^{19}$
G at RHIC/LHC energies \cite{Kharzeev:2007jp,Skokov:2009qp}. In the
future low-energy experiments at RHIC, NICA and FAIR which is targeted
to probe dense and cold nuclear matter, the magnetic field are also
expected to be very strong. Such a strong magnetic may bring some
significant effects to the matter, for example, the chiral magnetic
effect \cite{Kharzeev:2007jp,Fukushima:2008xe}. The strong magnetic
fields have significant effects on the quark pairings in the CSC phase
\cite{Ferrer:2005vd,Ferrer:2006vw,Ferrer:2007iw,Feng:2011fj}. Recently
the magnetic field tuning of the BEC-BCS crossover has been studied
\cite{Wang:2010uj}. 

One can extend the boson-fermion model discussed above to study a
system with oppositely charged fermions $\Psi^{T}=(\psi_{1},\psi_{2})$
and neutral scalar bosons \cite{Wang:2010uj}. An external magnetic
field is also included in the model and coupled with the fermions.
Then the fermion part and interaction part of Eq. (\ref{lagrangian1})
are 
\begin{eqnarray}
\mathcal{L}_{f} & = & \bar{\Psi}(i\gamma^{\mu}\partial_{\mu}+\mu\gamma^{0}-q\sigma_{3}\gamma^{\mu}A_{\mu}-m)\Psi,\\
\mathcal{L}_{I} & = & \varphi\bar{\Psi}_{C}(i\gamma^{5}\sigma_{2})\Psi+\varphi^{*}\bar{\Psi}(i\gamma^{5}\sigma_{2})\Psi_{C},\label{lagrangian-mag}
\end{eqnarray}
where $q$ denotes the charge of the fermion. $A_{\mu}$ is the vector
potential of the external magnetic field. As discussed in Sec. \ref{sec:bose-ferm},
the Lagrangian is invariant under the $\mathrm{U(1)_{B}}$ transformation
$\Psi\rightarrow\Psi'=e^{-i\alpha}\Psi$, $\varphi\rightarrow\varphi'=e^{i2\alpha}\varphi$.
Hence the simple relation $\mu_{b}=2\mu$ ensured by chemical equilibrium
remains. In order to describe the BEC of these molecules, we also
separate the zero-mode of the boson field $\varphi$ and replace it
by its expectation value $\phi\equiv\langle\varphi\rangle$, which
represents the electrically neutral difermion condensate. The mean-field
effective action is then 
\begin{eqnarray}
I^{B}(\overline{\psi},\psi) & = & \frac{1}{2}\int d^{4}x\, d^{4}y\,\overline{\Psi}_{\pm}(x){\cal S}_{(\pm)}^{-1}(x,y)\Psi_{\pm}(y)+\nonumber \\
 &  & +(4\mu^{2}-m_{b}^{2})\mid\phi\mid^{2}+\mid(\partial_{t}-2i\mu)\varphi\mid^{2}\nonumber \\
 &  & -\mid\nabla\varphi\mid^{2}-m_{b}^{2}\mid\varphi\mid^{2},\label{b-action}
\end{eqnarray}
 where the fermion inverse propagators of the Nambu-Gorkov positive
and negative charged fields $\Psi_{+}=(\psi_{2},\psi_{1C})^{T}$ and
$\Psi_{-}=(\psi_{1},\psi_{2C})^{T}$ are given by 
\begin{equation}
{\cal S}_{(\pm)}^{-1}=\left(\begin{array}{cc}
[G_{(\pm)0}^{+}]^{-1} & i\gamma^{5}\Delta^{*}\\
i\gamma^{5}\Delta & [G_{(\pm)0}^{-}]^{-1}
\end{array}\right),\label{inv-propg}
\end{equation}
 with 
\begin{equation}
[G_{(\pm)0}^{\pm}]^{-1}(x,y)=[i\gamma^{\mu}\Pi_{\mu}^{(\pm)}-m\pm\mu\gamma^{0}]\delta^{4}(x-y),\label{B-x-inv-prop}
\end{equation}
 and $\Pi_{\mu}^{(\pm)}=i\partial_{\mu}\pm qA_{\mu}$. Without loss
of generality, the magnetic field can be chosen along the z-axis with
$A^{\mu}=(0,Bx_{1},0,0)$. By using Ritus' transformation to momentum
space, the effective potential at zero temperature reads 
\begin{eqnarray}
\Omega & = & -\frac{qB}{2\pi^{2}}\sum_{e=\pm1}\sum_{k=0}^{\infty}d(k)\int_{0}^{\infty}dp_{3}\xi_{e}+\frac{(m_{b}^{2}-4\mu^{2})\Delta^{2}}{4}+\frac{1}{4\pi^{2}}\sum_{e=\pm1}\int_{0}^{\infty}\omega_{e}p^{2}dp,
\end{eqnarray}
in which $d(k)=(1-\frac{\delta_{k0}}{2})$ denote the spin degeneracy
of the Landau levels. $k=0,1,2...$ is the Landau level (LL). The
energy dispersion of fermions and bosons is given by 
\begin{eqnarray}
\epsilon_{e}(k) & = & \sqrt{(\xi_{k}-e\mu)^{2}+\Delta^{2}},\qquad e=\pm1\\
\omega_{e} & = & \sqrt{p^{2}+m_{b}^{2}}-2e\mu,\qquad e=\pm1,
\end{eqnarray}
respectively. $\xi_{k}=\sqrt{p_{3}^{2}+2|q|Bk+m^{2}}$ is the energy
of free fermions in the magnetic field. The parameters of the model
are the momentum cutoff and the fermion mass. The total fermion number
density can always be chosen to be at $x=0$ where the fermion number
fractions of fermions and bosons are equal. Here we choose: the momentum
cutoff is chosen to be a Gaussian type $\exp[-(p_{3}^{2}+2|q|Bk)/\Lambda^{2}]$
with $\Lambda=1$GeV, and the fermion mass $m=0.2\Lambda$. With the
effective potential, the gap equation $\partial\Omega/\partial\Delta=0$
and the density equation $-\partial\Omega/\partial\mu=n$ can be simultaneously
solved.

In this model there are two variable parameters: the bare boson mass
$m_{b}$ and the magnetic field $B$. From Fig. \ref{fig:crossover-mb-T0},
with $B=0$ and $m_{b}$ is small the system is in BEC regime, a large
$m_{b}$ corresponds to BCS regime. Varying $m_{b}$, the system makes
crossovers between these two regimes. In the following the boson mass
$m_{b}$ is fixed since the effect of magnetic field can be shown. 

\begin{figure}
\includegraphics[scale=0.42]{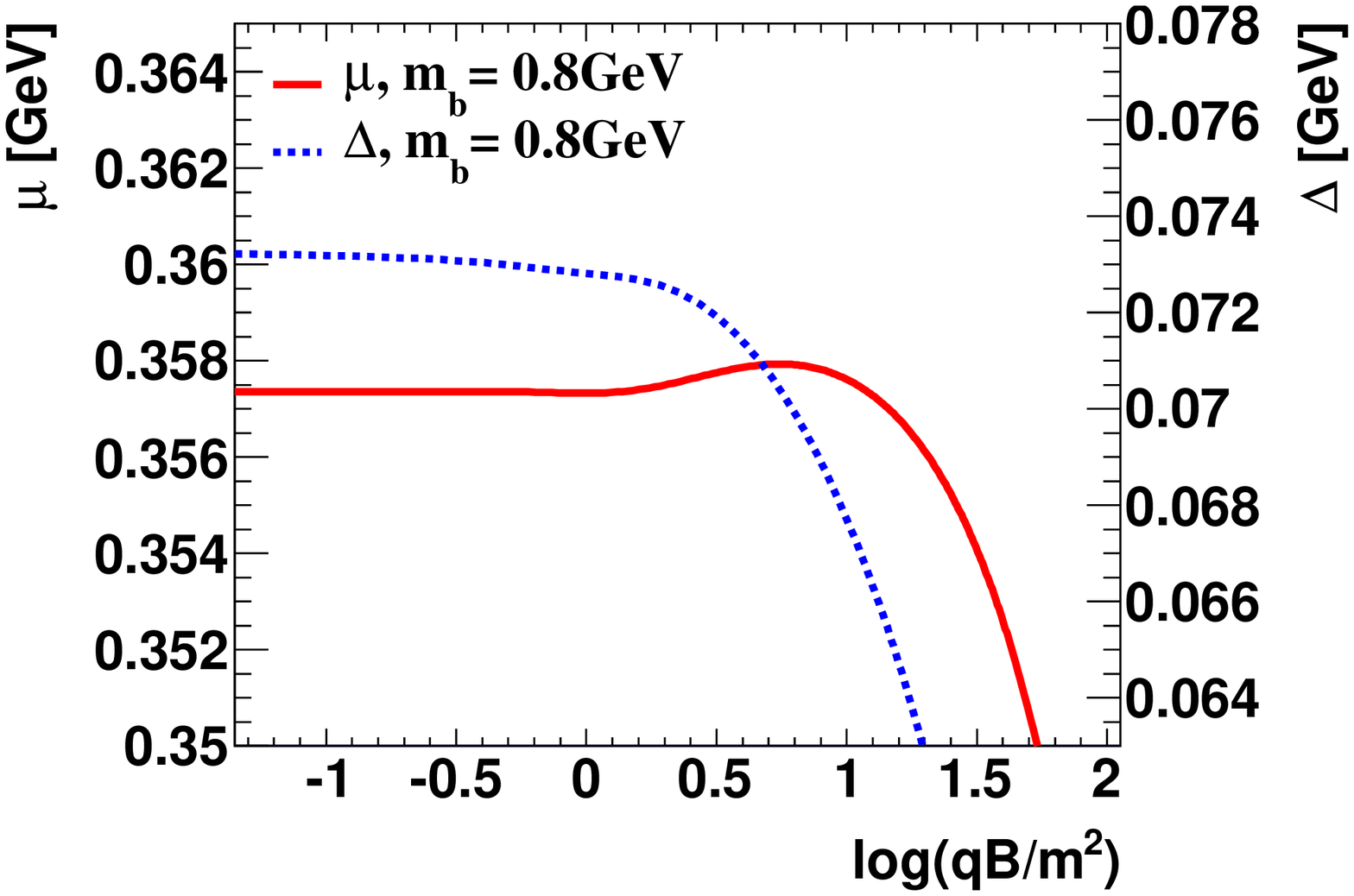} \includegraphics[scale=0.42]{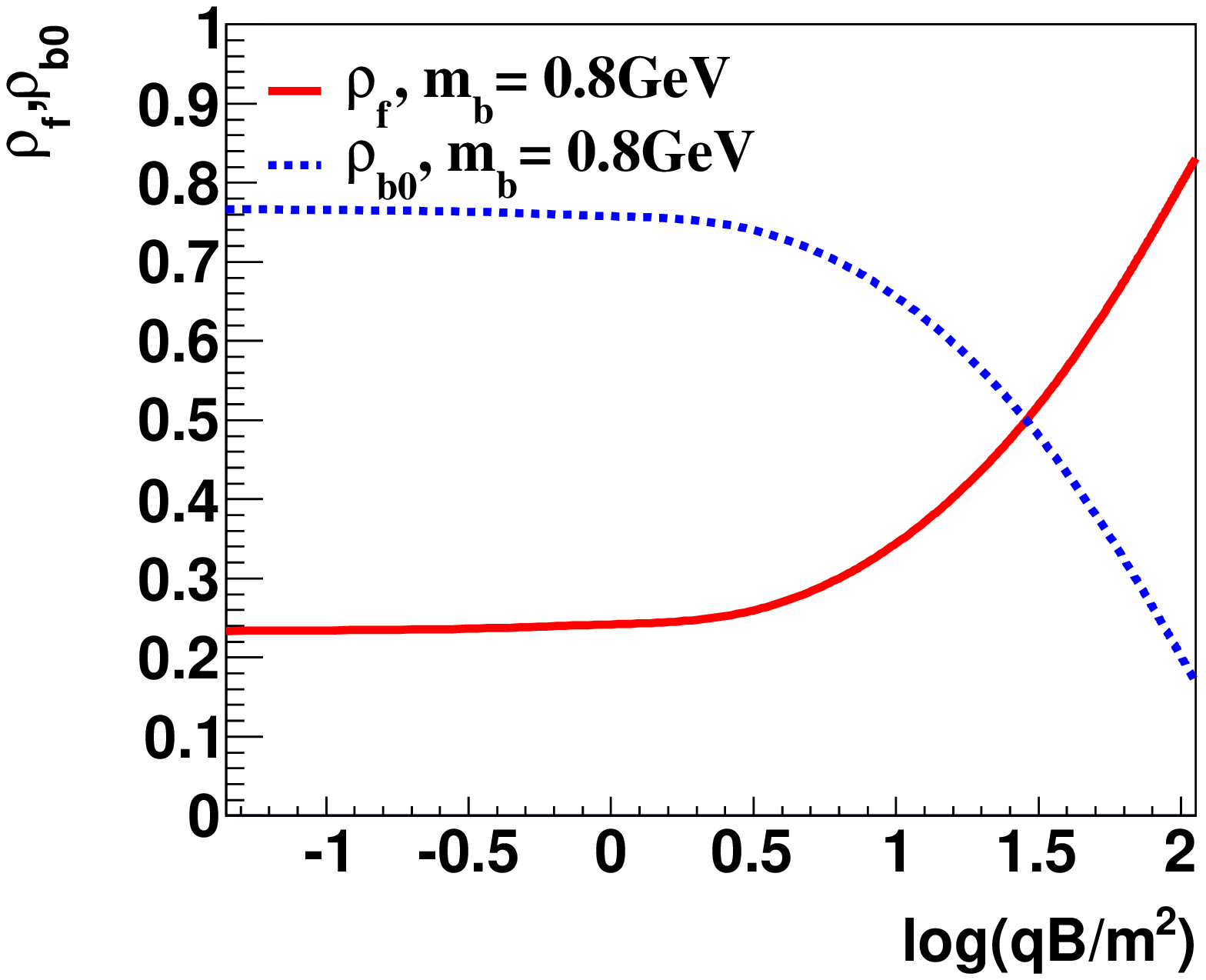}

\caption{\label{fig:crossover-b}The fermionic chemical potential and the gap
(upper panel), and fermion number fractions (lower panel) as functions
of $\ln(qB/m^{2})$. Starting from the BEC side where the BEC component
is much larger than the BCS component at small magnetic field on the
left. The system crossover to a pure BCS state at large magnetic fields
on the right.}
\end{figure}

Fig. \ref{fig:crossover-b} shows the chemical potential, the gap
and fermion number fraction of the condensate bosons and fermions
as functions of $B$ with $m_{b}=0.8$ GeV \cite{Wang:2010uj}. As
the magnetic field is weak on the left end of the panels, the system
is in BEC regime with the condensate fraction much larger than the
fermion fraction. Increasing $B$ till $\ln(qB/m^{2})\sim0.1$, a
de Haas van Alphen behavior will be present. But in this case there
is only one oscillation with small amplitude on the curves for $\mu$
and $\Delta$, since the system is in the BEC regime in weak magnetic
field with the gap about 73 MeV, while the de Haas van Alphen oscillation
favors small $B$ and the amplitude is suppressed by a large gap.
By setting a large enough value for the bare boson mass to let the
system start in BCS side, by tuning the magnetic field the oscillation
will be more obvious and one can find several crossovers in the fermion
number fraction. Continuously increasing the magnetic field, a pure
BCS state settles down on the right end of the panels. This process
is a BCS-BEC crossover by varying the magnetic field, but the origin
is different from that in non-relativistic case where magnetic field
is used to tune the Feshbach resonance. 

\begin{figure}
\includegraphics[scale=0.25]{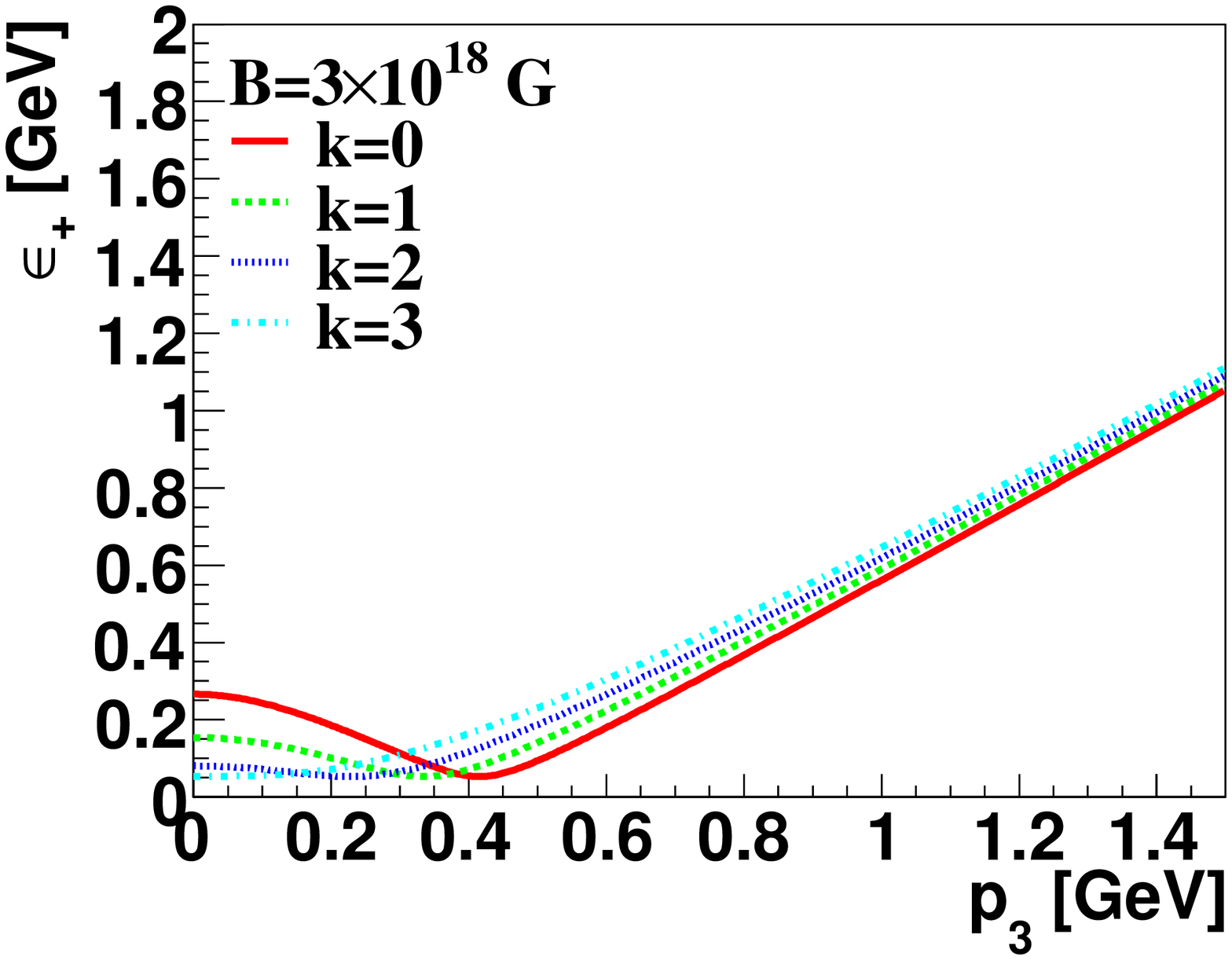} \includegraphics[scale=0.25]{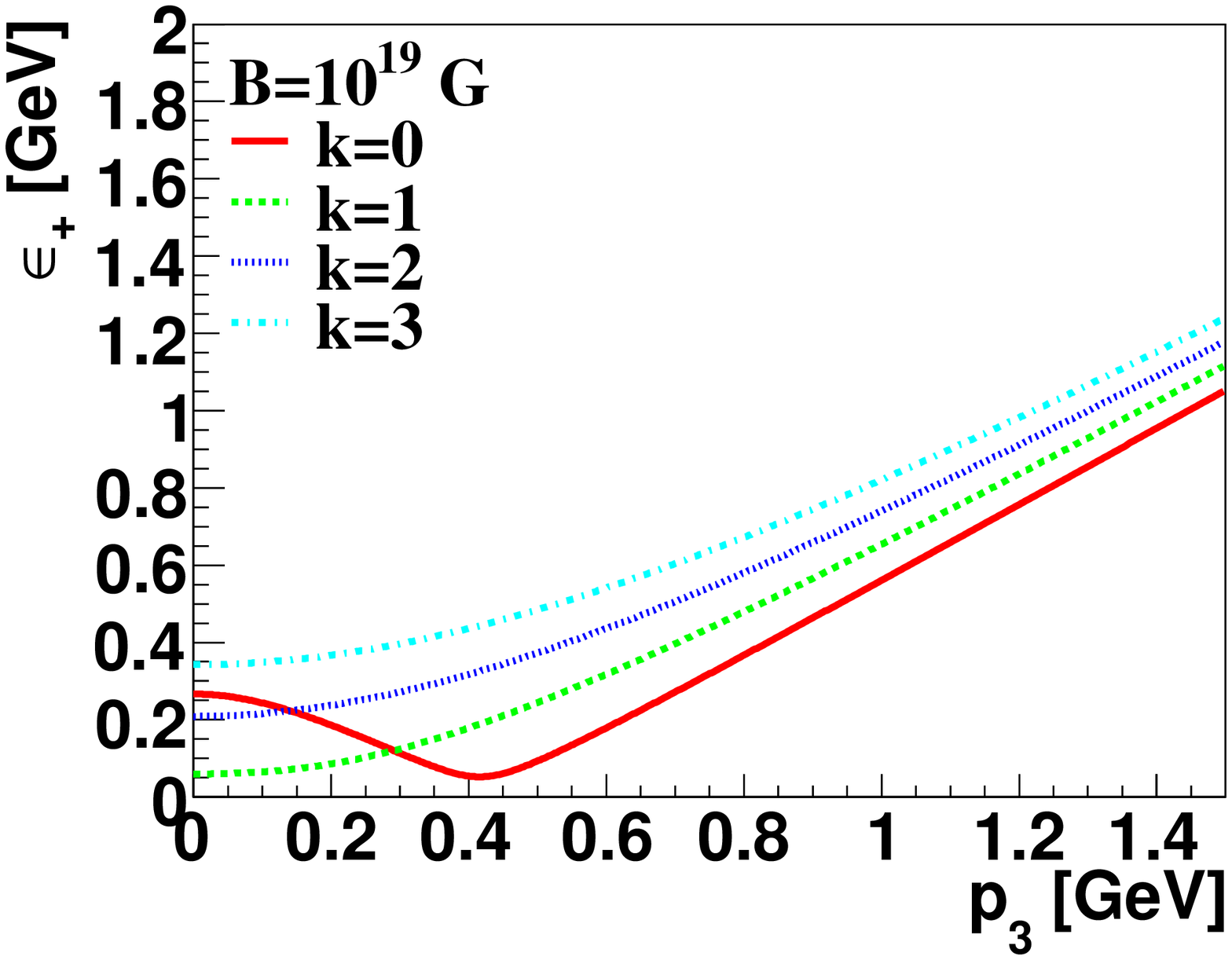}
\includegraphics[scale=0.25]{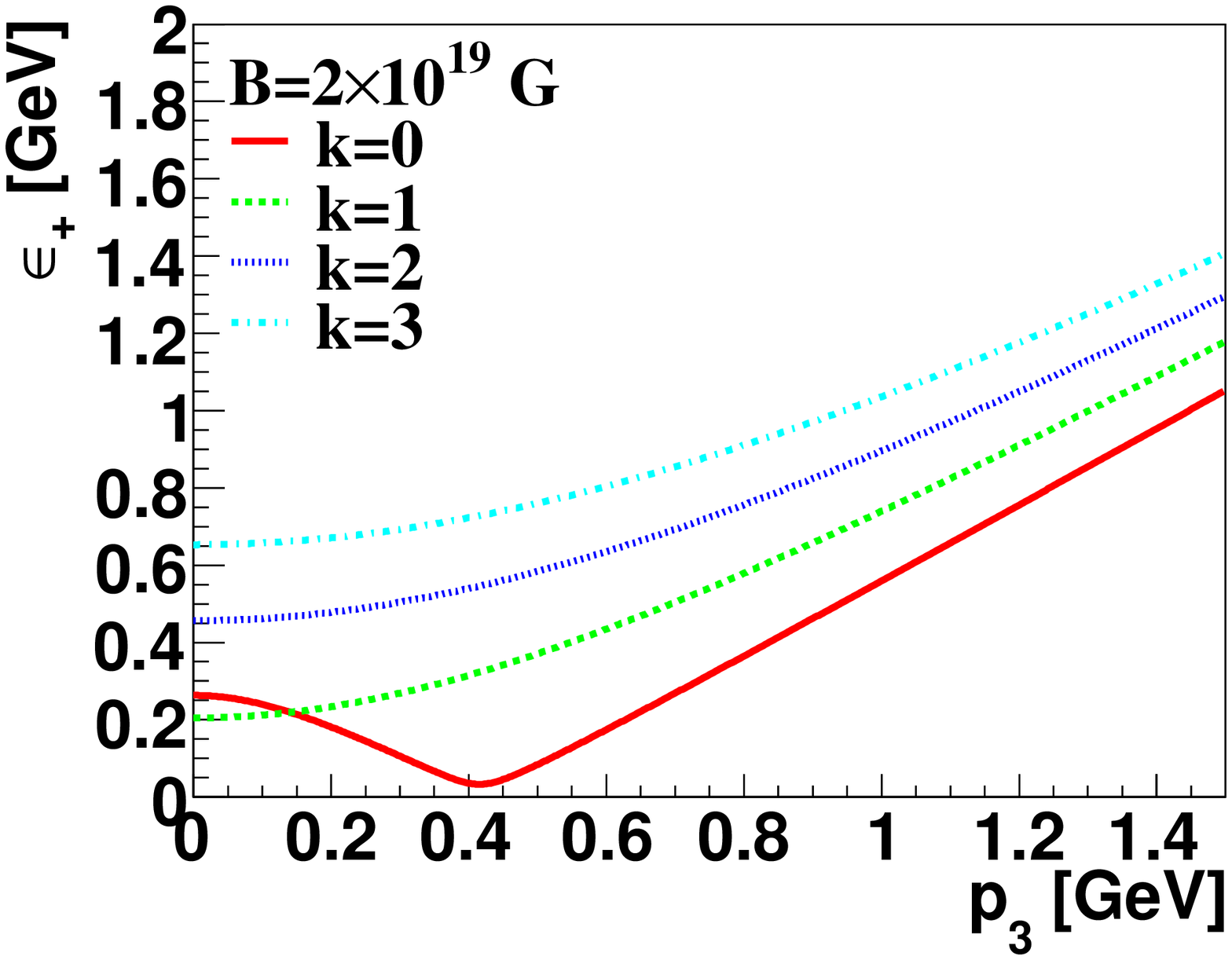}

\caption{\label{fig:dispersion-b}Positive energy component of fermion dispersion
relation with fixed $m_{b}$ and $B$. The LLs $k=0,1,2,3$.}
\end{figure}

The mechanism of the crossover can be explained by the energy dispersion
of the fermion shown in Fig. \ref{fig:dispersion-b} \cite{Wang:2010uj}.
In the left panel of Fig.\ref{fig:dispersion-b}, the LLs with $k<3$
contribute to the BCS component on which the minima is located at
$p_{3}\neq0$, while the one with $k=3$ contributes to the BEC one.
Since the fermion energy splitting between different LLs and the density
of states of each LL are all proportional to $\sqrt{eB}$, when the
field increases not only the energy levels become more separated,
as seen from the figure, they can also accommodate a larger number
of particles. As a consequence, when the field increases, the number
of occupied LLs reduces, or in other words, the magnetic field will
press the fermions to lower LLs. Hence, with increasing the field
the contribution from the lowest LL becomes more important and finally
dominant. The higher levels shown in the middle and lower panel of
Fig. \ref{fig:dispersion-b} are likely not to contribute in strong
fields. That is the reason why in a strong magnetic field the system
is in the BCS regime.

If the relativistic BCS-BEC theories discussed in the literature have
any relevance for the physics of neutron stars and the future low-energy
heavy ion collision experiments, we should study the effects of the
magnetic field on the BCS-BEC crossover within a more realistic model,
as extremely strong magnetic fields are expected to be present.

\section{Diquark properties and BCS-BEC crossover in dense quark matter}

The interaction between two quarks in the anti-triplet channel in
color space is attractive, leading to Cooper pairs of quarks at extremely
high densities and low temperatures. Due to asymptotic freedom of
QCD, the interaction is weak and the Cooper pair wave function has
a correlation length that exceeds the inter-particle distance. However,
as the density is lowered, the interaction strength increases and
the Cooper pair becomes more localized. Eventually, Cooper pairs will
form tightly bound molecular diquark states, then the diquark BEC
regime is settled. Since the interaction between quarks is strong,
it is argued that the diquark fluctuation is large around the critical
temperature $T_{c}$ and a pseudo-gap may be observed. When the temperature
rises, the pseudo-gap will becomes smaller and finally disappear as
the dissociation of diquark fluctuation. Recently there are some work
that focus on those issues within the NJL model.

One considers a 2SC case in the NJL model. The Lagrangian reads 
\begin{eqnarray}
\mathcal{L}_{NJL} & = & \bar{\psi}(i\gamma_{\mu}\partial^{\mu}-\hat{m}_{0}+\mu)\psi+G_{s}[(\bar{\psi}\psi)^{2}+(\bar{\psi}i\gamma_{5}\mathbf{\tau}\psi)^{2}]\nonumber \\
 &  & +G_{D}[\bar{\psi}i\gamma^{5}\tau_{2}J_{a}\psi_{C}][\bar{\psi}_{C}i\gamma^{5}\tau_{2}J_{a}\psi],\label{2SC-lagrangian}
\end{eqnarray}
where $\psi$ and $\psi_{C}$ are quark field and its charge conjugate
respectively. $\tau_{2}$ and $\mathbf{\tau}$ are the Pauli matrices
in flavor space, and $(J_{a})_{bc}=-i\epsilon_{abc}(a=1,2,3)$ denote
the antisymmetric color matrices. $G_{S}$/$G_{D}$ are coupling constants
for quark-anti-quark/quark-quark channel, which together with the
momentum cutoff $\Lambda$ and the bare quark mass $\hat{m}_{0}$
are the input parameters of the model. The hat of $\hat{m}_{0}$ means
the bare quark mass in flavor space.

To introduce the chiral condensate and the diquark degrees of freedom,
one can use the mean field approximation which is equivalent to the
Hubbard-Stratonovich transformation in non-relativistic case. The
diquark field can be decomposed into the condensate part $\Delta_{a}=2G_{D}\left\langle \bar{\psi}i\gamma^{5}\tau_{2}J_{a}\psi_{C}\right\rangle $
and the fluctuation part $\varphi_{a}$, then one gets $\bar{\psi}i\gamma^{5}\tau_{2}J_{a}\psi_{C}=\frac{\Delta_{a}+\varphi_{a}}{2G_{D}}$.
In the 2SC phase the superconducting gap are chosen as $\Delta_{1,2}=0$
and $\Delta_{3}\neq0$ without loss of generality. Then the Lagrangian
(\ref{2SC-lagrangian}) becomes
\begin{eqnarray}
\mathcal{L}_{NJL} & \thickapprox & -\frac{1}{2}\bar{\Psi}S^{-1}\Psi-\frac{1}{4G_{D}}\sum_{a}|\Delta_{a}|^{2}-G_{s}(\sigma_{u}+\sigma_{d})^{2}\nonumber \\
 &  & -\frac{1}{8G_{D}}(\varphi_{aR}^{2}+\varphi_{aI}^{2})+\frac{1}{2}\bar{\Psi}\varphi_{ai}\hat{\Gamma}_{ai}\Psi.\label{mf-lagrangian}
\end{eqnarray}
The quark fields can be expressed in the NG basis, $\Psi=(\psi,\psi_{C})^{T}$
and $\bar{\Psi}=(\bar{\psi},\bar{\psi}_{C})$. $\sigma_{u,d}$ are
chiral condensates and can also be derived in the mean field approach.
The inverse propagator then reads
\begin{equation}
S^{-1}=-\left(\begin{array}{cc}
P_{\mu}\gamma^{\mu}+\mu\gamma^{5}-\hat{m} & i\gamma^{5}\tau_{2}J_{a}\Delta^{*}\\
i\gamma^{5}\tau_{2}J_{a}\Delta & P_{\mu}\gamma^{\mu}-\mu\gamma^{5}-\hat{m}
\end{array}\right),\label{S-matr}
\end{equation}
where the quark mass in flavor space is $\hat{m}=(m_{0}+m_{q})\cdot\mathbf{1}_{f}$
with $m_{q}$ the chiral condensate $-2G_{s}(\sigma_{u}+\sigma_{d})$.
In present case only the scalar quark-quark channel is considered.
The original complex diquark field has been decomposed into two real
bosonic fields: the real part $\varphi_{aR}$ and the imaginary part
$\varphi_{aI}$ with $\varphi_{a}=\frac{1}{\sqrt{2}}(\varphi_{aR}+i\varphi_{aI})$.
The last term in Eq. (\ref{mf-lagrangian}) is a Yukawa type quark-quark-diquark
vertex with the index $i=I,R$, with which the diquark dynamic properties
can be studied. $\hat{\Gamma}_{aR,I}=\frac{i}{\sqrt{2}}\gamma^{5}\tau_{2}J_{a}\tau_{1,2}^{NG}$,
where $\tau_{1,2}^{NG}$ are the Pauli matrices in the NG space. There
are additional two tadpole terms: $\varphi_{a}\Delta^{*}$ and $\varphi_{a}^{*}\Delta$,
but in a self-consistent theory all the tadpole terms should cancel
themselves. One can prove that they are canceled by the tadpole terms
of the one-loop diagram generated by the terms $\bar{\psi}i\gamma^{5}\tau_{2}J_{a}\psi_{C}\varphi_{a}^{*}$
and $\varphi_{a}\bar{\psi}_{C}i\gamma^{5}\tau_{2}J_{a}\psi$, 
\begin{eqnarray}
T^{\varphi} & = & T_{0}^{\varphi}+T_{1-loop}^{\varphi}\nonumber \\
 & = & -\frac{i}{4G_{D}}\Delta^{*}+\frac{1}{4}\int_{K}\mathrm{Tr}\left[i\gamma_{5}\tau_{2}J_{a}S_{12}\right]\nonumber \\
 & \equiv & 0.\label{tadpole}
\end{eqnarray}
This relation is satisfied due to the gap equation in the mean field.
Similarly the tadpole terms corresponding to $\varphi_{a}^{*}$ can
be proved to cancel each other. For clarity the tadpole terms are
not included in the Lagrangian. The full diquark propagators are derived
via the Dyson-Schwinger type equation, 
\begin{equation}
D_{i,a}^{-1}(p_{0},\mathbf{p})=-\frac{1}{4G_{D}}-\Pi_{i,a}(p_{0},\mathbf{p}),\label{DSE-diquark}
\end{equation}
where $p_{0}=i2n\pi T$ are the Mastubara frequencies ($n=0,\pm1,\pm2...$)
and $\Pi_{i,a}$ are the diquark self-energies, which have the following
properties in the 2SC phase: $\Pi_{R/I,a}=\frac{1}{2}(\Pi_{0}^{a}\pm\Pi_{1}^{a})$,
and $\Pi_{0}^{1}=\Pi_{0}^{2}\neq\Pi_{0}^{3}$, $\Pi_{1}^{a}=\delta_{a3}\Pi_{1}^{3}$.
The expression for the $\Pi_{0,1}^{a}$ are
\begin{eqnarray}
\Pi_{0}^{1,2} & = & -g^{2}\int_{K}\mathrm{Tr}[S_{22}(K)\gamma^{5}\tau_{2}J_{1,2}S_{11}(P+K)\gamma^{5}\tau_{2}J_{1,2}]\nonumber \\
 & = & 2\int\frac{d^{3}k}{(2\pi)^{3}}\{\frac{e_{1}'\epsilon_{k}^{e'}+\xi_{k}^{e'}}{2e_{1}'\epsilon_{k}^{e'}}\frac{1-f(e_{1}'\epsilon_{k}^{e'})-f(\xi_{p+k}^{e})}{p_{0}-e_{1}'\epsilon_{k}^{e'}-\xi_{p+k}^{e}}\nonumber \\
 &  & +\frac{e_{1}\epsilon_{p+k}^{e}+\xi_{p+k}^{e}}{2e_{1}\epsilon_{p+k}^{e}}\frac{1-f(\xi_{k}^{e'})-f(e_{1}\epsilon_{p+k}^{e})}{p_{0}-\xi_{k}^{e'}-e_{1}\epsilon_{p+k}^{e}}\}c_{k,p+k},\label{pi012}
\end{eqnarray}
 
\begin{eqnarray}
\Pi_{0}^{3} & = & -g^{2}\int_{K}\mathrm{Tr}[S_{22}(K)\gamma^{5}\tau_{2}J_{3}S_{11}(P+K)\gamma^{5}\tau_{2}J_{3}]\nonumber \\
 & = & 4\int\frac{d^{3}k}{(2\pi)^{3}}\frac{e_{1}'\epsilon_{k}^{e'}+\xi_{k}^{e'}}{2e_{1}'\epsilon_{k}^{e'}}\frac{e_{1}\epsilon_{p+k}^{e}+\xi_{p+k}^{e}}{2e_{1}\epsilon_{p+k}^{e}}\frac{1-f(e_{1}'\epsilon_{k}^{e'})-f(e_{1}\epsilon_{p+k}^{e})}{p_{0}-e_{1}'\epsilon_{k}^{e'}-e_{1}\epsilon_{p+k}^{e}}c_{k,p+k},\label{pi03}
\end{eqnarray}
 
\begin{eqnarray}
\Pi_{1}^{1,2} & = & -g^{2}\int_{K}\mathrm{Tr}[S_{12}(K)\gamma^{5}\tau_{2}J_{1,2}S_{12}(P+K)\gamma^{5}\tau_{2}J_{1,2}]\nonumber \\
 & = & 0,\label{pi112}
\end{eqnarray}
 
\begin{eqnarray}
\Pi_{1}^{3} & = & -g^{2}\int_{K}\mathrm{Tr}[S_{12}(K)\gamma^{5}\tau_{2}J_{3}S_{12}(P+K)\gamma^{5}\tau_{2}J_{3}]\nonumber \\
 & = & -\int\frac{d^{3}k}{(2\pi)^{3}}\frac{\Delta_{3}^{2}}{e_{1}e_{1}'\epsilon_{k}^{e}\epsilon_{p+k}^{e'}}\frac{1-f(e_{1}\epsilon_{k}^{e})-f(e_{1}'\epsilon_{p+k}^{e'})}{p_{0}-e_{1}\epsilon_{k}^{e}-e_{1}'\epsilon_{p+k}^{e'}}c_{k,p+k},\label{pi13}
\end{eqnarray}
where the quasi-quark energies are $\xi_{k}^{e}=eE_{k}-\mu$, $\epsilon_{k}^{e}=\sqrt{(\xi_{k}^{e})^{2}+\Delta^{2}}$
with $E_{k}=\sqrt{k^{2}+m_{q}^{2}}$. $c_{k,p+k}$ is the product
of energy projectors, $c_{k,p+k}=[1+ee'\frac{\mathbf{k}\cdot(\mathbf{p+k})+m_{q}^{2}}{E_{k}E_{p+k}}]$,
where summation is implied over $e,e',e_{1},e'_{1}$. 

\begin{figure}
\includegraphics[scale=0.5]{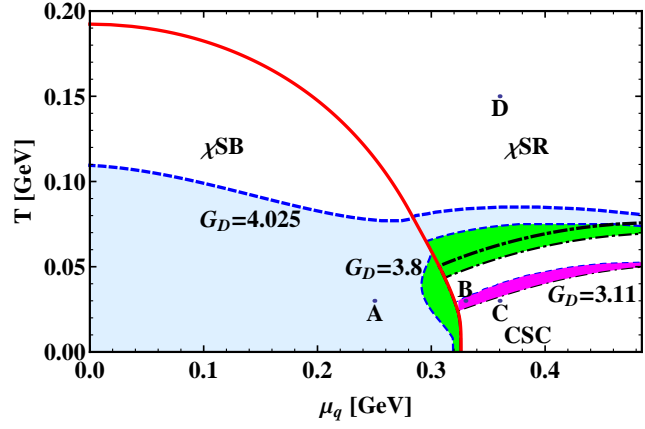}

\caption{\label{fig:phase-diag}The phase diagram obtained within the two flavor
NJL model. }
\end{figure}

With the full propagator of diquarks, one can study the stability
and dissociation properties of diquarks. In Fig. \ref{fig:phase-diag},
the red solid line separates the chiral symmetry broken phase from
the symmetric phase (indicated by $\chi SB/\chi SR$); CSC denotes
the color-superconducting phase. In the chiral symmetry broken phase
and CSC phase, with a strong diquark coupling $G_{D}$, due to nonzero
mass gap $m_{q}$ and the CSC gap $\Delta_{a}$, stable diquark poles
can be found in the window $(-2m_{q},2m_{q})$ and $(-2\Delta_{a},2\Delta_{a})$
respectively in $\mathbf{p}=0$ limit. In $\chi SB$ phase (with a
large $G_{D}$) and normal phase, there is a boundary below which
diquark pole equations $\frac{1}{4G_{D}}+\Pi_{i,a}(p_{0},\mathbf{p})=0$
have solutions. That line is defined as the diquark dissociation boundary,
see the blue dashed lines for three values of the diquark coupling
constant, $G_{D}=3.11,3.8,4.025$ (in units of $\mathrm{GeV}^{-2}$)
in Fig. \ref{fig:phase-diag}. The corresponding regions in Fig. \ref{fig:phase-diag}
are filled with light blue, green, and magenta color, respectively.
These poles also exist in the CSC phases, however, for the sake of
clarity there is no color in the CSC regions. The CSC phases for $G_{D}=3.11,3.8,4.025$
are bounded by the red solid line and the dash-dotted lines from bottom
to top, respectively. Note that the diquark coupling constants we
have chosen here are in the weak-coupling or BCS regime. As we increase
$G_{D}$, Bose-Einstein condensation of diquarks can take place in
the region below the dissociation lines, provided the bare quark mass
is nonzero \cite{Deng:2006ed,Kitazawa:2007zs,Deng:2008ah,Nishida:2005ds,Sun:2007fc,Brauner:2008td,Abuki:2010jq,Basler:2010xy}.
Note that in Ref. \cite{Kitazawa:2007zs}, a vanishing decay width
was imposed as an additional criterion for the location of the dissociation
boundary.

\begin{figure}
\includegraphics[scale=0.3]{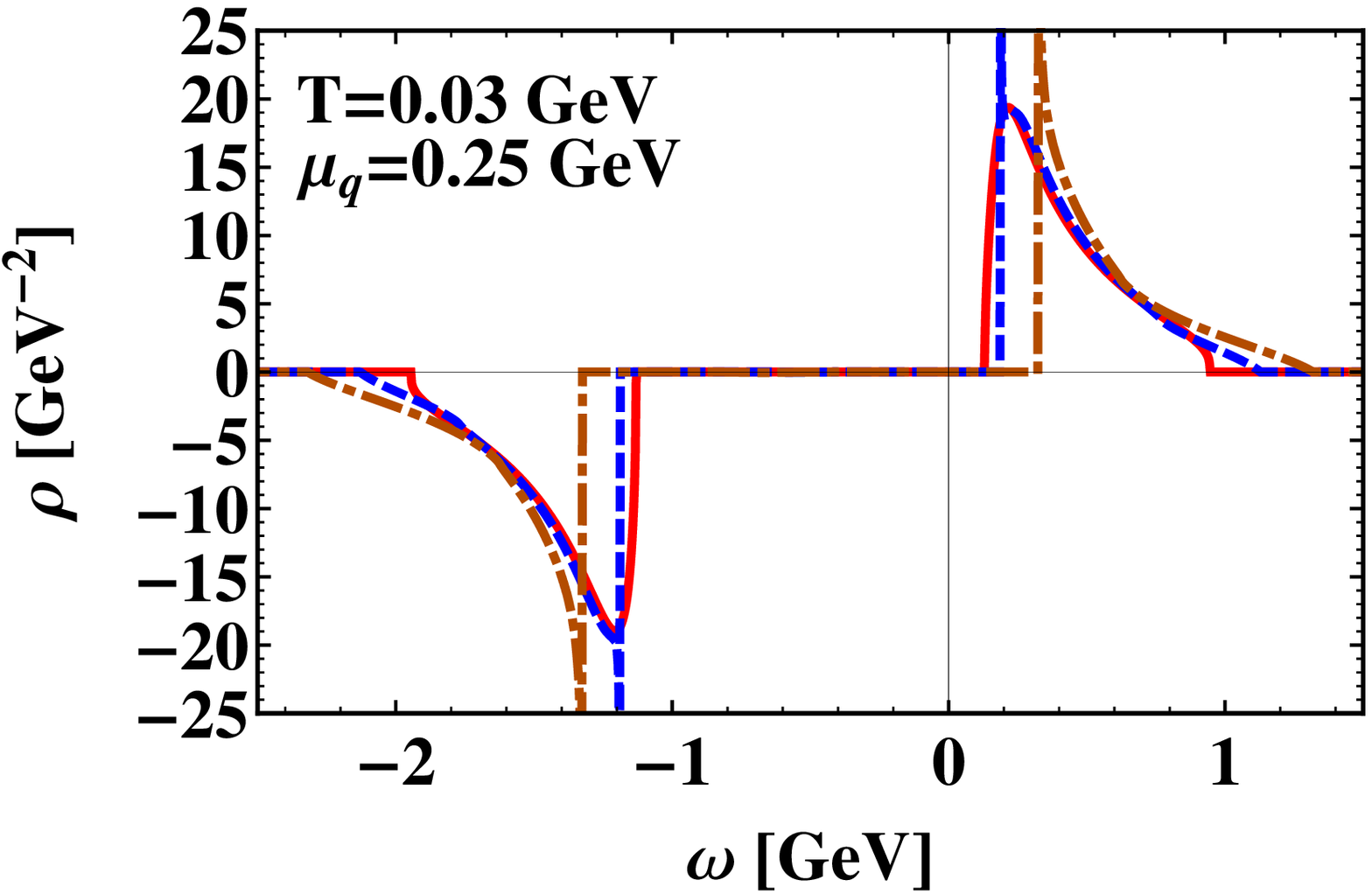}\includegraphics[scale=0.3]{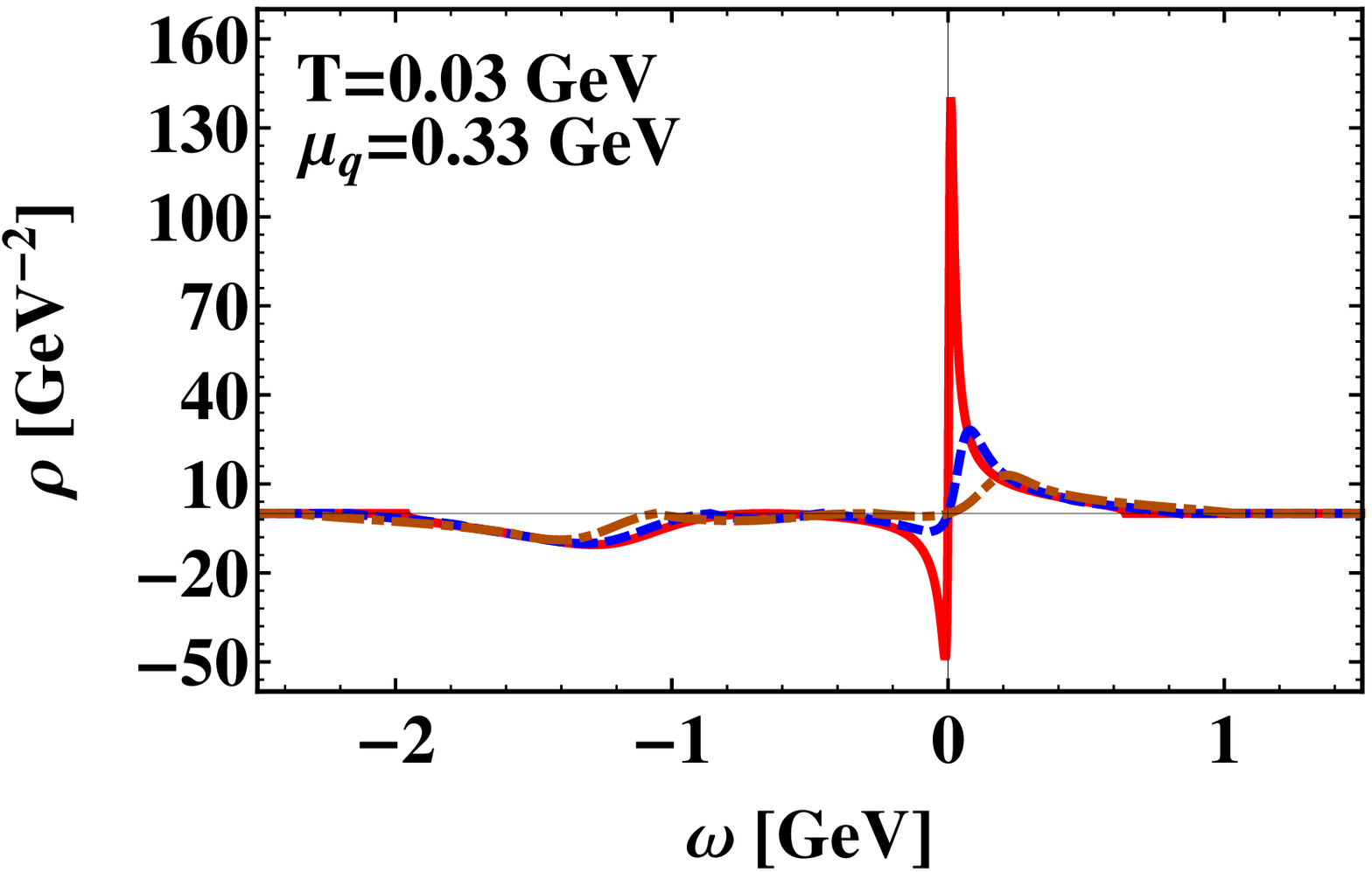}\includegraphics[scale=0.3]{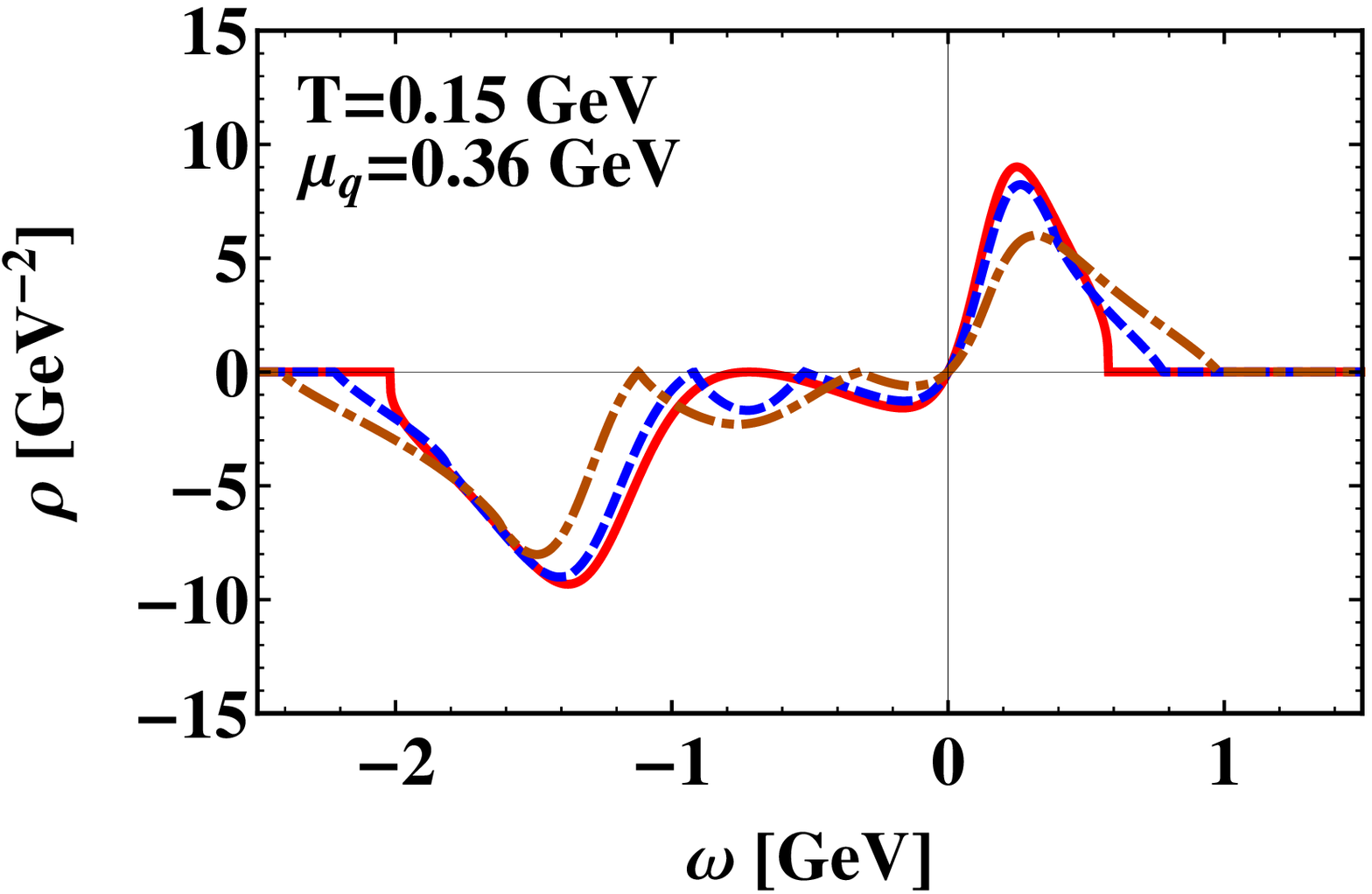}

\includegraphics[scale=0.3]{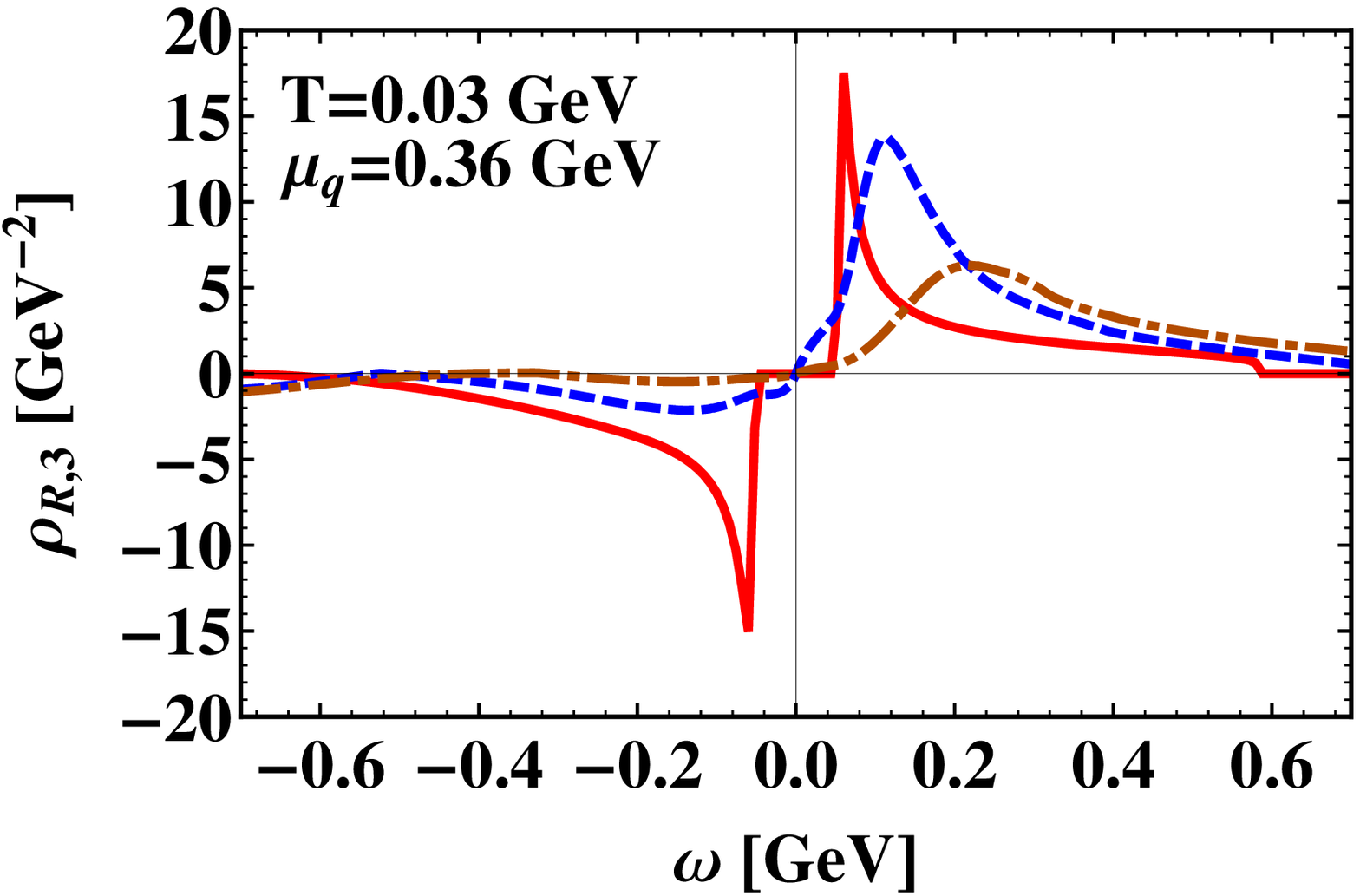}\includegraphics[scale=0.3]{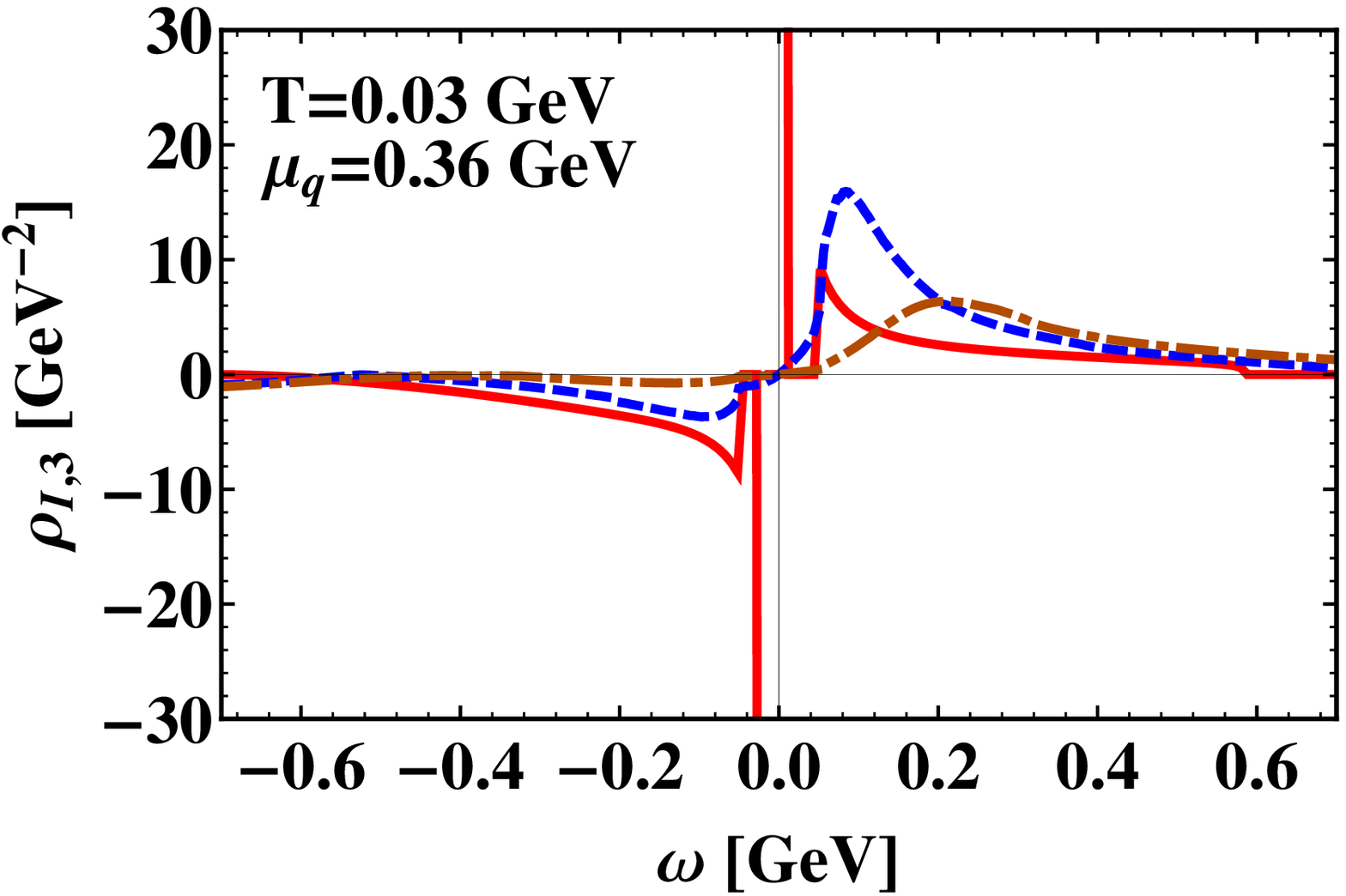}\includegraphics[scale=0.3]{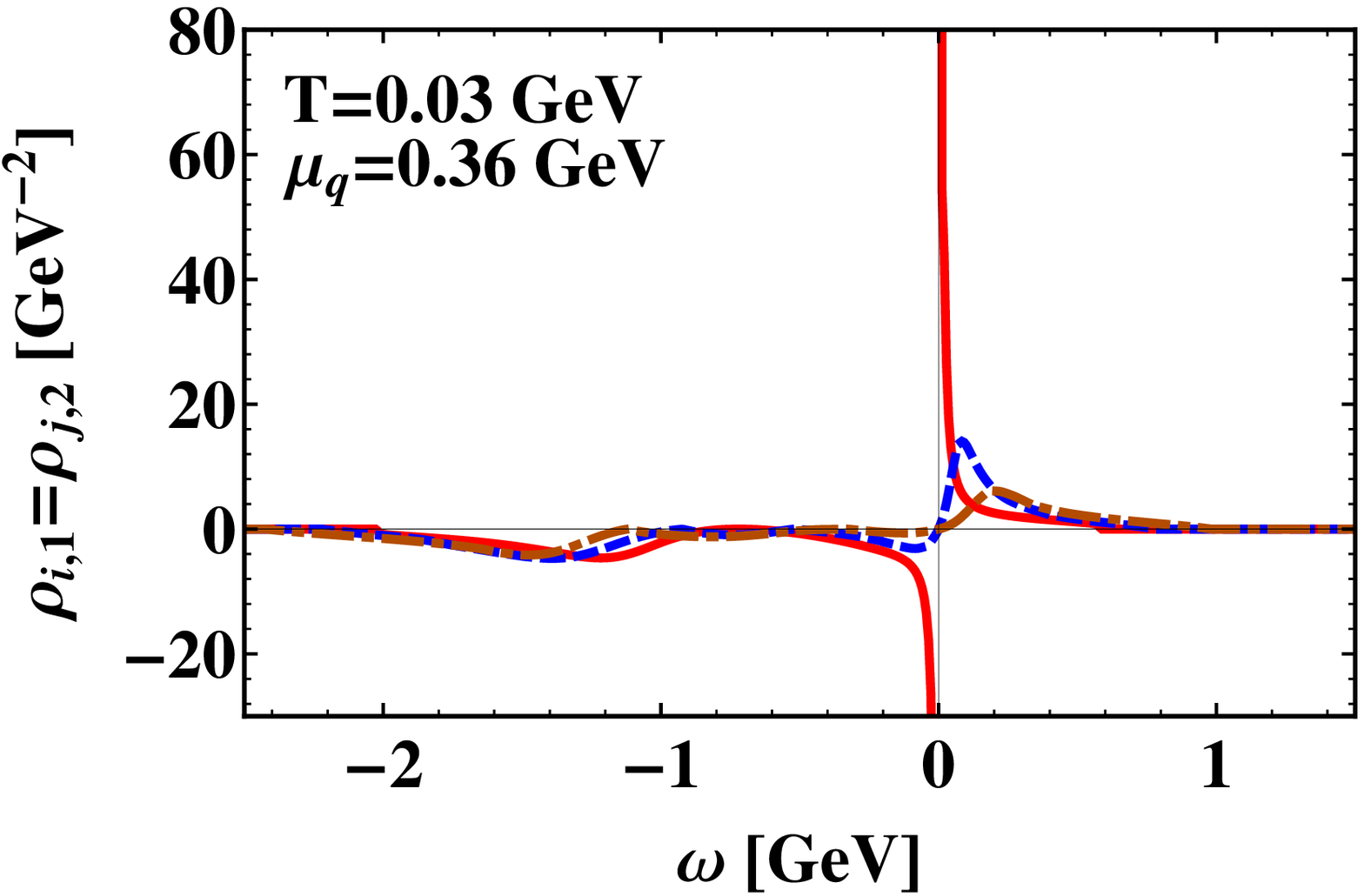}

\caption{\label{fig:diqu-spec}Spectral densities for different values of $T$
and $\mu_{q}$. The upper panels correspond to $(T,\mu_{q})=$(0.03,0.25),
(0.03,0.33), (0.15,0.36) GeV, respectively. The lower panels are for
$(T,\mu_{q})=(0.03,0.36)$ GeV (inside CSC phase). The indices in
the spectral density refer to I,R and colors. }
\end{figure}

In Fig. \ref{fig:diqu-spec}, the diquark spectral densities are presented
for four points: $A$ ($T=0.03$, $\mu=0.25$ GeV), $B$ ($T=0.03$,
$\mu=0.33$ GeV), $C$ ($T=0.03$, $\mu=0.36$ GeV) and $D$ ($T=0.15$,$\mu=0.36$)
in the phase diagram Fig. \ref{fig:phase-diag}. The spectral density
for diquarks is defined as following 
\begin{equation}
\rho_{i,a}(\omega,\mathbf{p})=\frac{1}{\pi}\frac{\mathrm{Im}D_{i,a}^{-1}(\omega+i\eta,\mathbf{p})}{[\mathrm{Re}D_{i,a}^{-1}(\omega+i\eta,\mathbf{p})]^{2}+[\mathrm{Im}D_{i,a}^{-1}(\omega+i\eta,\mathbf{p})]^{2}}\;.\label{eq:spectral-density}
\end{equation}
In each panel there are three curves corresponding to three momenta
$|\mathbf{p}|=0,0.2,0.4$ GeV. The diquark coupling constant is chosen
as a weak one $G_{D}=3.11\mathrm{GeV}^{-2}$. The upper three panels
of Fig. \ref{fig:diqu-spec} correspond to points $A$, $B$, $D$,
from left to right respectively. All the six components of spectral
density with indices $i=R,I$ and $a=1,2,3$ are identical outside
CSC phase. At point $A$ (in $\chi SB$ phase), with a weak diquark
coupling, there are only broad bumps above two times quark mass gap
($|\omega|>2m_{q}$). The point $B$ is located below the diquark
dissociation boundary in normal phase, where diquarks have poles corresponding
to stable diquark resonances. The point $D$ is above the dissociation
boundary in the normal phase, and no diquark pole exists. The broad
bumps in the diquark spectral density indicate unstable diquark resonances.
The lower three panels are for $\varphi_{I}$ and $\varphi_{R}$ diquark
field with color indices 1,2,3 at point $C$ in CSC region. The $I,R$
diquark field with color index 3 are gapped while the others are not.
From the expressions of diquark self-energy, five Nambu-Goldstone
modes are recognized, those are the $I/R$ fields with color indices
$1$ and $2$, and the $I$ field with the color index $3$. They
can be directly proved with the fact that the diquark pole is located
at $(\omega,\mathbf{p})=0$ GeV due to the gap equations for $\Delta_{a}$.
In the lower-middle (spectral density for the $\varphi_{I,3}$ field
) and lower-right panel (spectral density for the $\varphi_{I/R,1/2}$
field), one can find the five Nambu-Goldstone modes at $(\omega,\mathbf{p})=0$
GeV. The lower-left panel is for $\varphi_{R,3}$ field without Nambu-Goldstone
mode. 

\begin{figure}
\includegraphics[scale=0.48]{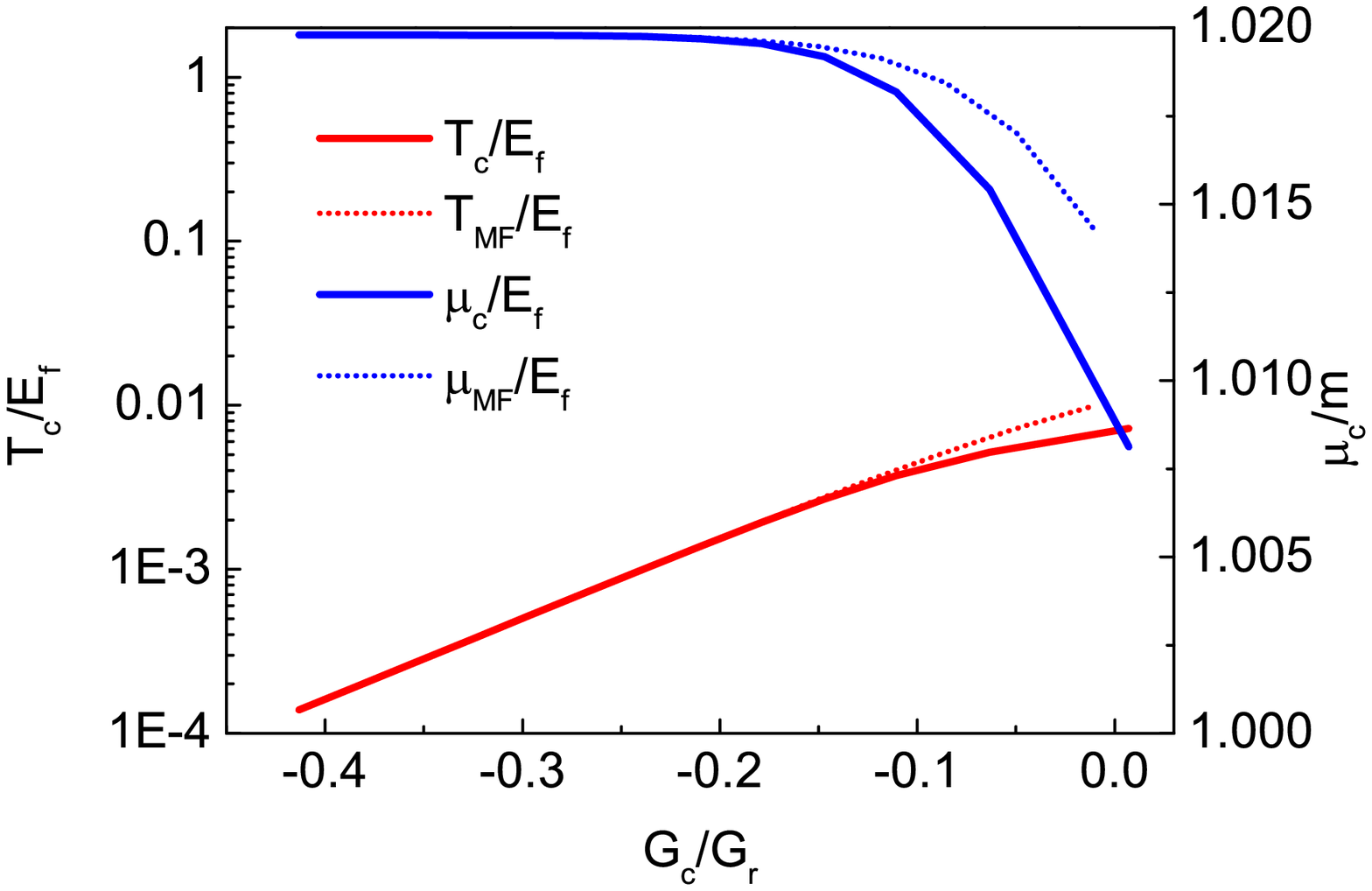} \includegraphics[scale=0.48]{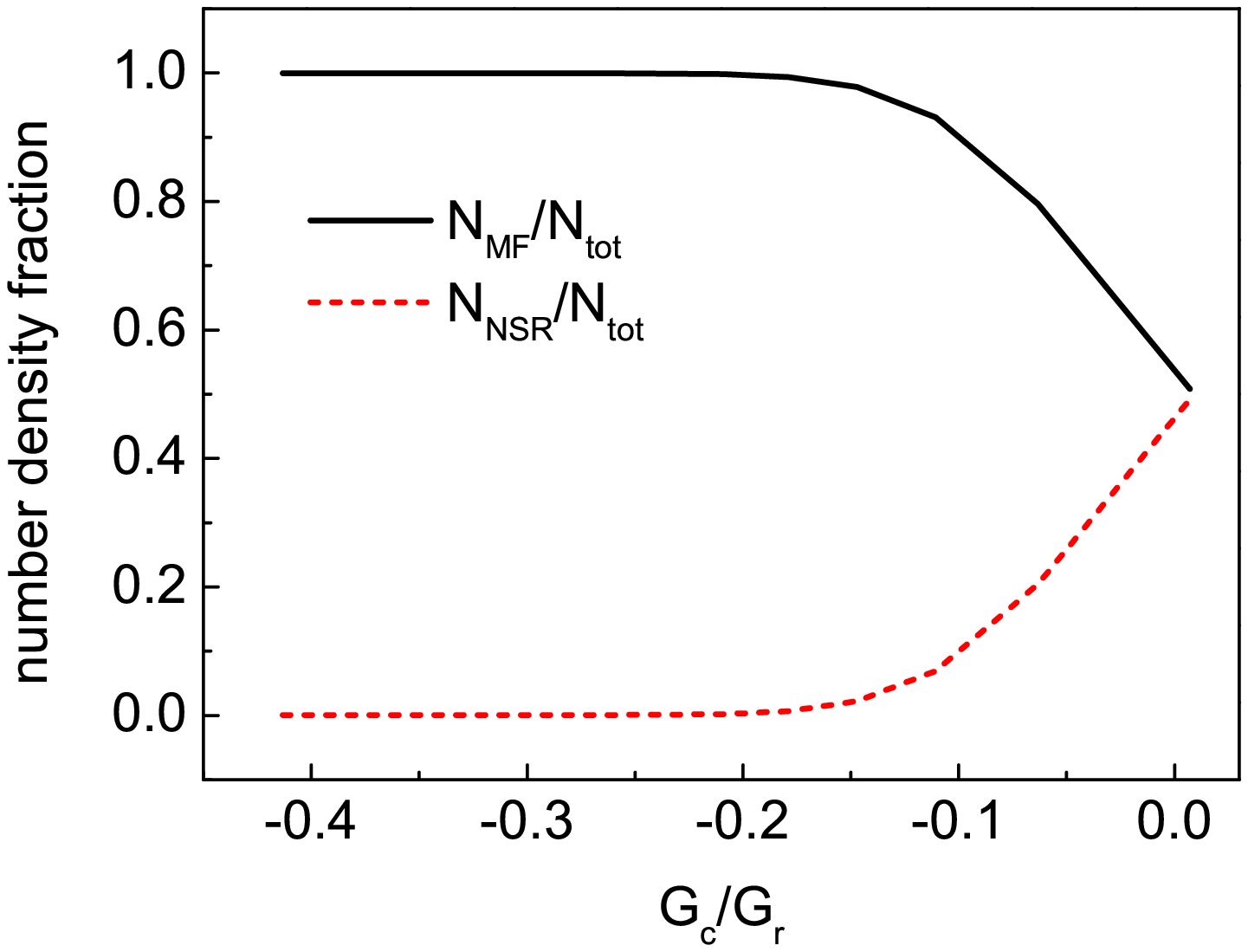}

\caption{\label{fig:njl_crossover}$T_{c}/E_{F}$ and $\mu_{c}/m$ (left panel)
and number density fraction (right panel) as as functions of normalized
diquark coupling constant $G_{c}/G_{r}$. Here $T_{c}$ and $\mu_{c}$
are the CSC critical temperature and chemical potential respectively. }
\end{figure}

To evaluate the thermal diquark contribution to the thermodynamic
potential, Abuki has developed a method to express the potential in
terms of diquark spectral density \cite{Abuki:2006dv}. The total
thermodynamic potential is decomposed into two terms $\Omega=\Omega_{MF}+\Omega_{fluc}$,
where $\Omega_{MF}$ is the mean field potential and $\Omega_{fluc}$
is the thermal diquark contribution. The full diquark propagator at
some coupling constant $4G_{D}={\cal {G}}$ is obtained with the dispersion
relation 
\begin{eqnarray}
D_{i,a}^{{\cal {G}}}(i\omega_{n},\mathbf{p}) & = & \frac{1}{-1/{\cal {G}}-\Pi_{i,a}}=\int_{-\infty}^{\infty}d\omega\frac{\rho_{i,a}^{{\cal {G}}}(\omega,\mathbf{p})}{\omega-i\omega_{n}}.\label{dispersion-relation}
\end{eqnarray}
The thermodynamic potential is given by 
\begin{eqnarray}
\Omega_{fluc} & = & \frac{1}{2}T\sum_{n,i,a}\int\frac{d\mathbf{p}}{(2\pi)^{3}}\log\left[\frac{1}{4G_{D}}+\Pi_{i,a}\right]-(T=\mu=0\text{ part})\nonumber \\
 & = & -\frac{1}{2}\int_{0}^{4G_{D}}\frac{d{\cal {G}}}{{\cal {G}}^{2}}\int\frac{d\mathbf{p}}{(2\pi)^{3}}T\sum_{n,i,a}\int_{-\infty}^{\infty}d\omega\frac{\rho_{i,a}^{{\cal {G}}}(\omega,\bold{p})}{\omega-i\omega_{n}}-(T=\mu=0\text{ part})\nonumber \\
 & = & -\frac{1}{2}\int_{0}^{4G_{D}}\frac{d{\cal {G}}}{{\cal {G}}^{2}}\int\frac{d\mathbf{p}}{(2\pi)^{3}}\int_{-\infty}^{\infty}d\omega\sum_{i,a}\rho_{i,a}^{{\cal {G}}}(\omega,\bold{p})[\frac{1}{2}+n_{B}(\omega)]-(T=\mu=0\text{ part}).\label{potential-spectral-represnetation}
\end{eqnarray}
The advantage of the propagator in terms of the dispersion relation
is that the summation over Matsubara frequency can be analytically
performed. The fluctuation part can be decomposed into two pieces:
the Nozières-Schmitt-Rink term 
\begin{eqnarray}
\Omega_{NSR} & = & -\frac{1}{2}\int_{0}^{4G_{D}}\frac{d{\cal {G}}}{{\cal {G}}^{2}}\int\frac{d\mathbf{p}}{(2\pi)^{3}}\int_{-\infty}^{\infty}d\omega\sum_{i,a}\rho_{i,a}^{{\cal {G}}}(\omega,\bold{p})[n_{B}(\omega)+\theta(-\omega)]\nonumber \\
 & = & -\frac{1}{2}\int\frac{d\mathbf{p}}{(2\pi)^{3}}\int_{-\infty}^{\infty}d\omega\sum_{i,a}\delta_{i,a}(\omega,\bold{p})[n_{B}(\omega)+\theta(-\omega)],\label{potential-nsr}
\end{eqnarray}
 and the quantum fluctuation term 
\begin{eqnarray}
\Omega_{qfl} & = & -\frac{1}{2}\int_{0}^{4G_{D}}\frac{d{\cal {G}}}{{\cal {G}}^{2}}\int\frac{d\mathbf{p}}{(2\pi)^{3}}\int_{-\infty}^{\infty}d\omega\sum_{i,a}\Delta\rho_{i,a}^{{\cal {G}}}(\omega,\bold{p})\frac{\mathrm{sgn}(\omega)}{2}\nonumber \\
 & = & -\frac{1}{2}\int\frac{d\mathbf{p}}{(2\pi)^{3}}\int_{-\infty}^{\infty}\frac{d\omega}{\pi}\sum_{i,a}\Delta\delta_{i,a}(\omega,\bold{p})\frac{\mathrm{sgn}(\omega)}{2},\label{potential-qfl}
\end{eqnarray}
 in which the integral over the spectral density is represented by
the phase shift defined as 
\begin{equation}
\int_{0}^{4G_{D}}\frac{d{\cal {G}}}{{\cal {G}}^{2}}\rho_{i,a}^{{\cal {G}}}=\frac{i}{2}\log\left[\frac{-\frac{1}{4G_{D}}-\Pi_{i,a}(\omega+i\eta,\mathbf{p})}{-\frac{1}{4G_{D}}-\Pi_{i,a}(\omega-i\eta,\mathbf{p})}\right]=\delta_{i,a}(\omega,\mathbf{p}).\label{phase-shift}
\end{equation}
Here $\Delta\rho_{i,a}^{{\cal {G}}}=\rho_{i,a}^{{\cal {G}}}-(T=\mu=0\text{ part})$
and $\Delta\delta_{i,a}=\delta_{i,a}-(T=\mu=0\text{ part})$. $n_{B}$
is the bosonic distribution function. $\Omega_{NSR}$ is the contribution
from the thermal fluctuation which vanishes at zero temperature, while
$\Omega_{qfl}$ remains at zero T. In further calculation the quantum
fluctuation contribution is neglected, and only the thermal effect
is considered. The charge conjugation is maintained without $\Omega_{qfl}$.
There are two special values of $G_{D}$ which should be noticed.
The first one is $1/4G_{0}=-\Pi_{i,a}(2m,\mathbf{0})|_{T=\mu=\Delta=0}$,
with $G_{D}>G_{0}$ the stable diquark bound state can be found, and
with $G_{D}<G_{0}$ there is only unstable diquark resonance. The
diquark self-energy in the normal phase can be separated into the
vacuum part and the matter part as 
\begin{eqnarray}
\Pi(p_{0},\mathbf{p}) & = & \Pi^{mat}(p_{0},\mathbf{p})+\Pi_{T=\mu=0}(p_{0}+2\mu,\mathbf{p})\nonumber \\
 & = & 4\int\frac{d^{3}k}{(2\pi)^{3}}\frac{1-(e+e')/2-f(\xi_{k}^{e'})-f(\xi_{p+k}^{e})}{p_{0}-\xi_{k}^{e'}-\xi_{p+k}^{e}}c_{k,p+k}\nonumber \\
 &  & +4\int\frac{d^{3}k}{(2\pi)^{3}}\frac{1-\Theta(-e')-\Theta(-e)}{p_{0}+2\mu-e'E_{k}-eE_{p+k}}c_{k,p+k},\label{pimat}
\end{eqnarray}
where the first term is a convergent matter term since the integrand
function contains a Fermi distribution function for positive energy.
To remove the divergence in the vacuum term, the self-energy is subtracted
by $\Pi(2m,\mathbf{0})$. Then one can defined a renormalized self-energy
as $\Pi^{\mathrm{ren}}(p_{0},\mathbf{p})=\Pi(p_{0},\mathbf{p})-\Pi(2m,\mathbf{0})$.
Meanwhile the coupling constant is renormalized as 
\begin{equation}
-\frac{1}{4G_{R}}=\frac{1}{4G_{D}}-\frac{1}{4G_{0}},\label{gr}
\end{equation}
where $G_{R}$ is related to the scattering length as $1/4G_{R}=m/4\pi a_{s}$
by taking the low energy limit $p=k\ll m$. The other limit for the
coupling is $G_{c}=-\Pi_{i,a}(0,\mathbf{0})|_{T=\mu=\Delta=0}$. As
$G_{D}$ approaching $G_{c}$ the mass of diquark bound state can
become zero. If $G_{D}>G_{c}$ the vacuum becomes unstable. Considering
a system with total baryonic number density fixed, one obtain with
the fermion number conservation, 
\begin{eqnarray}
N_{tot}(\mu,T) & = & N_{MF}(\mu,T)+N_{NSR}(\mu,T)\nonumber \\
 & = & N_{MF}(\mu,T)+\frac{1}{2}\int\frac{d^{3}\mathbf{p}}{(2\pi)^{3}}\int_{-\infty}^{\infty}\frac{d\omega}{\pi}\sum_{i,a}\frac{\partial\delta_{i,a}}{\partial\mu}(\omega,\bold{p})[n_{B}(\omega)+\theta(-\omega)],\label{density-eq_njl}
\end{eqnarray}
where $N_{MF}$ is the fermion number density from free quarks, the
total quark number density is given by $N_{tot}=N_{c}N_{f}\frac{k_{F}^{3}}{3\pi^{2}}$
with $k_{F}$ being the effective Fermi momentum. Together with the
Thouless condition 
\begin{equation}
-\frac{1}{4G_{D}}-\Pi_{i,a}(0,\mathbf{0})=0,\label{Thouless-condition}
\end{equation}
by tuning $G_{D}$ in the range $[0,G_{0}]$ for weak couplings and
in the range $[G_{0},G_{c}]$ for strong couplings, one can look at
the BEC-BCS crossover along the CSC boundary $(\mu_{c},T_{c})$ with
fixed $k_{F}$. Because in the CSC boundary the gap is always zero,
the six components of the diquark self-energy and the phase shift
in above equations with indices $i=R/I$ and $a=1,2,3$ are the same.
Then the diquark number density can be evaluated as 
\begin{eqnarray}
N_{NSR}(\mu,T) & \approx & -N_{c}\int\frac{p^{2}dp}{2\pi^{2}}\int_{-\infty}^{\infty}d\omega\frac{\partial\mathrm{Re}\Pi}{\partial\mu}\rho(\omega,\bold{p})[n_{B}(\omega)+\theta(-\omega)].\label{nnsr-appr1}
\end{eqnarray}
When in strong couplings, $G_{R}>0$, the diquark spectral density
$\rho(\omega,\bold{p})=\rho_{c}(\omega,\bold{p})+\rho_{\delta}(\omega,\bold{p})$
consists of two parts, the continuous part (unbound part) and the
pole part (bound part), with $\rho_{\delta}=Z_{B}\delta(\omega+2\mu-E_{Bp})+Z_{\bar{B}}\delta(\omega+2\mu+E_{\bar{B}p})$
where $Z_{B/\bar{B}}=-\frac{1}{\partial\mathrm{Re}\Pi/\partial\mu}\mid_{\omega=\pm E_{B/\bar{B}p}-2\mu}$.
$E_{B/\bar{B}p}$ is the energy of the diquark and anti-diquark bound
state with three momentum $p$. As a consequence the diquark number
density is also divided into two parts, the unbound diquark contribution
$N_{un}$ and the bound diquark $N_{B}$ contribution. In the weak
coupling case $G_{R}<0$, only the unbound part is left. In Fig. \ref{fig:njl_crossover},
the parameters are set to $\Lambda=0.7$ GeV, $\Lambda_{B}=0.65$
GeV, $m=0.2\Lambda$, and the total fermion number density is fixed
at $k_{F}=0.2m$. With these parameters, one studies the diquark fluctuation
effect near the unitary limit. The left panel shows the CSC critical
temperature and chemical potential as functions of diquark normalized
coupling constant. As the coupling becomes stronger, $T_{c}$ rises
and $\mu_{c}$ decreases. The effect of the diquark fluctuation is
found to lower $T_{c}$ and $\mu_{c}$ (solid curves) comparing to
the mean field results $T_{MF}$ and $\mu_{MF}$ (dotted curves).
At a very weak coupling, the CSC boundary can be well approximated
by the mean field approximation. The right panel shows fermion number
density fraction of quarks and diquarks. The diquark component increases
with the diquark coupling constant. In the strong coupling side with
$G_{R}>0$, $T_{c}$($\mu_{c}$) continuously increases (decreases)
with the coupling constant. Both the free fermion fraction $N_{MF}/N_{tot}$
and unbound diquark fraction $N_{un}/N_{tot}$ decrease with the coupling
constant once the effective chemical potential of the system becomes
negative $\mu-m<0$, while the diquark bound state contribution becomes
nonzero and finally dominant.

In this section, we summarize recent results on the diquark spectral
densities in different regions of the phase diagram. The quark mass
and CSC gap serve as boundaries for stable diquarks. In the 2SC phase,
there are gap structures in the spectral density of the thermal diquark
with red and green colors. The infinite peaks at $(\omega,\mathbf{p})=(0,\mathbf{0})$
GeV indicate five Goldstone modes. In the NJL model, the scattering
length $a_{s}$ of fermions can be determined by taking the non-relativistic
limit for the $T$ matrix, which is an advantage in comparison with
the boson-fermion model beyond leading order calculation. Considering
a total baryonic number density conserved system, tuning the scattering
length along the CSC boundary where the Thouless condition is satisfied,
the boundary of the CSC and the number density fraction of free quarks
and diquarks are calculated. When the $1/a_{s}$ is negative and small,
the diquark contribution to the number density is negligible and the
system is in the pure BCS regime. The diquark unstable resonance has
a remarkable contribution when the absolute value of the effective
chemical potential is small. When $1/a_{s}$ is positive, the diquark
bound state will form. Once the effective chemical potential becomes
negative, increasing $1/a_{s}$ the number density fraction of diquark
bound states becomes nonzero and quickly dominates and the system
settles in a deep BEC regime.

\section{Baryon as quark-diquark collective modes \label{sec:baryon}}

Baryon can be studied as a bound state of three constituent quarks
\cite{Oettel:2001kd,Oettel:2000jj,Cloet:2005pp,Eichmann:2007nn,Eichmann:2008ef,Eichmann:2009qa}.
Quarks and gluons carrying color charges are confined inside baryons
and mesons. The static properties of baryons in vacuum can be studied
with the relativistic Faddeev equation \cite{Pepin:1999hs,Ishii:1995bu}.
In this approach, the baryon is assumed to be stable with a separable
form of the T- matrix. The original Faddeev equation can be reduced
to the Bethe-Salpeter equation (BSE), which is an eigen-equation for
the baryonic vertex. But with a nonzero temperature and baryon density,
the Faddeev equation or the BSE is hard to solve numerically. The
way around is to take the static approximation for the intermediate
quark propagator in the Faddeev equation. Then the baryon can be studied
at nonzero temperature by a two step process: first, the thermal diquark
propagator is simulated by the Dyson-Schwinger equation, and second,
the diquark is coupled with another quark to form a baryon. In previous
works, the diquark is always assumed to be stable, but this is not
necessarily true. Baryon can be a bound state of a quark and an unstable
diquark, which is like a borromean state \cite{Zhukov:1993aw} in
nuclear and atom physics.

To add the baryon field into the Lagrangian (\ref{mf-lagrangian}),
a coupling term of the quark-quark-diquark-diquark is introduced as
\begin{eqnarray}
{\cal L}_{B} & = & G_{B}\varphi_{a}^{\dagger}\bar{\psi}_{a}\psi_{b}\varphi_{b}\nonumber \\
 & \simeq & -\frac{1}{2G_{B}}\overline{\mathbf{B}}\mathbf{B}+\frac{1}{2}\overline{\mathbf{B}}\widehat{\Gamma}_{Bi}\Psi_{a}\varphi_{ai}+\frac{1}{2}\varphi_{ai}\overline{\Psi}_{a}\widehat{\Gamma}_{Bi}^{*}\mathbf{B}\;.\label{lag-b}
\end{eqnarray}
Here, $\psi_{a}\varphi_{a}=\left\langle \psi_{a}\varphi_{a}\right\rangle +\beta_{a}$,
and the baryonic field is defined as $B=G_{B}\left\langle \psi_{a}\varphi_{a}\right\rangle $.
The terms of order $O(\beta_{a}^{2})$ is neglected. Actually this
is equivalent to take the static approximation in Faddeev equation.
The baryonic fields in the NG basis are then denoted by $\mathbf{B}=(B,B_{c})^{T}$
and $\overline{\mathbf{B}}=(\overline{B},\overline{B}_{c})$. The
baryon-quark-diquark vertices are $\widehat{\Gamma}_{BR}=\frac{1}{\sqrt{2}}1_{NG}$
and $\widehat{\Gamma}_{BI}=i\frac{1}{\sqrt{2}}\tau_{3}^{NG}$, respectively.
The sum of the Lagrangians (\ref{mf-lagrangian}) and (\ref{lag-b})
is the starting point for the further treatment. 

The 11-component in the NG space of the inverse baryon propagator
is $S_{B}^{-1}=-1/(2G_{B})-\Sigma$, where 
\begin{equation}
\Sigma(P)=-\frac{1}{4}\sum_{a}\int_{K}S_{11}^{a}(P-K)[D_{R,a}(K)+D_{I,a}(K)]\label{eq:baryon-selfen}
\end{equation}
is the 11-component of the baryon self-energy. The quark propagator
in the NG space, $S_{11}^{a}$, is diagonal in color space. In presence
of a non-vanishing diquark condensate, $S_{11}^{1}=S_{11}^{2}\neq S_{11}^{3}$.
If the diquark condensate vanishes, $S_{11}^{1}=S_{11}^{2}=S_{11}^{3}$
and $D_{R,a}=D_{I,b}$ for any $a,b$. Inserting the spectral density
form of the diquark full propagator Eq. (\ref{dispersion-relation})
into Eq. (\ref{eq:baryon-selfen}), the summation over Matsubara frequency
can be handled. Then the positive energy component of the baryon full
propagator can be extracted with energy projectors, $S_{B,+}^{-1}(p_{0},\mathbf{p}=\mathbf{0})=\frac{1}{2}\mathrm{Tr}\left[S_{B}^{-1}\Lambda_{\mathbf{p=0}}^{+}\gamma^{0}\right]$,
where $\Lambda_{\mathbf{p}}^{s}$ is the energy projector $\Lambda_{\mathbf{p}}^{s}=\frac{1}{2}\left[1+s\left(\gamma_{0}\gamma\cdot\mathbf{p}+\gamma_{0}M_{B}\right)/E_{p}\right]$,
with $E_{p}=\sqrt{p^{2}+M_{B}^{2}}$ and $s=\pm1$. In the homogeneous
limit, $\mathbf{p}=\mathbf{0}$, the energy projector assumes a simple
form, $\Lambda_{\mathbf{p=0}}^{s}=\frac{1}{2}(1+s\gamma_{0})$, which
is independent of $M_{B}$. The full expression of the positive energy
component of the baryon propagator is,

\begin{eqnarray}
\mathcal{S}_{B,+}^{-1}(l_{0},\mathbf{0}) & = & \frac{1}{2}\mathrm{Tr}\left\{ [-1/(2G_{B})-\Sigma^{R}]\Lambda_{+}^{\mathbf{0}}\gamma^{0}\right\} \nonumber \\
 & = & -1/(2G_{B})+\int\frac{p^{2}dp}{4\pi^{2}}\int_{-\infty}^{\infty}\frac{d\omega}{\pi}\{4\rho_{R/I}^{1}(\omega,\mathbf{p})\frac{e_{1}\epsilon_{\mathbf{p}}^{e}+\xi_{\mathbf{p}}^{e}}{2e_{1}\epsilon_{\mathbf{p}}^{e}}\frac{f(e_{1}\epsilon_{\mathbf{p}}^{e})+n(\omega)}{l_{0}-\omega+e_{1}\epsilon_{\mathbf{p}}^{e}}\nonumber \\
 &  & +[\rho_{R}^{3}(\omega,\mathbf{p})+\rho_{I}^{3}(\omega,\mathbf{p})]\frac{f(\xi_{\mathbf{p}}^{e})+n(\omega)}{l_{0}-\omega+\xi_{\mathbf{p}}^{e}}\}(1+e\frac{m_{q}}{E_{\mathbf{p}}}).
\end{eqnarray}
where summation is implied over $e,e_{1}$. 

\begin{figure}
\includegraphics[scale=0.32]{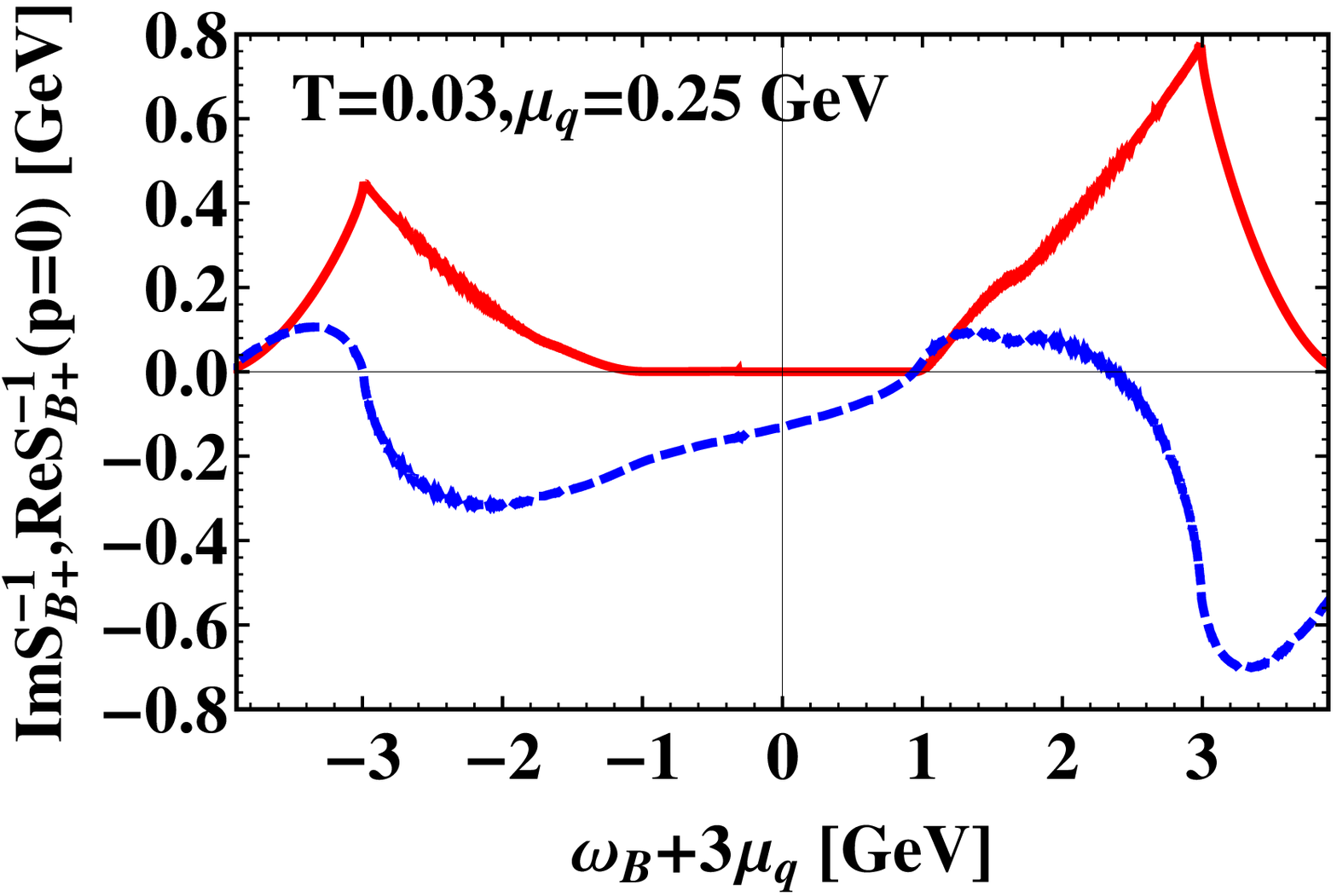} \includegraphics[scale=0.32]{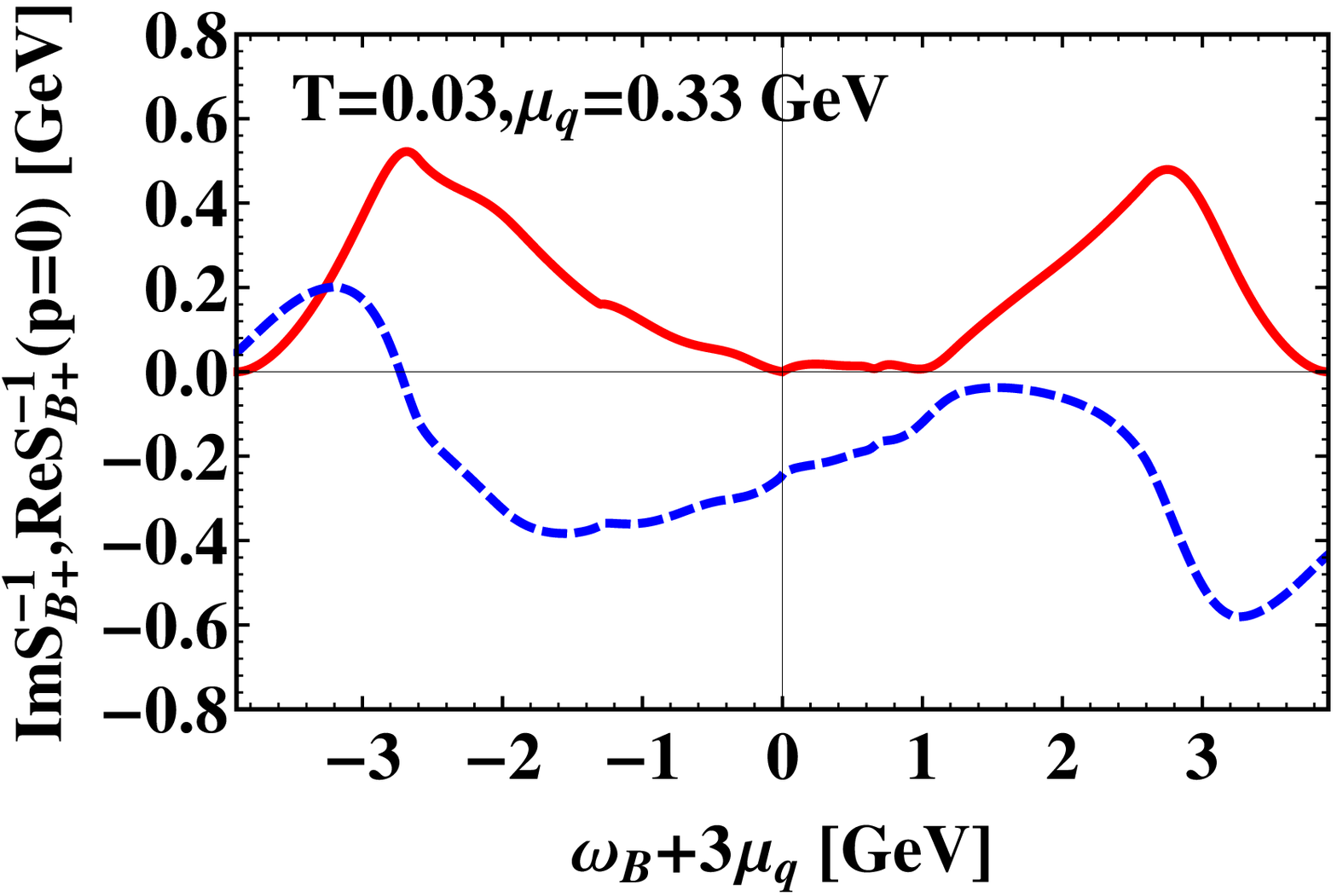}

\includegraphics[scale=0.32]{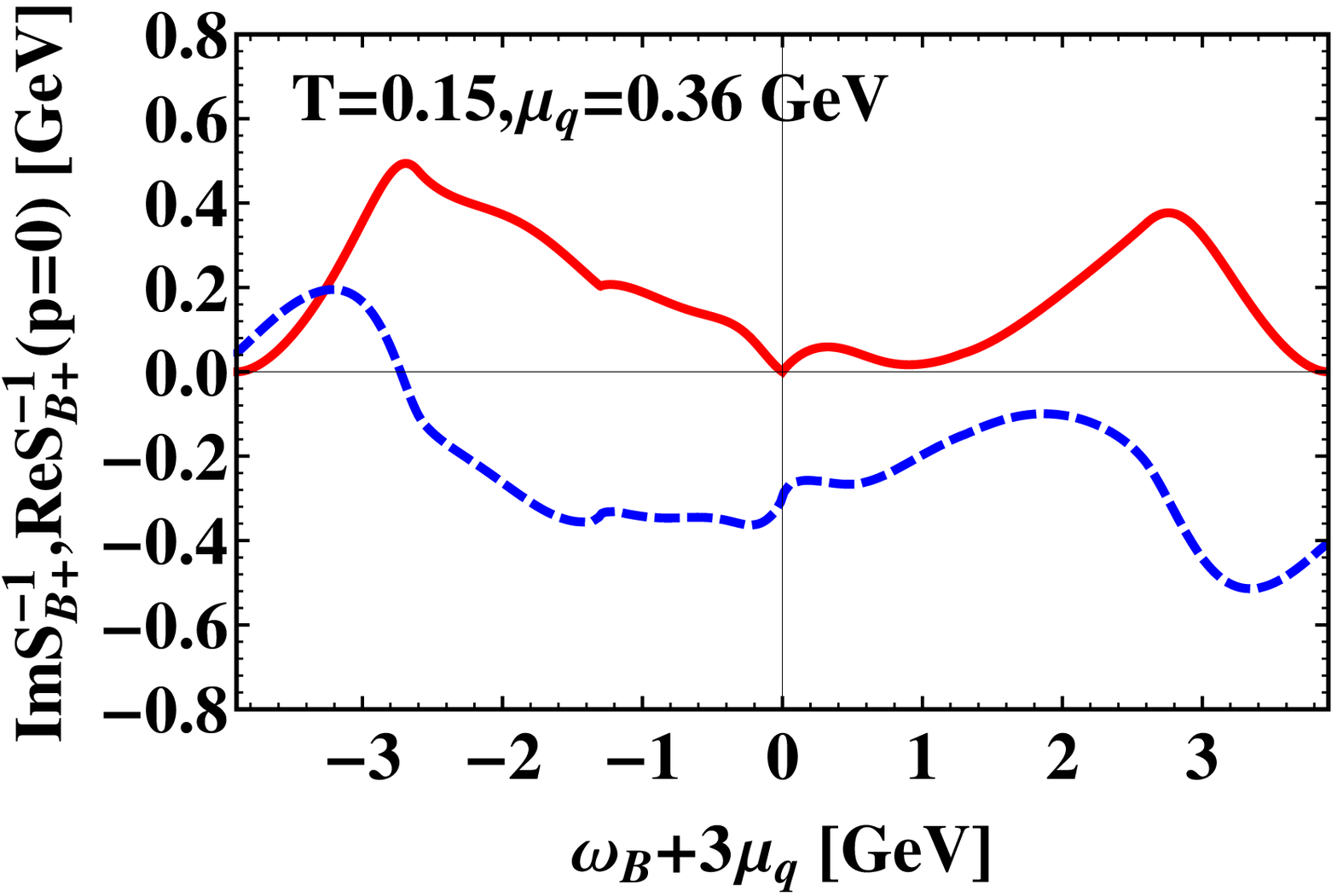} \includegraphics[scale=0.32]{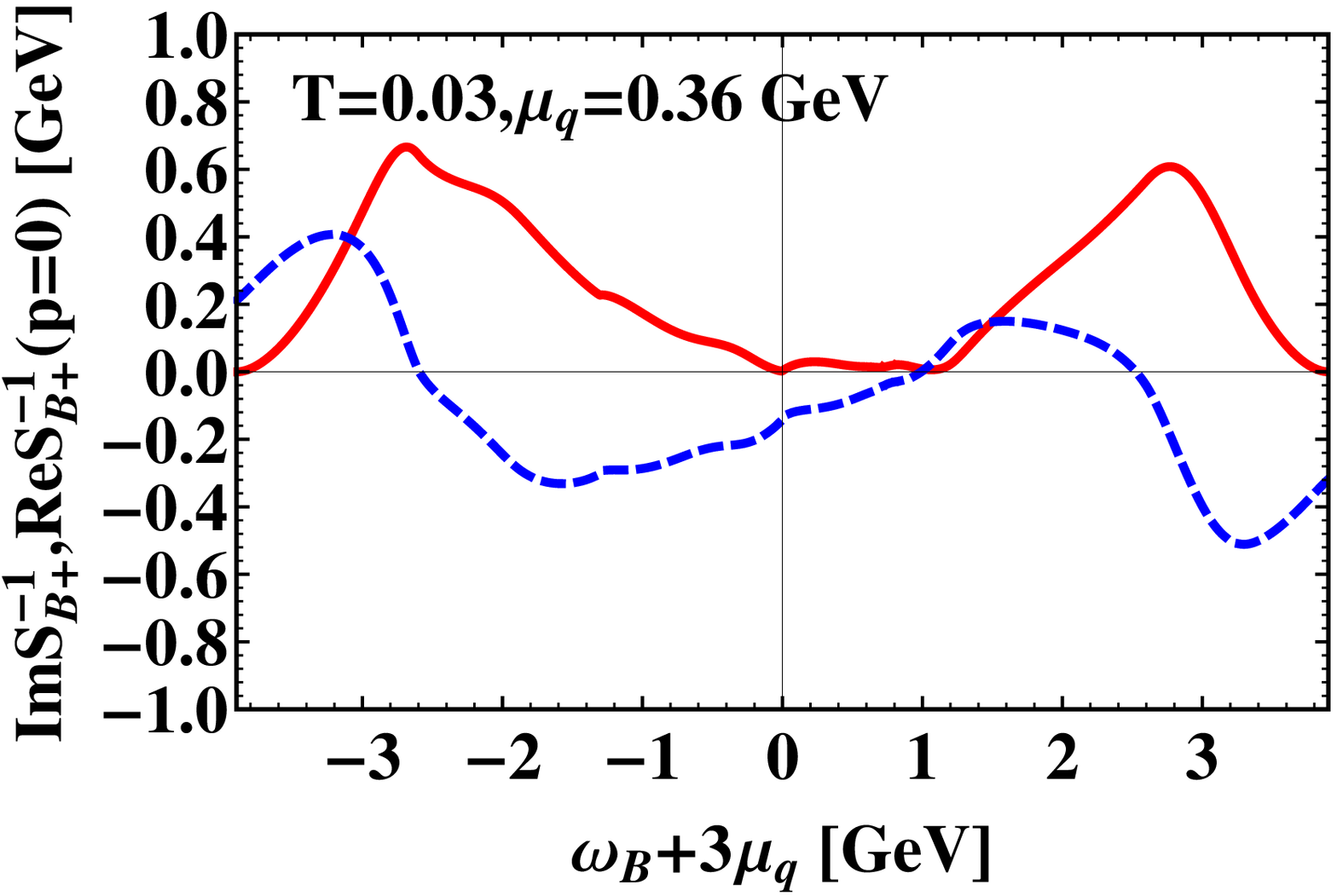}

\caption{\label{fig:re-im-pole-l}The real (blue dashed) and imaginary (red
solid) parts of the inverse propagators for baryons as functions of
energy $\omega$ at different $T$ and $\mu_{q}$. From top to bottom,
the first panel: $T=0.03$ and $\mu_{q}=0.25$ (point A). The second
panel: $T=0.03$ and $\mu_{q}=0.33$ (point B). The third panel: $T=0.15$
and $\mu_{q}=0.36$ (point D). The fourth panel: $T=0.03$ and $\mu_{q}=0.36$
(point C). All units in GeV. }
\end{figure}

Fig. \ref{fig:re-im-pole-l} shows the real and imaginary parts of
the inverse retarded Greens function for baryons (positive energy
component), again at points A,B,C, and D in the phase diagram of Fig.
\ref{fig:phase-diag}. The diquark coupling constant is taken to be
weak, $G_{D}=3.11$ GeV$^{-2}$. The constant $G_{B}=10.04$ GeV$^{-1}$
is chosen to make the baryon mass of $0.94$ GeV in vacuum. In the
chiral symmetry broken phase with $m_{q}\neq0$ and $\Delta=0$ (point
A), there are no diquark condensates or resonances but stable baryon
resonances: in the upper-left panel, one can see that $\mathrm{Re}S_{B+}^{-1}(\omega_{B},\mathbf{0})=0$
has a solution at $\omega_{B}+3\mu_{q}\approx0.94$ GeV, i.e., close
to the nucleon rest mass. There is a region of $\omega_{B}\in[-3(m_{q}+\mu_{q}),3(m_{q}-\mu_{q})]$
or $M_{B}\in[-3m_{q},3m_{q}]$, where the imaginary part $\mathrm{Im}S_{B+}^{-1}(\omega_{B},\mathbf{0})$
is very small (smaller than $10^{-6}$ GeV) in the homogeneous limit.
The position is just inside this region, i.e., $M_{B}<3m_{q}$: the
baryon weighs less than its constituents. It is therefore stable,
although its constituents by themselves are unbound, like in a Borromean
state in atomic or nuclear physics. The upper-right panel shows the
case with diquark resonances but outside the CSC phase (point B).
There is no positive energy baryon pole in this case. In the region
of higher temperatures and quark chemical potentials where chiral
symmetry is restored and where there are neither diquark condensates
nor resonances (point D), there are also no baryon resonances and
the absolute value of $\mathrm{Im}S_{B+}^{-1}$ is very large. This
case is shown in the third panel. In the CSC phase (point C), there
are baryon poles but with nonzero imaginary parts, indicating unstable
baryon resonances, as shown in the fourth panel.

There are a lot of works using the simplified Faddeev equation by
static approximation to study baryon properties in vacuum and nuclear
matter, see, for example, Ref. \cite{Bentz:2001vc,Bentz:2002um}.
But the issues the authors focus on are different. To give the diquark
propagator they used the proper time regularization method which introduces
an effective confinement, but the method is not applicable to nonzero
temperature case. The diquark T-matrix is approximated by the constant
term $\frac{1}{4G_{D}}$ plus the pole terms, which is equivalent
to the assumption of a stable form diquark propagator, while as presented
in Fig. \ref{fig:re-im-pole-l}, the baryon can also be formed by
an unstable diquark and a quark. The pole approximation for the diquark
propagator will miss some of important physics like Borromean state.
In further calculation the physics mass of baryon bound state or baryon
resonance can be obtained at any given $T$ and $\mu$ on the phase
diagram based on the mean field approximation by the NJL. In the present
approach, at low temperatures, the baryon mass have only a slight
decrease when the chemical potential rises, but in Ref. \cite{Bentz:2001vc,Bentz:2002um}
baryon mass decreases significantly. The reason is that the vector
meson is not included and a large baryon number density could not
be obtained by increasing the chemical potential. Also the confinement
mechanism at finite temperatures should be incorporated. In Ref. \cite{Gastineau:2005wm},
the static approximation and the stable diquark are used. The authors
considered a three-flavor NJL model, where the baryon mass is found
to decrease by $25\%$ at normal nuclear matter densities. These issues
can also be considered in the present framework. 

In summary, diquark propagating modes are derived with an NJL-type
model in different regions of the phase diagram of strongly interacting
matter. Baryon formation and dissociation in dense nuclear and quark
matter is then studied via the baryon poles and spectral densities,
incorporating the previously obtained diquark propagator. The stable
baryon resonances with zero width are present in the phase of broken
chiral symmetry, where the diquarks could be an unstable resonance.
This indicates that the baryon can be a Borromean like bound state.
There are no baryon poles in the chirally symmetric phase. In the
CSC phase, baryon poles exist, but they are found to be unstable due
to a sizable width.

\section{Nonlocal extension of NJL model and its applications}

The NJL-type model applied to quarks is a successful schematic effective
theory for QCD, in which the interaction between quarks are described
by point-like couplings. The model can be used to study the spontaneous
chiral symmetry breaking and quark pairings. But the basic version
of the NJL model have some shortcomings. For example, the NJL model
can not exhibit confinement as in QCD. The other one is the constituent
quark mass is independent of momentum which is in conflict with the
lattice data and the Dyson-Schwinger equation from QCD. The second
point can be studied by a nonlocal extension of the point-like coupling
in the NJL model. The nonlocal quark current reads, 
\begin{equation}
J_{M}=\int d^{4}x_{1}d^{4}x_{2}f(x_{1})f(x_{2})\bar{q}(x-x_{1})\Gamma_{M}q(x+x_{2}),\label{nonlocal_current}
\end{equation}
in which $M=\sigma,\pi$ stand for the scalar meson and the pion.
$\Gamma_{\sigma/\pi}$ are $1$ and $\gamma^{5}\tau_{a}$ respectively
with flavor indices $a=1,2,3$. $f$ is a form factor in momentum
space, usually taken as a Gaussian type $f^{2}(p^{2})=\exp(-p^{2}/\Lambda^{2})$.
The basic NJL model is non-renormalizable and some form of ultraviolet
regularization is necessary with a cutoff parameter which is a part
of the model. But in the nonlocal NJL model the momentum integration
are automatically convergent in the loop diagrams and no additional
regularization method is needed \cite{Radzhabov:2010dd}. 

The parameters of the model including the coupling constant $G$,
the quark current mass $m_{c}$ and the momentum cutoff $\Lambda$
are determined by fitting the pion mass $M_{\pi}$ and the decay constant
$f_{\pi}$ in vacuum. The meson sectors are defined by introducing
the auxiliary scalar field $\tilde{\sigma}=G\left<J_{\sigma}(x)\right>$
and pseudo-scalar $\pi^{a}=G\left<J_{\pi}(x)\right>$. The dressed
quark propagator in the mean field approximation is determined by
the following equation 
\begin{eqnarray}
S(p)^{-1} & = & \slashed{p}-m_{c}-\Sigma(p),\label{nonlocal_qk_prop}
\end{eqnarray}
where $\Sigma(p)$ is the self-energy which turn out to be $\Sigma(p)=m_{d}f^{2}(p^{2})$,
with $m_{d}=iG\Gamma_{M}\int_{K}\mathrm{Tr}[\Gamma_{M}S(k)f^{2}(k^{2})]$
a momentum independent constant serving as an order parameter for
the dynamical chiral phase transition. The 1PI diagram of the self-energy
vanishes due to the integration for $x_{1},x_{2}$ goes from $-\infty$
to $+\infty$. Comparing to the classic NJL model, the constituent
quark mass now depends on the three momentum by a Gaussian factor.
In Dyson-Schwinger equation from QCD, the dressed quark propagator
has the form $S(p)=Z(p^{2})[\slashed{p}-M(p^{2})]$. The renormalization
function $Z(p^{2})$ can also be obtained in the nonlocal NJL framework
by considering the thermal meson correlation beyond mean field approximation
\cite{Radzhabov:2010dd}.

The meson propagator is given by RPA for the Bethe-Salpeter equation
$D^{M-1}(p)=-G^{-1}+\Pi^{M}(p)$, with $\Pi^{M}(p)$ polarization
function in the mean field approximation \cite{Radzhabov:2010dd},
\begin{equation}
\Pi^{M}(p)=i\int_{K}f^{2}[(k+p/2)^{2}]f^{2}[(k-p/2)^{2}]\mathrm{Tr}[S(k+p/2)\Gamma_{M}S(k-p/2)\Gamma_{M}].
\end{equation}
The pion mass is obtained by the pole condition $-G^{-1}+\Pi^{\pi}(p^{2}=m_{\pi}^{2})=0$.
Then the quasi-particle propagator can be written as $D^{\pi}(p)=\frac{g_{\pi}^{2}}{p^{2}-m_{\pi}^{2}}$
with $g_{\pi}=\frac{1}{\partial{\Pi^{\pi}(p)}/\partial p^{2}}\mid_{p^{2}=m_{\pi}^{2}}$.

To calculate the pion weak decay constant, the weak current is introduced
by a delocalization procedure for the quark fields \cite{Radzhabov:2010dd},
that is
\begin{equation}
q(y)\rightarrow Q(x,y)=E(x,y)q(y),\label{delocalization}
\end{equation}
where $E(x,y)=\mathcal{P}\exp{i\int_{x}^{y}dz^{\mu}[\mathcal{V}_{\mu}^{a}(z)+\mathcal{A}_{\mu}^{a}(z)\gamma^{5}]T^{a}}$
is the Schwinger phase factor, $\mathcal{V}_{\mu}^{a}(z)$ and $\mathcal{A}_{\mu}^{a}(z)$
are vector and axial-vector gauge fields respectively. The nonlocal
current is modified as 
\begin{equation}
J_{M}=\int d^{4}x_{1}d^{4}x_{2}f(x_{1})f(x_{2})\bar{Q}(x-x_{1},x)\Gamma_{M}Q(x,x+x_{2}).\label{nonlocal_current_delo}
\end{equation}
Then the weak vertices are introduced in the present nonlocal NJL
model of two types: the weak current coupled with a quark $\Gamma^{5}$
and the weak current coupled with the quark meson vertex $\Gamma^{5M}$.
Both give rise to the bubble diagrams which contribute to pion weak
decay. 

The nonlocal extension of the quark NJL model is inspired and therefore
reflects some important features of QCD. The momentum dependence of
the constituent quark mass can be considered and the original sharp
momentum cutoff is replaced by a smooth one. Then all the loop diagrams
are automatically convergent and hence the cutoff dependence of the
model is highly weaken. The last point can be looked at from the calculation
of the meson loop effect in the $1/N_{c}$ expansion. In the nonlocal
study \cite{Radzhabov:2010dd}, the next to leading order contribution
to the quark condensate turn out to be positive in vacuum, in contrast
to the result obtained from the local NJL model \cite{Oertel:2000jp},
in which the beyond mean field correction is a nonlinear function
of the momentum cutoff $\Lambda_{M}$.

\section{Ginzburg-Landau effective theory in three-flavor dense quark matter
with axial anomaly}

The QCD phases in high density regime are controlled by the chiral
and diquark condensates $\phi=\left\langle \overline{q}q\right\rangle $
and $d=\left\langle qq\right\rangle $. The interplay between the
Nambu-Goldstone (NG) and color superconducting (CSC) phases can be
described in an model-independent way by the Ginzburg-Landau (GL)
theory \cite{Matsuura:2003md,Giannakis:2001wz}. It has been predicted
in the GL theory that a new critical point and smooth crossover arise
from the coupling between chiral and diquark condensates induced by
axial anomaly at the low temperature in the QCD phase diagram \cite{Hatsuda:2006ps,Yamamoto:2007ah,Schmitt:2010pf}.
Such a coupling is also related to the continuity between the quark
and hadronic matter \cite{Schafer:1998ef}. The new critical point
can also be confirmed in the three-flavor NJL model with axial anomaly
\cite{Steiner:2005jm,Abuki:2010jq,Zhang:2011xi}. 

The form of the GL free energy to the sixth order in the fields can
be constrained by the symmetry, 
\begin{equation}
\mathcal{G}=SU(3)_{L}\otimes SU(3)_{R}\otimes U(1)_{B}\otimes U(1)_{A}\otimes SU(3)_{C}
\end{equation}
where subscripts $L$, $R$, $B$, $A$ and $C$ stand for left-handed
flavor, right-handed flavor, baryon number, axial charge, and color
symmetry. The left-handed and right-handed quark fields transformed
under $\mathcal{G}$ as 
\begin{eqnarray}
q_{\lambda} & \rightarrow & e^{-i\lambda\theta_{A}}e^{-i\theta_{B}}V_{\lambda}V_{C}q_{\lambda},
\end{eqnarray}
where $\lambda=L(+),R(-)$, $V_{L/R/C}$ are rotational matrices of
$SU(3)_{L/R/C}$, and $\theta_{A/B}$ are rotational angles of $U(1)_{A/B}$. 

The chiral fields are defined and transformed as 
\begin{eqnarray}
\phi_{ij} & \sim & \left\langle \overline{q}_{R\alpha}^{j}q_{L\alpha}^{i}\right\rangle ,\nonumber \\
\phi & \rightarrow & e^{-2i\theta_{A}}V_{L}\phi V_{R}^{\dagger},
\end{eqnarray}
where $i,j$ denote the flavor indices and $\alpha$ denotes the color
indices. We see that $\phi$ is invariant under $Z(2)_{A}\subset U(1)_{A}$.
Then we have transformation rules for these field quantities, 
\begin{eqnarray}
\phi\phi^{\dagger} & \rightarrow & V_{L}\phi\phi^{\dagger}V_{L}^{\dagger},\nonumber \\
\phi^{\dagger}\phi & \rightarrow & V_{R}\phi\phi^{\dagger}V_{R}^{\dagger},\nonumber \\
\det\phi & \rightarrow & e^{-6i\theta_{A}}\det\phi.
\end{eqnarray}
We see that $\det\phi$ is invariant under $Z(6)_{A}\subset U(1)_{A}$. 

The diquark fields are defined as 
\begin{eqnarray}
\left(d_{\lambda}\right)_{i\alpha} & \sim & \epsilon_{\alpha\beta\gamma}\epsilon_{ijk}\left\langle \left(\overline{q}_{\lambda}\right)_{\beta}^{j}\left(q_{C\lambda}\right)_{\gamma}^{k}\right\rangle =\epsilon_{\alpha\beta\gamma}\epsilon_{ijk}\left\langle \left(\overline{q}_{\lambda}\right)_{\beta}^{j}C\left(\overline{q}_{\lambda}^{T}\right)_{\gamma}^{k}\right\rangle ,\nonumber \\
\left(d_{\lambda}^{\dagger}\right)_{\alpha i} & \sim & \epsilon_{\alpha\beta\gamma}\epsilon_{ijk}\left\langle \left(\overline{q}_{C\lambda}\right)_{\beta}^{j}\left(q_{\lambda}\right)_{\gamma}^{k}\right\rangle =\epsilon_{\alpha\beta\gamma}\epsilon_{ijk}\left\langle \left(q_{\lambda}^{T}\right)_{\beta}^{j}C\left(q_{\lambda}\right)_{\gamma}^{k}\right\rangle ,
\end{eqnarray}
where $C=i\gamma^{2}\gamma_{0}$. Under $\mathcal{G}$, the diquark
fields transform as 
\begin{eqnarray}
d_{\lambda} & \rightarrow & e^{2i\lambda\theta_{A}}e^{2i\theta_{B}}V_{\lambda}d_{\lambda}V_{C}^{T},\nonumber \\
d_{\lambda}^{\dagger} & \rightarrow & e^{-2i\lambda\theta_{A}}e^{-2i\theta_{B}}V_{C}^{*}d_{\lambda}^{\dagger}V_{\lambda}^{\dagger}.
\end{eqnarray}
The color singlet quantities in diquark fields transform as 
\begin{eqnarray}
d_{\lambda}d_{\lambda}^{\dagger} & \rightarrow & V_{\lambda}d_{\lambda}d_{\lambda}^{\dagger}V_{\lambda}^{\dagger},\nonumber \\
d_{\lambda}d_{-\lambda}^{\dagger} & \rightarrow & e^{4i\lambda\theta_{A}}V_{\lambda}d_{\lambda}d_{-\lambda}^{\dagger}V_{-\lambda}^{\dagger},\nonumber \\
\det d_{\lambda} & \rightarrow & e^{6i\lambda\theta_{A}}e^{6i\theta_{B}}\det d_{\lambda}.
\end{eqnarray}
Then the most general form of the GL free energy which is invariant
under the transformation of $\mathcal{G}$ read
\begin{eqnarray}
\Omega(\phi,d_{L},d_{R}) & = & \Omega_{\chi}(\phi)+\Omega_{d}(d_{L},d_{R})+\Omega_{\chi d}(\phi,d_{L},d_{R}),\nonumber \\
\Omega_{\chi}(\phi) & = & \frac{a_{0}}{2}\mathrm{Tr}\phi^{\dagger}\phi+\frac{b_{1}}{4!}(\mathrm{Tr}\phi^{\dagger}\phi)^{2}+\frac{b_{1}}{4!}\mathrm{Tr}(\phi^{\dagger}\phi)^{2}-\frac{c_{0}}{2}(\det\phi+\det\phi^{\dagger})\nonumber \\
\Omega_{d}(d_{L},d_{R}) & = & \alpha_{0}\mathrm{Tr}[d_{L}d_{L}^{\dagger}+d_{R}d_{R}^{\dagger}]+\beta_{1}([\mathrm{Tr}(d_{L}d_{L}^{\dagger})]^{2}+[\mathrm{Tr}(d_{R}d_{R}^{\dagger})]^{2})\nonumber \\
 &  & +\beta_{2}[\mathrm{Tr}(d_{L}d_{L}^{\dagger})^{2}+\mathrm{Tr}(d_{R}d_{R}^{\dagger})^{2}]+\beta_{3}\mathrm{Tr}[(d_{R}d_{L}^{\dagger})(d_{L}d_{R}^{\dagger})]\nonumber \\
 &  & +\beta_{4}\mathrm{Tr}(d_{L}d_{L}^{\dagger})\mathrm{Tr}(d_{R}d_{R}^{\dagger}),\nonumber \\
\Omega_{\chi d}(\phi,d_{L},d_{R}) & = & \gamma_{1}\mathrm{Tr}(d_{R}d_{L}^{\dagger}\phi+d_{L}d_{R}^{\dagger}\phi^{\dagger})+\lambda_{1}\mathrm{Tr}(d_{L}d_{L}^{\dagger}\phi\phi^{\dagger}+d_{R}d_{R}^{\dagger}\phi^{\dagger}\phi)\nonumber \\
 &  & +\lambda_{2}\mathrm{Tr}(d_{L}d_{L}^{\dagger}+d_{R}d_{R}^{\dagger})\cdot\mathrm{Tr}(\phi^{\dagger}\phi)+\lambda_{3}[\det\phi\cdot\mathrm{Tr}(d_{L}d_{R}^{\dagger}\phi^{\dagger})+H.c.].\label{eq:pot-xi-d}
\end{eqnarray}

For three massless flavor case, the most symmetric condensate is in
the form 
\begin{eqnarray}
\phi & = & \mathrm{diag}(\sigma,\sigma,\sigma),\nonumber \\
d_{L} & = & -d_{R}=\mathrm{diag}(d,d,d).
\end{eqnarray}
Then the GL free energy reads,
\begin{eqnarray}
\Omega_{3F} & = & \frac{a}{2}\sigma^{2}-\frac{c}{3}\sigma^{3}+\frac{b}{4}\sigma^{4}+\frac{f}{6}\sigma^{6}\nonumber \\
 &  & +\frac{\alpha}{2}d^{2}+\frac{\beta}{4}d^{4}-\gamma d^{2}\sigma+\lambda d^{2}\sigma^{2},
\end{eqnarray}
where coefficients $a,b,c,f,\alpha\beta\gamma\lambda$ come from those
in Eq. (\ref{eq:pot-xi-d}). Here $a,\alpha$ are two essential parameters
to drive the phase transition. $b$ can change sign with $T,\mu$,
so a positive $\sigma^{6}$ term ($f>0$) is introduced to stabilize
the system for negative $b$. $\beta$ is positive definite from effective
theories and weak-coupling QCD. The $\sigma^{3}$ and $d^{2}\sigma$
terms are from axial anomaly so the coefficients $c$ and $\gamma$
are related and are all positive. $\lambda$ is also positive from
the NJL model and weak-coupling QCD. 

There are four phases: the normal (NOR) phase with $\sigma=0$ and
$d=0$, the CSC phase with $\sigma=0$ and $d\neq0$, the Nambu-Goldstone
(NG) phase with $\sigma\neq0$ and $d=0$, and the coexistence (COE)
phase with $\sigma\neq0$ and $d\neq0$. The phases at a specific
set of parameters can be determined by comparing the global minima
of the free energies $\Omega_{NOR}$, $\Omega_{CSC}$, $\Omega_{NG}$
and $\Omega_{COE}$. From NOR or CSC phase with $\sigma=0$ to the
NG phase with $\sigma\neq0$, there is a first-order transition phase
for the chiral symmetry breaking or restoration. The transition between
the NOR phase with $d=0$ and the CSC phase $d\neq0$ is a second-order
one with a discontinuity of $\frac{\partial s}{\partial T}$.

\section{NJL model for three-flavor dense quark matter with axial anomaly}

The effective Lagrangian for three-flavor dense quark matter with
axial anomaly can also be derived from the NJL model \cite{Abuki:2010jq,Basler:2010xy}.
The NJL Lagrangian reads 
\begin{eqnarray}
\mathcal{L} & = & \bar{q}(i\gamma^{\mu}\partial_{\mu}-m_{q}+\mu\gamma^{0})q+\mathcal{L}^{(4)}+\mathcal{L}^{(6)},
\end{eqnarray}
where $\mathcal{L}^{(4)}=\mathcal{L}_{\chi}^{(4)}+\mathcal{L}_{d}^{(4)}$.
The four-fermion interaction term reads
\begin{eqnarray}
\mathcal{L}_{\chi}^{(4)} & = & G\sum_{f=0}^{N_{\text{f}}^{2}-1}\left[(\bar{q}\tau_{f}q)^{2}+(\bar{q}i\gamma_{5}\tau_{f}q)^{2}\right],\nonumber \\
\mathcal{L}_{\mathrm{d}}^{(4)} & = & H\sum_{A,A'=2,5,7}\left[(\bar{q}i\gamma_{5}\tau_{A}\lambda_{A'}q_{C})(\bar{q}_{C}i\gamma_{5}\tau_{A}\lambda_{A'}q)+(\bar{q}\tau_{A}\lambda_{A'}q_{C})(\bar{q}_{C}\tau_{A}\lambda_{A'}q)\right],
\end{eqnarray}
where $q_{C}=C\bar{q}^{T}$ and $\bar{q}_{C}=q^{T}C$ with $C=i\gamma^{2}\gamma_{0}$.
The six-fermion interaction term is $\mathcal{L}^{(6)}=\mathcal{L}_{\mathrm{H}}^{(6)}+\mathcal{L}_{\text{mixing}}^{(6)}$,
where $\mathcal{L}_{\mathrm{H}}^{(6)}$ is the standard 't Hooft term,
\begin{eqnarray}
\mathcal{L}_{\mathrm{H}}^{(6)} & = & -K\left\{ \det_{\text{flaovr}}[\bar{q}(1+\gamma_{5})q]+\det_{\text{flavor}}[\bar{q}(1-\gamma_{5})q]\right\} \nonumber \\
 & = & -K\varepsilon_{ijk}\left\{ [\bar{q}_{1}(1+\gamma_{5})q_{i}][\bar{q}_{2}(1+\gamma_{5})q_{j}][\bar{q}_{3}(1+\gamma_{5})q_{k}]\right.\nonumber \\
 &  & +\left.[\bar{q}_{1}(1-\gamma_{5})q_{i}][\bar{q}_{2}(1-\gamma_{5})q_{j}][\bar{q}_{3}(1-\gamma_{5})q_{k}]\right\} ,
\end{eqnarray}
where $q_{1,2,3}$ denote $u$, $d$ and $s$ respectively. Here we
introduce a mixing term $\mathcal{L}_{\text{mixing}}^{(6)}$ for the
coupling between the chiral and diquark condensates, 
\begin{eqnarray}
\mathcal{L}_{\text{mixing}}^{(6)} & = & \frac{1}{8}K'\sum_{A,B,A'=2,5,7}\left\{ [\bar{q}P_{BA}(1+\gamma_{5})q][\bar{q}\tau_{B}\lambda_{A'}(1+\gamma_{5})q^{\text{C}}][\bar{q}^{\text{C}}\tau_{A}\lambda_{A'}(1+\gamma_{5})q]\right.\nonumber \\
 &  & +\left.[\bar{q}\tau_{A}\lambda_{A'}(1-\gamma_{5})q^{\text{C}}][\bar{q}^{\text{C}}\tau_{B}\lambda_{A'}(1-\gamma_{5})q][\bar{q}P_{AB}(1-\gamma_{5})q]\right\} ,
\end{eqnarray}
where $P_{AB}$ are matrices in flavor space,
\begin{eqnarray}
P_{22} & = & \left(\begin{array}{ccc}
0 & 0 & 0\\
0 & 0 & 0\\
0 & 0 & 1
\end{array}\right),\; P_{55}=\left(\begin{array}{ccc}
0 & 0 & 0\\
0 & 1 & 0\\
0 & 0 & 0
\end{array}\right),\; P_{77}=\left(\begin{array}{ccc}
1 & 0 & 0\\
0 & 0 & 0\\
0 & 0 & 0
\end{array}\right),\nonumber \\
P_{25} & = & P_{52}^{\dagger}=\left(\begin{array}{ccc}
0 & 0 & 0\\
0 & 0 & 1\\
0 & 0 & 0
\end{array}\right),\; P_{57}=P_{75}^{\dagger}=\left(\begin{array}{ccc}
0 & 1 & 0\\
0 & 0 & 0\\
0 & 0 & 0
\end{array}\right),\; P_{72}=P_{27}^{\dagger}=\left(\begin{array}{ccc}
0 & 0 & 0\\
0 & 0 & 0\\
1 & 0 & 0
\end{array}\right).
\end{eqnarray}

Since we can decompose the chiral and diquark condensates in the following
form 
\begin{eqnarray}
\Phi & = & -q\bar{q}=(\phi_{\alpha}^{\text{color}}\lambda_{\alpha})\otimes(\phi_{f}^{\text{flavor}}\tau_{f})\otimes(\phi_{0}^{\text{spin}}+\phi_{5}^{\text{spin}}\gamma_{5}+\phi_{\mu}^{\text{spin}}\gamma^{\mu}+\phi_{\mu5}^{\text{spin}}\gamma^{\mu}\gamma_{5}+\phi_{\mu\nu}^{\text{spin}}S^{\mu\nu}),\nonumber \\
D & = & -q\bar{q}^{\text{C}}=(d_{\alpha}^{\text{color}}\lambda_{\alpha})\otimes(d_{f}^{\text{flavor}}\tau_{f})\otimes(d_{0}^{\text{spin}}+d_{5}^{\text{spin}}\gamma_{5}+d_{\mu}^{\text{spin}}\gamma^{\mu}+d_{\mu5}^{\text{spin}}\gamma^{\mu}\gamma_{5}+d_{\mu\nu}^{\text{spin}}S^{\mu\nu}).
\end{eqnarray}
Then we have 
\begin{eqnarray}
\bar{q}\tau_{f}q & = & -\text{Tr}\left[q\bar{q}\tau_{f}\right]=2\phi_{0}^{f},\nonumber \\
\bar{q}\gamma_{5}\tau_{f}q & = & -\text{Tr}\left[q\bar{q}\gamma_{5}\tau_{f}\right]=2\phi_{5}^{f},\nonumber \\
\bar{q}^{\text{C}}i\gamma_{5}\tau_{A}\lambda_{A'}q & = & -\text{Tr}\left[q\bar{q}^{\text{C}}i\gamma_{5}\tau_{A}\lambda_{A'}\right]=2id_{5}^{A'A},\nonumber \\
\bar{q}i\gamma_{5}\tau_{A}\lambda_{A'}q^{\text{C}} & = & (\bar{q}^{\text{C}}i\gamma_{5}\tau_{A}\lambda_{A'}q)^{\dagger}=-2i(d_{5}^{A'A})^{\dagger},\nonumber \\
\bar{q}^{\text{C}}\tau_{A}\lambda_{A'}q & = & -\text{Tr}\left[q\bar{q}^{\text{C}}\tau_{A}\lambda_{A'}\right]=2d_{0}^{A'A},\nonumber \\
\bar{q}\tau_{A}\lambda_{A'}q^{\text{C}} & = & (\bar{q}^{\text{C}}\tau_{A}\lambda_{A'}q)^{\dagger}=2(d_{0}^{A'A})^{\dagger}.
\end{eqnarray}

In the mean-field approximation, considering the CFL channel and dropping
the $0^{-}$ state, we have 
\begin{eqnarray}
\langle\phi_{0}^{f}\rangle\tau_{f} & = & \sigma,\;\langle\phi_{5}^{f}\rangle\tau_{f}=0,\nonumber \\
\langle d_{5}^{A'A}\rangle & = & \frac{1}{2}d\delta_{A'A},\;\langle d_{0}^{A'A}\rangle=0.
\end{eqnarray}
Now the Lagrangian becomes
\begin{eqnarray}
\mathcal{L}_{\chi}^{(4)} & \rightarrow & 4G\sigma\bar{q}q-6G\sigma^{2},\nonumber \\
\mathcal{L}_{\text{d}}^{(4)} & \rightarrow & H\left[d^{*}(\bar{q}^{\text{C}}\gamma_{5}\tau_{A}\lambda_{A}q)-(\bar{q}\gamma_{5}\tau_{A}\lambda_{A}q^{\text{C}})d\right]-3H|d|^{2},\nonumber \\
\mathcal{L}^{(6)} & \rightarrow & -2K\sigma^{2}\bar{q}q+4K\sigma^{3},\nonumber \\
\mathcal{L}_{\text{mixing}}^{(6)} & \rightarrow & -\frac{K'}{4}|d|^{2}\bar{q}q-\frac{K'}{4}\sigma\left[d^{*}(\bar{q}^{\text{C}}\tau_{A}\lambda_{A}\gamma_{5}q)-(\bar{q}\tau_{A}\lambda_{A}\gamma_{5}q^{\text{C}})d\right]+\frac{3K'}{2}\sigma|d|^{2}.
\end{eqnarray}
Thus the Lagrangian can be expressed in the Nambu-Gorkov basis $\Psi=\frac{1}{\sqrt{2}}(q,q^{\text{C}})^{\text{T}}$,
\begin{eqnarray}
\mathcal{L} & = & \frac{1}{2}\bar{\Psi}S^{-1}\Psi-U,
\end{eqnarray}
where 
\begin{eqnarray}
S^{-1}(p) & = & \left(\begin{array}{cc}
p_{\mu}\gamma^{\mu}+\mu\gamma^{0}-M & \Delta\gamma_{5}\tau_{A}\lambda_{A}\\
-\Delta^{*}\gamma_{5}\tau_{A}\lambda_{A} & p_{\mu}\gamma^{\mu}-\mu\gamma^{0}-M
\end{array}\right),\nonumber \\
U & = & 6G\sigma^{2}+3H|d|^{2}-4K\sigma^{3}-\frac{3}{2}K'\sigma|d|^{2}.
\end{eqnarray}
with
\begin{eqnarray}
M & = & m_{\mathrm{q}}-4G+2K\sigma^{2}+\frac{1}{4}K'|d|^{2},\nonumber \\
\Delta & = & \frac{K'}{2}\sigma d-2H.
\end{eqnarray}

Thus we obtain the thermodynamic potential as
\begin{eqnarray}
\Omega & = & -\int\frac{d^{3}\mathbf{p}}{(2\pi)^{3}}\sum_{\pm}\left\{ \left[16T\ln\left(1+\exp(-\frac{\omega_{8}^{\pm}}{T})\right)+8\omega_{8}^{\pm}\right]\right.\nonumber \\
 &  & +\left.\left[2T\ln\left(1+\exp(-\frac{\omega_{1}^{\pm}}{T})\right)+\omega_{1}^{\pm}\right]\right\} +U(\sigma,d).
\end{eqnarray}
where $\omega_{1}^{\pm}=\sqrt{(E_{\mathbf{p}}\pm\mu)^{2}+|\Delta|^{2}}$,
$\omega_{8}^{\pm}=\sqrt{(E_{\mathbf{p}}\pm\mu)^{2}+4|\Delta|^{2}}$
with $E_{\mathbf{p}}=\sqrt{\mathbf{p}^{2}+M^{2}(\sigma,d)}$. Thus
the gap equations are
\begin{eqnarray}
\frac{\partial\Omega}{\partial\sigma} & = & -\int\frac{d^{3}\mathbf{p}}{(2\pi)^{3}}\sum_{\pm}\left\{ \left[-16f(\omega_{8}^{\pm})+8\right]\frac{\partial\omega_{8}^{\pm}}{\partial\sigma}+\left[-2f(\omega_{1}^{\pm})+1\right]\frac{\partial\omega_{1}^{\pm}}{\partial\sigma}\right\} +\frac{\partial U}{\partial\sigma}=0,\nonumber \\
\frac{\partial\Omega}{\partial d} & = & -\int\frac{d^{3}\mathbf{p}}{(2\pi)^{3}}\sum_{\pm}\left\{ \left[-16f(\omega_{8}^{\pm})+8\right]\frac{\partial\omega_{8}^{\pm}}{\partial d}+\left[-2f(\omega_{1}^{\pm})+1\right]\frac{\partial\omega_{1}^{\pm}}{\partial d}\right\} +\frac{\partial U}{\partial d}=0.
\end{eqnarray}

The gap equations without mixing 't Hooft term has been solved with
parameters in Table \ref{tab:parameters} \cite{Abuki:2010jq}. There
are three phases in the phase diagram: the NOR phase with $\sigma=0$,
$d=0$; the NG phase with $\sigma\neq0$, $d=0$; the CSC phase with
$\sigma=0$, $d\neq0$. If quarks are massive $m_{\mathrm{q}}\neq0$,
there is a critical point at high temperature which is called the
Asakawa-Yazaki point, which is the endpoint of the first-order phase
transition line in the phase diagram leading to a crossover. 

\begin{table}
\caption{The parameters for the gap equations without 't Hooft term used in
Ref. \cite{Abuki:2010jq}. \label{tab:parameters}}

\centering{}%
\begin{tabular}{|c|c||c||c||c||c||c|}
\hline 
 & $m_{\mathrm{q}}$ {[}MeV{]}  & $G\Lambda^{2}$  & $H\Lambda^{2}$  & $K\Lambda^{5}$  & $M$ {[}MeV{]}  & $\sigma^{1/3}$ {[}MeV{]}\tabularnewline
\hline 
\hline 
I  & 0  & 1.926  & 1.74  & 12.36  & 355.2  & -240.4\tabularnewline
\hline 
\hline 
II  & 5.5  & 1.918  & 1.74  & 12.36  & 367.6  & -241.9\tabularnewline
\hline 
\end{tabular}
\end{table}

When the mixing term is introduced with a strong enough chiral-diquark
interplay $K'=4.2K$, a new critical point at low temperature will
emerge just as predicted in the Ginzburg-Landau approach \cite{Hatsuda:2006ps,Yamamoto:2007ah}.
The interaction between chiral and diquark condensates weakens the
chiral symmetry spontaneous breaking and leads to the COE phase with
$\sigma\neq0$, $d\neq0$ at low temperature. As a consequence, the
first-order phase transition between the NG and CSC phase becomes
a crossover. This is a new critical endpoint at the other end of the
first-order phase transition line. 

At the same time, the mixing term also induces a BEC-BCS crossover
\cite{Nishida:2005ds,Abuki:2006dv,Deng:2006ed,Sun:2007fc,He:2007yj,Kitazawa:2007zs}.
Here a new criterion in the dispersion relation is used to define
a BEC state. For $\mu>M$, there will be non-vanishing momentum $p=\sqrt{\mu^{2}-M^{2}}$
to give the minimum energy and it is a BCS state. But for $\mu<M$,
the minimum energy has to at $p=0$ which means the system is in a
BEC state. Since we have the COE phase with non-vanishing chiral condensate,
the BEC state occurs in phase diagram on the left side of the curve
$\mu=M(\mu,T)$. Physically, the BEC state can also be explained as
a compound particle including two strong coupling quarks because the
diquark channel is strengthened by the mixing term just as $H'=H+\frac{1}{4}K'\sigma$
for sufficiently large $K'$. 

However, it was pointed out that the 2SC is present if $K'$ is sufficiently
strong \cite{Basler:2010xy}. This is due to that the axial anomaly
induces a mutual amplifi{}cation of the strange chiral condensate
and the non-strange diquark condensate. As the consequence, the critical
point found in Ref. \cite{Abuki:2010jq} only survives for very narrow
parameter space of $K'$ otherwise most parts are covered by the 2SC
phase \cite{Basler:2010xy}.

\section{Ginzburg-Landau approach to collective modes in spin-one CSC}

The Ginzburg-Landau approach can be used to study the collective modes
in spin-one color superconductors \cite{Brauner:2008ma,Brauner:2009df,Pang:2010wk}.
The spin-one color superconductor involves pairing of quarks of same
flavor. The diquark condensate or the order parameter $\Delta$ is
then a color anti-triplet and spin triplet, so it is a $3\times3$
complex matrix and transform as

\begin{eqnarray}
\Delta & \rightarrow & U\Delta R\label{eq:transformation}
\end{eqnarray}
where $U=\exp(i\theta_{a}\lambda_{a})\in\text{U}(3)=\text{SU}(3)_{\text{c}}\times\text{U}(1)_{\text{B}}$
and $R=\exp(i\alpha_{i}J_{i})\in SO(3)$ are transformation matrices.
Here $\lambda_{a}$ are eight Gell-Mann matrices and $\lambda_{0}$
is normalized unit matrix, $(J_{i})_{jk}=-i\epsilon_{ijk}$ are generators
of $\text{SO}(3)_{\text{R}}$, $\theta_{a}(a=0,...,8)$ and $\alpha_{i}(i=1,2,3)$
are rotation angles in $\text{U}(3)_{\text{L}}$ and $\text{SO}(3)_{\text{R}}$
group space. There are 18 real parameters in $\Delta$, among which
12 parameters are carried by the $\text{U}(3)_{\text{L}}\times\text{SO}(3)_{\text{R}}$
transformation making a 12-dimensional degenerate vacuum manifold.
Then $\Delta$can be parametrized by the remaining 6 real parameters
which characterize different vacuum states,
\begin{eqnarray}
\Delta & = & \left(\begin{array}{ccc}
\Delta_{1} & i\delta_{3} & -i\delta_{2}\\
-i\delta_{3} & \Delta_{2} & i\delta_{1}\\
i\delta_{2} & -i\delta_{1} & \Delta_{3}
\end{array}\right).\label{eq:Delta_parameterization}
\end{eqnarray}

\subsection{Ginzburg-Landau free energy and ground states}

Up to fourth order in $\Delta$ and two derivatives, the most general
$\text{U}(3)_{\text{L}}\times\text{SO}(3)_{\text{R}}$ and parity
invariant Ginzburg-Landau free energy density functional can be written
as
\begin{eqnarray}
\mathcal{F}[\Delta] & = & a_{1}\text{Tr}(\partial_{i}\Delta\partial_{i}\Delta^{\dagger})+a_{2}(\partial_{i}\Delta_{ai})(\partial_{j}\Delta_{aj}^{*})+b\text{Tr}(\Delta\Delta^{\dagger})\nonumber \\
 &  & +d_{1}[\text{Tr}(\Delta\Delta^{\dagger})]^{2}+d_{2}\text{Tr}(\Delta\Delta^{\dagger}\Delta\Delta^{\dagger})+d_{3}\text{Tr}[\Delta\Delta^{\text{T}}(\Delta\Delta^{\text{T}})^{\dagger}].\label{eq:Free_energy}
\end{eqnarray}
The time-dependent GL functional, or Lagrangian, is then written as
\begin{eqnarray}
\mathcal{L} & = & ic_{1}\text{Tr}[\Delta^{\dagger}\partial_{0}\Delta]+c_{2}\text{Tr}[(\partial_{0}\Delta^{\dagger})(\partial_{0}\Delta)]-\mathcal{F}[\Delta],\label{eq:Lagrangian from GL}
\end{eqnarray}
The ground state is found by minimizing $\mathcal{F}[\Delta]$. The
phase structure, or orientation in the field space, of $\Delta_{0}$
depends on $d_{2}$ and $d_{3}$. Here, only the following four ground
states represented by the following four matrices occupy a part of
the phase diagram,
\begin{eqnarray}
\Delta_{\text{CSL}}=\frac{1}{\sqrt{3}}\left(\begin{array}{ccc}
1 & 0 & 0\\
0 & 1 & 0\\
0 & 0 & 1
\end{array}\right), & \Delta_{\text{poloar}}=\left(\begin{array}{ccc}
0 & 0 & 0\\
0 & 0 & 0\\
0 & 0 & 1
\end{array}\right),\nonumber \\
\Delta_{\text{A}}=\frac{1}{\sqrt{2}}\left(\begin{array}{ccc}
0 & 0 & 0\\
0 & 0 & 0\\
1 & i & 0
\end{array}\right), & \Delta_{\epsilon}=\left(\begin{array}{ccc}
0 & 0 & 0\\
0 & 0 & \beta\\
\alpha & i\alpha & 0
\end{array}\right),\label{eq:4_phase}
\end{eqnarray}
where $\alpha=\sqrt{(d_{2}+d_{3})/[2(2d_{2}+d_{3})]}$ and $\beta=\sqrt{d_{2}/(2d_{2}+d_{3})}$.
The pattern of spontaneous symmetry breaking determines the low-energy
spectrum of the system, i.e., the NG bosons. While some of the NG
bosons are associated with the generators of the color $\text{SU}(3)_{\text{c}}$
group and are thus eventually absorbed in gluons via the Higgs-Anderson
mechanism, those stemming from spontaneous breaking of baryon number
or rotation symmetry remain in the spectrum as physical soft modes.
As we will now see, some of the phases exhibit the unusual type-II
NG bosons, in accordance with general properties of spontaneously
broken symmetries in quantum many-body systems.

\subsection{CSL phase}

When $d_{2}+d_{3}>0$ and $d_{2}>d_{3}$, the ground state is the
CSL phase in which the spin and color are coupled in the pairing,
so the symmetry breaking pattern is $\text{U}(3)_{\text{L}}\times\text{SO}(3)_{\text{R}}\rightarrow\text{SO}(3)_{\text{V}}$.
There are 9 broken generators leading to 9 NG bosons as follows,
\begin{itemize}
\item $\lambda_{0}\otimes\mathbbm1$. Type-I NG singlet, $E^{2}\sim(a_{1}+a_{2})k^{2}$.
\item $\sqrt{\frac{1}{2}}(\lambda_{7}\otimes\mathbbm1-\mathbbm1\otimes J_{1}),\sqrt{\frac{1}{2}}(-\lambda_{5}\otimes\mathbbm1-\mathbbm1\otimes J_{2}),\sqrt{\frac{1}{2}}(\lambda_{2}\otimes\mathbbm1-\mathbbm1\otimes J_{3})$.
Type-I NG triplet, $E^{2}\sim(a_{1}+a_{2})k^{2}$.
\item $\lambda_{\alpha}\otimes\mathbbm1,\alpha=1,3,4,6,8$. Type-I NG 5-plet,
$E^{2}\sim(a_{1}+a_{2})k^{2}$.
\end{itemize}

\subsection{Polar phase}

When $d_{3}<0$ and $d_{2}+d_{3}<0$, the ground state is the polar
phase. The symmetry breaking pattern is $\text{U}(3)_{\text{L}}\times\text{SO}(3)_{\text{R}}\rightarrow\text{U}(2)_{\text{L}}\times\text{SO}(2)_{\text{R}}$.
There are 7 broken generators which, however, give rise only to 5
NG bosons, organized in the following multiplets,
\begin{itemize}
\item $\sqrt{2}\mathcal{P}_{3}\otimes\mathbbm1$, where $\mathcal{P}_{3}=\frac{1}{\sqrt{6}}(\lambda_{0}-\sqrt{2}\lambda_{8})=\text{diag}(0,0,1)$
is the projector onto the third color. Type-I NG singlet, $E^{2}\sim a_{1}k_{\perp}^{2}+(a_{1}+a_{2})k_{3}^{2}$.
\item $\mathbbm1\otimes J_{j},j=1,2$. Type-I NG doublet, $E^{2}\sim(a_{1}+a_{2})k_{\perp}^{2}+a_{1}k_{3}^{2}$.
\item $\lambda_{\alpha}\otimes\mathbbm1,\alpha=4,5,6,7$. Type-II NG doublet,
$E^{2}\sim a_{1}^{2}k_{\perp}^{4}+(a_{1}+a_{2})^{2}k_{3}^{4}$.
\end{itemize}
The presence of type-II NG bosons is due to nonzero color density
of the polar ground state.

\subsection{A-phase}

When $d_{3}>0$ and $d_{2}<0$, the ground state is the A-phase. The
symmetry breaking pattern is $\text{U}(3)_{\text{L}}\times\text{SO}(3)_{\text{R}}\rightarrow\text{U}(2)_{\text{L}}\times\text{SO}(2)_{\text{V}}$.
Unlike in the polar phase, the diquark spin is now circularly polarized.
Among 7 broken generators, there is only one giving rise to a type-I
NG mode,
\begin{itemize}
\item $\sqrt{\frac{2}{3}}(\mathcal{P}_{3}\otimes\mathbbm1+\mathbbm1\otimes J_{3})$.
Type-I NG singlet, $E^{2}\sim(a_{1}+a_{2})k_{\perp}^{2}+a_{1}k_{3}^{2}$.
\end{itemize}
The rest 6 generators produce only 3 type-II NG bosons due to non-zero
color and spin density of the A-phase vacuum,
\begin{itemize}
\item $\lambda_{\alpha}\otimes\mathbbm1,\alpha=4,5,6,7$. Type-II NG double,
$E^{2}\sim(a_{1}+a_{2})^{2}k_{\perp}^{4}+a_{1}^{2}k_{3}^{4}$.
\item $\mathbbm1\otimes\frac{1}{\sqrt{2}}(J_{1}\pm iJ_{2})$. Type-II NG
singlet, $E^{2}\sim a_{1}^{2}k_{\perp}^{4}+(a_{1}+a_{2})^{2}k_{3}^{4}$.
\end{itemize}

\subsection{$\epsilon$-phase}

When $d_{3}>d_{2}>0$, the ground state is the $\epsilon$-phase.
The symmetry breaking pattern is $\text{U}(3)_{\text{L}}\times\text{SO}(3)_{\text{R}}\rightarrow\text{U}(1)_{L}\times\text{SO}(2)_{\text{V}}$.
The spin of the second diquark color is longitudinal polarized, while
that of third color is circularly polarized. Out of the 10 broken
generators only two correspond to type-I NG modes:
\begin{itemize}
\item $\sqrt{2}\mathcal{P}_{2}\otimes\mathbbm1$, where $\mathcal{P}_{2}=\text{diag}(0,1,0)$
is the projector onto the second color. Type-I NG singlet, $E^{2}\sim a_{1}k_{\perp}^{2}+(a_{1}+a_{2})k_{3}^{2}$.
\item $\sqrt{\frac{2}{3}}(\mathcal{P}_{3}\otimes\mathbbm1+\mathbbm1\otimes J_{3})$.
Type-I NG singlet, $E^{2}\sim(a_{1}+a_{2})k_{\perp}^{2}+a_{1}k_{3}^{2}$.
\end{itemize}
The remaining 8 generators give rise to 4 type-II NG modes due to
nonzero color and spin density of the $\epsilon$ vacuum,
\begin{itemize}
\item $\frac{1}{\sqrt{2}}(\lambda_{1}\pm i\lambda_{2})\otimes\mathbbm1$.
Type-II NG singlet, $E^{2}\sim a_{1}^{2}k_{\perp}^{4}+(a_{1}+a_{2})^{2}k_{3}^{4}$.
\item $\frac{1}{\sqrt{2}}(\lambda_{4,6}\pm i\lambda_{5,7})\otimes\mathbbm1$.
Type-II NG singlet, $E^{2}\sim(a_{1}+a_{2})^{2}k_{\perp}^{4}+a_{1}^{2}k_{3}^{4}$.
\item $\mathbbm1\otimes\frac{1}{\sqrt{2}}(J_{1}\pm iJ_{2})$. Type-II NG
singlet, $E^{2}\sim a_{1}^{2}k_{\perp}^{4}+(a_{1}+a_{2})^{2}k_{3}^{4}$. 
\end{itemize}
In summary the low-energy physics of spin-one CSC is analyzed in terms
of the NG excitations within the Ginzburg-Landau theory. The four
phases that appear in the phase diagram possess different NG modes
of the spontaneously broken color, baryon number and rotational symmetry.
Those stemming from the color symmetry will eventually be absorbed
into gluons, making them massive by the Anderson-Higgs mechanism.
The other NG bosons will remain in the spectrum as physical soft modes.
Unlike in all the other phases, in the isotropic CSL phase all quarks
can be gapped so that the NG bosons are the only truly gapless states
in the spectrum.

\section{Summary}

We give an overview on recent progress in quark pairings in dense
quark matter. These progress include the BCS-BEC crossover in the
CSC with and without external magnetic field, baryon formation and
dissociation in quark/nuclear matter, Ginzburg-Landau effective theory
on dense quark/nuclear matter with anomaly, and collective and Nambu-Goldstone
modes in spin-one CSC. 

The boson-fermion model in the cold atom system can be extended to
relativistic case to describe the relativistic BEC-BCS crossover in
the CSC. In a charge conserved system, by tuning the bare boson mass
the effective coupling between fermions changes, and the fermion number
fraction and fermionic chemical potential as functions of the effective
coupling also change indicating a crossover between BCS and BEC states.
The pseudo-gap and effects of thermal bosons can also be systematically
studied within the CJT formalism. The BCS-BEC crossover can also be
dealt with in a pure fermionic model in which the fermion scattering
length can be derived in comparison with the result at the low momentum
limit for the T-matrix. The bosonic degrees of freedom are introduced
by a bosonization procedure. In dense quark matter with fixed total
baryonic number density, tuning the coupling constant along the CSC
boundary, the baryon number fraction of free quarks and thermal diquarks
vary with the scattering length, indicating a typical BCS-BEC crossover.
Both the double channel (boson-fermion model) and single channel (pure
fermionic model) models can describe BCS-BEC crossover in relativistic
case equally well. The strong magnetic fields which exist in pulsars
and non-central heavy ion collisions can tune the BCS-BEC crossover
in a charged fermion system, but the origin is different from that
in a cold atom system where the magnetic field is used to change the
fermion coupling via Feshbach resonance. A strong magnetic field up
to the energy scale of QCD have impact on the fermion energy dispersion.
Sitting in a BEC state, varying the magnetic field can induce an oscillation
of the fermion number fractions which makes the system be in the BCS-
and BEC-dominant region. As the magnetic field becomes even stronger
all fermions are pressed to the lowest Landau-level, hence the system
settles down in the BCS regime.

Baryons can also be regarded as a bound state of quark-diquark coupling
in the NJL-type model, which is a simplification for the Faddeev equation.
The diquark spectral density can be obtained by solving the DSE, with
which the full diquark propagator can be written in a spectral density
form and the summation over Matsubara frequencies can be analytically
done. Then coupling the diquark with a quark within DSE, the baryon
spectral density can be calculated in different phases. The formation
and dissociation properties of baryon can then be investigated. In
previous studies, diquarks are always assumed to be stable or as quasi-particles.
This is not necessarily true because a baryon can also be a stable
bound state of a quark and an unstable diquark, which bears some similarities
to Borromean state in nuclear physics. The saturation of nuclear matter
is still a challenge in current model. First one should include the
vector meson channel to obtain a large enough baryonic number density.
On the other hand, an effective confinement should also be included.
In the normal quark NJL model, the absence of confinement is obvious
since the quark mass is momentum independent. The nonlocal extension
of the NJL model may provide a mechanism to include the confinement,
in which the constituent quark mass is a momentum function. The dynamic
quark mass and renormalization factor calculated within the nonlocal
NJL model have a great consistency with the DSE analysis and the lattice
results. The other advantage of the extended model is that in the
bubble diagrams there always exist a Gaussian form factor which makes
the integrals automatically converge.

The phase diagram at low temperatures near the confinement and CSC
boundary is still not clear. Recently with the three-flavor NJL model
including the axial anomaly, a low temperature critical point near
the baryon chemical potential axis is found in a Ginzburg-Landau analysis
as a result of the interplay between the chiral and diquark condensates.
One can derive the Ginzburg-Landau effective potential with the axial
anomaly term within the NJL model by introducing the six-fermion mixing
term of the diquark-diquark-quark-anti-quark coupling. When coupling
constant of the mixing term is strong enough, a new critical point
at low temperature can emerge just as predicted in the Ginzburg-Landau
approach. But it has been argued that the 2SC is present instead of
the critical point if the coupling of the mixing term is strong. This
is due to that the axial anomaly induces a mutual amplifi{}cation
of the strange chiral condensate and the non-strange diquark condensate. 

The low-energy physics of spin-one CSC is analyzed in terms of the
NG excitations within the Ginzburg-Landau effective theory. The NG
modes of the spontaneously broken color, baryon number and rotational
symmetry are analyzed for four typical phases. Those stemming from
the color symmetry will eventually be absorbed into gluons, making
them massive by the Anderson-Higgs mechanism. The other NG bosons
will remain in the spectrum as physical soft modes. Unlike in all
the other phases, in the isotropic CSL phase all quarks can be gapped
so that the NG bosons are the only truly gapless states in the spectrum. 

Acknowledgment: QW thanks D. Blaschke, T. Brauner, V. de la Incera,
and H.-C. Ren for critically reading the manuscript. QW was supported
by the Kavli Institute for Theoretical Physics China during the program
{}``AdS/CFT and Novel Approaches to Hadron and Heavy Ion Physics'',
and is supported in part by the National Natural Science Foundation
of China under grant 10735040. 

\bibliographystyle{apsrev}
\bibliography{ref-wang}

\end{document}